\documentclass[prd,nofootinbib,preprint, 10 pt]{revtex4}
\usepackage[a4paper, total={7.0in, 10in}]{geometry}
\usepackage{mathrsfs}
\usepackage{amsmath}
\usepackage{amssymb}
\usepackage{epsfig}
\usepackage{epstopdf}
\usepackage{graphicx}
\usepackage{booktabs}
\usepackage{float}
\usepackage{amsmath}
\usepackage{color}
\usepackage{multirow}
\usepackage{array}
\newcolumntype{M}[1]{>{\centering\arraybackslash}m{#1}}
\makeatletter
\def\mathcolor#1#{\@mathcolor{#1}}
\def\@mathcolor#1#2#3{%
  \protect\leavevmode
  \begingroup\color#1{#2}#3\endgroup
}
\makeatother
\usepackage{multirow}
\usepackage{bm}
\usepackage{appendix}
\usepackage{slashed}
\usepackage{cancel}
\PassOptionsToPackage{unicode}{hyperref}
\PassOptionsToPackage{naturalnames}{hyperref}
\usepackage[colorlinks=true,urlcolor=blue,citecolor=blue,linkcolor=red,linktocpage=true,pdfproducer=medialab]{hyperref}
\usepackage{morefloats}
\usepackage{siunitx}

\begin{document}
\title{Lepton polarization dependent angular observables and the polarization asymmetries in the four-fold $\Lambda_b \rightarrow \Lambda(\rightarrow N \pi)  \ell^+\ell^-$ decay}
\author{Rana Khan$^{1, 2}$\footnote[1]{raanaakhan97@gmail.com}, Faisal Munir Bhutta$^1$\footnote[2]{faisal.munir@sns.nust.edu.pk}, Ishtiaq Ahmed$^2$\footnote[3]{ishtiaq@ncp.edu.pk}, M. Jamil Aslam$^{3}$\footnote[4]{jamil@qau.edu.pk}\vspace{0.3cm}} 

\affiliation{
$^1$ School of Natural Sciences, Department of 
			Physics and Astronomy, National University of Sciences and 
			Technology (NUST), Sector H-12, Islamabad, Pakistan\\
$^2$ National Centre for Physics, Quaid-i-Azam University Campus, 45320, Islamabad, Pakistan\\ 
$^3$ Physics Department, Quaid-i-Azam University, 45320, Islamabad, Pakistan\vspace{0.5cm}} 
\begin{abstract}
\vspace{0.3cm}
The rare decays mediated by flavor-changing neutral current processes, such as 
$b \to s \ell^{+}\ell^{-}$, provide powerful probes of the Standard Model and potential 
windows into new physics. Particularly, the angular observables in these exclusive decays are 
valuable because of their sensitivity to short-distance dynamics and their reduced dependence on 
hadronic uncertainties, which mainly arise from form factors. In this work, we analyze the 
$\Lambda_b \to \Lambda(\to N\pi)\ell^{+}\ell^{-}$ (with $N\pi=\{p\pi^-,n\pi^0\}$) decay with polarized final-state lepton 
and derive the corresponding four-fold differential decay distributions. For the longitudinal, 
normal, and transverse polarization states, we systematically identify the additional angular 
coefficients that emerge relative to the unpolarized case. We find that the longitudinal polarization 
preserves the structure of the unpolarized distribution, while the normal and transverse polarizations 
introduce some new additional angular coefficients. The analytical expressions of all polarized and unpolarized angular coefficients are explicitly derived in terms of the helicity and transversity amplitudes. To compare the variation in the polarized and unpolarized angular observables, we have plotted them against
the square of the momentum transfer $q^2$. Additionally, the Standard Model predictions of the polarization asymmetry observables are provided and their sensitivity to new physics is explored under different new physics scenarios. The obtained results, in the current study, for longitudinal and transverse polarization cases, provide a baseline for the lepton polarization dependent observables, which may serve as sensitive probes to test the Standard Model in these decays.
\end{abstract}
\maketitle

\section{Introduction}
The Standard Model (SM) of electroweak interactions provides an excellent framework for fundamental interactions. Despite all its successes, it cannot be regarded as a complete theory because of numerous reasons, \textit{e.g.}, it lacks in explaining the observed matter-antimatter asymmetry of the universe. Further, the particle spectrum of the SM does not have any candidate of dark matter, and there are a number of unpredicted parameters in the SM inserted by hand or extracted from the experimental data. To overcome these challenges, one needs to look for the signatures of possible physics beyond the SM (BSM), also known as the new physics (NP). Experimental searches for BSM physics are broadly categorized into two approaches: direct searches, which attempt to detect new particles produced in high-energy collisions at the Large Hadron Collider (LHC) \cite{ATLAS:2012yve, CMS:2012qbp, ATLAS:2015yey}. The LHC has yet to observe new particles, and their absence may be attributed to the possibility that they are too massive to be produced at the current LHC energies. In this situation, the ongoing experiments such as the LHCb and Belle II gain importance because of their ability to explore the NP through precision measurements of the decay properties involving the transitions of the $b$ quark \cite{Belle-II:2018jsg, LHCb:2013ghj, LHCb:2022vje, Li:2018lxi}. These are known as indirect searches, where the flavor-changing-neutral-current transitions (FCNC), $b\to s\ell^
{+}\ell^{-}$ \cite{LHCb:2022qnv, Altmannshofer:2008dz, Descotes-Genon:2013vna, Altmannshofer:2013foa, Das:2018sms, Zaki:2023mcw}, are at the forefront to probe the NP contributions through precise theoretical predictions and measurements of decay rates and angular observables.

Several observables in these FCNC transitions deviate from their SM predictions, \textit{e.g.}, the branching ratio of $B_s\to \phi \mu^{+}\mu^{-}$ differs by $3.5\sigma$ \cite{LHCb:2015wdu, LHCb:2021zwz}, and the measured value of angular observable $P^\prime_5$ in $B\to K^{*}\mu^{+}\mu^{-}$ in the intermediate $q^2\in [4.0,6.0]\,\text{GeV}^2$ bin discrepant by $3\sigma$ \cite{LHCb:2015svh, LHCb:2020lmf}. These anomalies are attributed to possible BSM in these $b\to s\mu^{+}\mu^{-}$ processes. Leaving aside the old measurements of the Lepton-flavor-universality (LFU) ratio, $R_{K}$ and $R_{K^*}$ the LHCb has updated their results \cite{LHCb:2022qnv, LHCb:2022vje}, which are now aligned with the SM predictions, indicating the universality of lepton couplings. To accommodate the various anomalies observed in these decays, different theoretical frameworks have been proposed as extensions of the SM. Among them, the pertinent ones include supersymmetric models \cite{Haber:2017aci, Kazakov:2000ra, Ali:1999mm}, the $Z'$ model \cite{Nasrullah:2018puc, Sheng:2021tom}, the Higgs doublet model \cite{Hu:2017qxj, Celis:2014cva} and the lepto-quark model \cite{Chowdhury:2022dps}. By using the effective field theory approach, these measurements were also analyzed in a model-independent way, where the NP can be accommodated in the form of new vector and axial-vector operators, see \textit{e.g.}, \cite{Alok:2010zd, Alok:2011gv, Descotes-Genon:2013wba, Hurth:2013ssa, Datta:2019zca, Kumar:2019qbv, Alok:2019ufo, Carvunis:2021jga, Alguero:2021anc, Alguero:2023jeh, Geng:2021nhg, Hurth:2021nsi, Ishaq:2013toa, Bhutta:2024zwj, Salam:2024nfv, Farooq:2024owx, Aarfi:2025qcp, Yasmeen:2024cki, Angelescu:2021lln, Alok:2022pjb, Ciuchini:2022wbq, SinghChundawat:2022ldm, Munir:2015gsp, Alok:2023yzg, Alok:2024cyq, MunirBhutta:2020ber, Altmannshofer:2021qrr}. Although NP in the form of scalar/pseudoscalar (S/P) or tensor (T) operators is not ruled out, they can not accommodate the current anomalies at their own. However, through a specific combination of V/A, the S/P or T operators can provide a moderate fit to the data \cite{Hurth:2023jwr, Vardani:2024bae}. 

If there is any NP in $b \to s \ell^{+}\ell^{-}$ transitions, it should also be evident in other decay modes induced by the same quark-level transition. There are several such decay modes, and $\Lambda_b \to \Lambda \ell^{+}\ell^{-}$ is one of them. 
The first investigation of CP violation in baryonic decays was carried out by the LHCb collaboration through the observation of decay \( \Lambda^{0}_{b} \rightarrow p K^- \mu^+\mu^- \) \cite{LHCb:2017slr}. In addition, the LHCb Collaboration has performed a test of LFU in the decay 
\(\Lambda^{0}_{b} \rightarrow p K^- \ell^+\ell^-\), 
reporting results that are consistent with the SM predictions \cite{LHCb:2019efc}. The $\Lambda_b$ baryon decays have been studied, see \textit{e.g.}, \cite{CDF:2011buy, Detmold:2012vy, Das:2018iap, Wang:2008ni, Beneke:2004dp, Boer:2014kda, Feldmann:2011xf, Yan:2019tgn, Roy:2017dum, Das:2023kch, Mahmoudi:2026aul}. One of the advantages of $\Lambda_b \to \Lambda\left(\to p\pi\right)\ell^{+}\ell^{-}$ decay is that even with the initial-state baryon $(\Lambda_b)$ unpolarized, the $\Lambda$ baryon spin can not only be used to understand the helicity structure of weak effective Hamiltonian \cite{Mannel:1997xy, Hiller:2007ur, Wang:2008sm, Chen:2001ki, Gutsche:2013pp, Gutsche:2013oea, Nasrullah:2018vky, Nasrullah:2020glp, Detmold:2016pkz}, but it can also reveal new angular coefficients in the cascade decay which provide more insight about the helicity structure of the decay. In Ref. \cite{Detmold:2016pkz}, high-precision lattice QCD calculations of the $\Lambda_b \to \Lambda$ form factors are performed, and the SM predictions for the differential branching fraction and various angular observables in the four-fold decay $\Lambda_b \to \Lambda(\to p^{+}\pi^{-})\mu^{+}\mu^{-}$ are presented, with both the initial-state $\Lambda_b$ and the final-state muons unpolarized. Additionally, the full angular distribution of the initial-state polarized
$\Lambda_b \to \Lambda\left(\to p\pi\right)\ell^{+}\ell^{-}$ decay has also been investigated \cite{Blake:2017une, Das:2020qws}.

With this motivation, one of the key objectives of this study is to systematically analyze the impact of final-state lepton polarization on the four-fold differential decay distribution in the $\Lambda_b \to \Lambda\left(\to N\pi\right)\ell^{+}\ell^{-}$ decay, with unpolarized initial-state $\Lambda_b$. Previously, lepton polarization effects and the polarization asymmetry observables, such as the single and double lepton polarization asymmetries and the corresponding polarized lepton forward-backward asymmetries in the $\Lambda_b \to \Lambda\ell^{+}\ell^{-}$ decay have been studied \cite{Aslam:2008hp, Aliev:2002hj, Bashiry:2007pd, Aliev:2004yf, Aliev:2012ac}, but to our knowledge, a complete analysis of the four-fold decay $\Lambda_b \to \Lambda\left(\to N\pi\right)\ell^{+}\ell^{-}$, with lepton polarization effects, is missing. In this context, we use the lattice QCD $\Lambda_b \to \Lambda$ form factors from \cite{Detmold:2016pkz} and, instead of considering a single four-fold decay distribution with unpolarized leptons, derive the three separate four-fold angular decay distributions which appear when one of the final state leptons is longitudinally, normally or transversely polarized. To achieve this goal, for the three polarizations of the final state lepton, we systematically identify the additional angular 
coefficients that emerge relative to the unpolarized leptons case. We find that for the longitudinal polarization case, the structure of the four-fold angular distribution remains the same as in the unpolarized distribution case. However, the helicity amplitudes contributions that appear when the lepton is longitudinally polarized are encapsulated in the different angular coefficients of the distribution. In contrast to the longitudinal polarization case, the contributions of normal and transverse polarization of the lepton modify the structure of the four-fold angular decay distributions along with the introduction of some new additional angular coefficients. The analytical expressions of all these polarized and unpolarized angular coefficients, including the lepton mass effects, are obtained in terms of both the helicity and transversity amplitudes. 

In addition, for a given polarization, we also examine the separate contributions from the lepton spin $+1/2$ and $-1/2$ states on various angular observables in the four-body decay process \( \Lambda_b \rightarrow \Lambda (\rightarrow N \pi) \mu^+ \mu^- \). Furthermore, to compare the variation in the polarized and unpolarized angular observables, we have plotted them together against the square of the momentum transfer $q^2$. For this purpose, we adopted the following strategy in this work:   

\begin{itemize}
\item In the first step of our analysis, we polarized the final-state lepton along the longitudinal (L), normal (N), and transverse (T) directions. For each polarization scenario, we derived the corresponding four-fold angular distribution and examined the contributions from the lepton spin states \( +\tfrac{1}{2} \) and \( -\tfrac{1}{2} \) (see Eqs. (\ref{eq:lon}), (\ref{eq:nor}), and (\ref{eq:trans}) for the L, N, and T cases, respectively). Notably, the analysis of the normal and transverse polarizations reveals the emergence of additional angular structures that are absent in both the unpolarized and longitudinal cases.

   \item For the longitudinal polarization ($\text{L}$), the spin–dependent terms naturally appear within the already established angular observables given in Eqs.~(\ref{DDRpolL})--(\ref{ACoefpolL}). The polarized predictions for these observables, including the individual contributions from the lepton spin states, are compared with their unpolarized counterparts within the SM.
   Based on these results, we construct the corresponding polarization asymmetries for each observable shown in FIG. \ref{PrimeK}.
   \item 
For the normal polarization ($\text{N}$), six additional spin–dependent coefficients emerge. Among them, two are real, while four are imaginary. Importantly, two of the imaginary terms do contribute to physical observables, i.e., differential branching ratio and hadron forward backward asymmetry, although their impact remains small within the SM. The other two imaginary contributions do not directly appear. Thus, the overall effect of $\text{N}$ polarization is expected to be suppressed in the SM.
    \item For the transverse (T) polarized scenario, we find twelve new angular coefficients. Among them, eight are real and four are imaginary. We further analyze angular observables, i.e., the differential branching ratios, forward-backward asymmetry of leptons, hadron and that of combined lepton-hadron modifications with new angular coefficients. We also plot the remaining new real angular coefficient within the SM. 

   \item To investigate the impact of the vector and axial-vector NP Wilson coefficients (WCs), we employ the global fit results of Ref. \cite{Alguero:2023jeh}, and focus on the NP scenarios that modify the polarization asymmetries arising from L and T lepton polarizations, while comparing them with the SM predictions.  
\item In contrast to the longitudinal L case, the polarization asymmetries of the additional spin-dependent individual angular coefficients associated with N and T polarization cases are suppressed within the SM, and the potential NP contributions are indistinguishable from the SM expectations. Consequently, for the N and T polarization, only the SM polarized asymmetry results for the additional real angular coefficients are presented.
\end{itemize}
The remainder of this paper is organized as follows. In Section~\ref{EH}, we present the weak effective Hamiltonian (WEH) governing the decay $\Lambda_b \to \Lambda \ell^{+}\ell^{-}$ in and beyond the SM. The decay amplitude in the helicity formalism and the helicity amplitudes are discussed in Section~\ref{HME}. The cascade decay $\Lambda \to N\pi$, is presented in Section~\ref{cascade-amp1}, followed by the full four-fold decay amplitude in Section~\ref{full-amp123}. In Sections~\ref{unpol-Rate} and \ref{ANGO}, we derive the four-fold differential decay distributions for the cases of unpolarized and polarized final-state lepton, respectively. From these distributions, a comprehensive set of physical observables is extracted, with their analytical expressions expressed in terms of the angular coefficients, as summarized in Section~\ref{physObs}. Section~\ref{Phenom} is devoted to a detailed numerical analysis of the angular observables and their associated asymmetries, both within the SM and in representative NP scenarios. Finally, the conclusions of this work are presented in Section~\ref{concl}.

\section{Theoretical Framework}\label{framework}
\subsection{Effective Hamiltonian}\label{EH}
The baryonic decay $\Lambda_b\to \Lambda \ell^{+}\ell^{-}$ occurs through the FCNC transitions in the SM, and the corresponding WEH is
\begin{align}\label{H1}
\mathcal{H}_{\text{eff}}^{\text{SM}}=-\frac{4 G_{F}}{\sqrt{2}}V_{tb}V^{\ast}_{ts}&\Bigg[C_{7}^{\text{eff}}(\mu)\mathcal{O}_{7}(\mu)+C_{9}^{\text{eff}}(\mu)\mathcal{O}_{9}(\mu)+C_{10}(\mu)\mathcal{O}_{10}(\mu)\Bigg],
\end{align}
where the effective operators $\left(\mathcal{O}_{i=7,\;9,\;10}\right)$ are defined as:
\begin{align}\label{op1}
\mathcal{O}_{7}(\mu) &=\frac{e}{16\pi ^{2}}m_{b}\left( \bar{s}\sigma _{\nu\rho }P_{R}b\right) F^{\nu \rho},
& \mathcal{O}_{9}(\mu) &=\frac{e^{2}}{16\pi ^{2}}(\bar{s}\gamma _{\nu }P_{L}b)(\bar{\ell}\gamma^{\nu }\ell),
& \mathcal{O}_{10}(\mu) &=\frac{e^{2}}{16\pi ^{2}}(\bar{s}\gamma _{\nu }P_{L}b)(\bar{\ell} \gamma ^{\nu }\gamma _{5} \ell).
\end{align}
Where $\ell$ is the electron, muon or tauon and e, $G_F$, $\lambda_t=V_{tb}V^{\ast}_{ts}$, $\alpha_{em}=\frac{e^2}{4\pi}$ are the electric charge, Fermi coupling constant,  CKM matrix elements and the  fine structure
constant respectively. $P_{L/R} = \frac{1\mp \gamma_5}{2}$ are the chiral projection operators, and $\mu$ is the renormalization scale, which is chosen typically $\mu\simeq m_b$, resulting in the elimination of all large logarithms from the amplitude calculation. In the operator $O_7$, $m_b$ is treated as the running quark mass in the modified minimal-subtraction $(\overline{\text{MS}})$-scheme. The contributions of the
factorizable quark-loop corrections to current-current and
penguin operators are absorbed in the effective WCs $C_{7,9}^{\text{eff}}(q^2)$ \cite{Blake:2016olu, Du:2015tda, Beneke:2001at,  Asatryan:2001zw, Greub:2008cy}. The explicit expressions of these Wilson coefficients are presented in Appendix \ref{WCsC7C9}. The numerical values of the WCs, $C_1,\ldots,C_{10}$, used in our numerical analysis are written in TABLE \ref{tab:Nvalues}. 
\begin{table*}[htp!]
\caption{The SM values of the WCs with next-to-next-leading logarithmic accuracy, evaluated at the renormalization scale $\mu\simeq m_{b}$.}\label{tab:Nvalues}
\centering
    \begin{tabular}{|ccccc|}
           \hline\hline
		\, $C_1=-0.294$ ,&  $C_2=1.017$ ,&  $C_3=-0.0059$,&  $C_4=-0.087$,&
    $C_5 =0.0004$ \\ 
             $C_6=0.0011$,&  \, $C_7= -0.324$,&  $C_8=-0.176$,&  $C_9=4.114$,&  $C_{10}=-4.193$\, \\
            \hline\hline
	\end{tabular}
\end{table*}

In the above Hamiltonian, the NP can be induced in two different ways:
\begin{itemize}
    \item The first is the modification of the short-distance WCs $C_{7}^{\text{eff}}, C_{9}^{\text{eff}}$ and $C_{10}$, which govern the leading-order contributions to the FCNC decay processes.
    \item The second is to introduce contributions from additional operators that are typically suppressed or absent in the SM.  These include chirality-flipped counterparts of $\mathcal{O}_{7}, \mathcal{O}_{9}$ and $\mathcal{O}_{10}$, denoted by $\mathcal{O}^{\text{NP}}_{7^{(')}}, \mathcal{O}^{\text{NP}}_{9^{(')}}$ and $\mathcal{O}^{\text{NP}}_{10^{(')}}$.
    Moreover, the operator basis can be enlarged to accommodate scalar, pseudoscalar, and tensor operators, which often arise in different NP models.
\end{itemize}
In our analysis, we employ both approaches. The new-physics contributions are parameterized by the WCs $C_{9}^{\text{NP}}$, $C_{10}^{\text{NP}}$, and their chirality-flipped counterparts $C_{9'}^{\text{NP}}$, $C_{10'}^{\text{NP}}$. In the second approach, we restrict ourselves to the vector and axial-vector operators, together with their chirality-flipped counterparts, which are motivated in several NP scenarios.
Furthermore, the WCs can be categorized according to their lepton-flavor structure. In particular, one distinguishes between LFU NP contributions and lepton-flavor-Universality-violating (LFUV) contributions. Since short-distance effects are independent of the initial hadronic state and its kinematics, they can be treated within this model-independent framework.  
After incorporating these modifications, the effective Hamiltonian $\mathcal{H}_{\text{eff}}$ takes the form:
\begin{align}
\mathcal{H}_{\text{eff}}& = \frac{G_F \alpha_{em}}{2\sqrt{2}\pi} V_{tb} V_{ts}^{*} \Bigg\{ 
    \; C_{7}^{\text{eff}}  \frac{m_{b}}{e} 
    \left( \overline{s} \sigma_{\nu\rho}(1 + \gamma_{5}) b \right) F^{\nu\rho}+ C_9^{\text{eff}}  (\overline{s} b)_{\mathrm{V-A}} (\overline{\ell} \ell)_{V} 
     + C_{9\ell}^{\text{NP}} (\overline{s} b)_{\mathrm{V-A}} (\overline{\ell} \ell)_{V} \nonumber \\
&  +C_{9^\prime\ell}^{\text{NP}}  (\overline{s}\, b)_{\mathrm{V+A}} (\overline{\ell} \ell)_{V}+ C_{10} (\overline{s}\, b)_{\mathrm{V-A}} (\overline{\ell} \ell)_{A}
     + C_{10\ell}^{\text{NP}}  (\overline{s}\, b)_{\mathrm{V-A}} (\overline{\ell} \ell)_{A} + C_{10^\prime\ell}^{\text{NP}} (\overline{s}\, b)_{\mathrm{V+A}} (\overline{\ell} \ell)_{A}
\Bigg\}\;.
\label{Heff main}
\end{align}
By separating the vector and axial-vector leptonic currents, the decay amplitude for $\Lambda_b\to \Lambda \ell^+\ell^-$ is given as
\begin{eqnarray}
\mathcal{M}_1^{s_{1},s_2}\left(s_{\Lambda_b}, s_{\Lambda}\right)=\frac{G_{F}\alpha_{em}}{2\sqrt{2}\pi}V_{tb}V^{\ast}_{ts}\,g^{\mu^{\prime}\mu}\Big\{T^{1}_{\mu}\left(s_{\Lambda_b}, s_{\Lambda}\right)L_{V,\mu^{\prime}}^{s_1,s_2}
+T^{2}_{\mu}\left(s_{\Lambda_b}, s_{\Lambda}\right)L_{A,\mu^{\prime}}^{s_1,s_2}\Big\},\label{Amp1ag}
\end{eqnarray}
where
\begin{eqnarray}
L_{V,\mu^{\prime}}^{{s_1,s_2}}&=&\left\langle l^+\left(p_{\ell^+},s_2\right)l^-\left(p_{\ell^-},s_1\right)\left|\bar{\ell}\gamma_{\mu^{\prime}}{\ell}\right|0\right\rangle = \bar u \left({p}_{\ell^-},s_1\right)\gamma_{\mu^{\prime}}\,v\left({p}_{\ell^+},s_2\right),\label{Amp1alep}
\\
L_{A,\mu^{\prime}}^{{s_1,s_2}}&=&\left\langle l^+\left(p_{\ell^+},s_2\right)l^-\left(p_{\ell^-},s_1\right)\left|\bar{\ell}\gamma_{\mu^{\prime}} \gamma_{5}{\ell}\right|0\right\rangle=\bar u \left({p}_{\ell^-},s_1\right)\gamma_{\mu^{\prime}}\gamma_{5} \,v\left({p}_{\ell^+},s_2\right),\label{Amp1blep}
\end{eqnarray}
with $s_1 (s_2)$ and $p_{\ell^-} (p_{\ell^+})$ specify the spin and momentum of lepton (anti-lepton), respectively. The hadronic terms $ T_{\mu}^{1}\left(s_{\Lambda_b}, s_{\Lambda}\right)$ and $T_{\mu}^{2}\left(s_{\Lambda_b}, s_{\Lambda}\right)$ are defined explicitly as
    \begin{align}
T^{1}_{\mu}\left(s_{\Lambda_b}, s_{\Lambda}\right) =& \left( C_{9}^{\text{eff}} + C_{9\ell}^{\text{NP}} \right) 
\left[ \left\langle \Lambda(k, s_{\Lambda}) \left| \overline{s} \gamma_{\mu}(1 - \gamma_{5}) b \right| \Lambda_b(p, s_{\Lambda_b}) \right\rangle \right]+ C_{9'\ell}^{\text{NP}} 
\left[ \left\langle \Lambda(k, s_{\Lambda}) \left| \overline{s} \gamma_{\mu}(1 + \gamma_{5}) b \right| \Lambda_b(p, s_{\Lambda_b}) \right\rangle \right] \nonumber \\
&- \frac{2m_b}{q^2} C_{7}^{\text{eff}} 
\left[ \left\langle \Lambda(k, s_{\Lambda}) \left| \overline{s} i \sigma_{\mu \nu} q^{\nu}(1 + \gamma_{5}) b \right| \Lambda_b(p, s_{\Lambda_b}) \right\rangle \right], \label{effH} \\
T^{2}_{\mu}\left(s_{\Lambda_b}, s_{\Lambda}\right) = & \left( C_{10} + C_{10\ell}^{\text{NP}} \right) 
\left[ \left\langle \Lambda(k, s_{\Lambda}) \left| \overline{s} \gamma_{\mu}(1 - \gamma_{5}) b \right| \Lambda_b(p, s_{\Lambda_b}) \right\rangle \right] + C_{10'\ell}^{\text{NP}} 
\left[ \left\langle \Lambda(k, s_{\Lambda}) \left| \overline{s} \gamma_{\mu}(1 + \gamma_{5}) b \right| \Lambda_b(p, s_{\Lambda_b}) \right\rangle \right]\;,
\label{effH1}
\end{align}
where $p$ and $k$ are the four momentum of  $ \Lambda_b$ and $ \Lambda$, respectively. As an exclusive process, the matrix elements for \( \Lambda_b \rightarrow \Lambda \) transition for different quark currents can be parameterized in terms of the ten helicity form factors (FFs). In the case of vector, axial vector, tensor, and pseudo-tensor currents, the corresponding matrix elements for \( \Lambda_b \rightarrow \Lambda \) are reported in \cite{Feldmann:2011xf,Detmold:2016pkz}, and collected in Appendix \ref{appendHME}.
\subsection{Helicity formalism and helicity amplitudes for $ \Lambda_{b} \rightarrow \Lambda \ell^+\ell^-$ decay}
\label{HME}
To write the decay amplitude for $\Lambda_b\to \Lambda \ell^+\ell^-$ decay using the helicity formalism, we use the  completeness property of helicity basis $\epsilon^{\mu}(m)$,
\begin{eqnarray}
\sum_{m, m^{\prime}=t, +, -, 0}\epsilon^{\mu^{\prime}}(m)\epsilon^{\ast\mu}(m^\prime)g_{mm^{\prime}}=g^{\mu^{\prime}\mu},\label{C22}
\end{eqnarray}
where $\epsilon^{\mu}$ represents the polarization vector of the virtual $j_{\text{eff}}^\mu$ (see FIG. \ref{fig:kinematica}) and  $m=t,+,-,0$ corresponds to the time-like, transverse, and longitudinal components of it with $g_{mm^{\prime}}=\text{diag}(+, -, -, -)$. Using Eq. (\ref{C22}) in Eq. (\ref{Amp1ag}), we can write the decay amplitude as
\begin{eqnarray}
\mathcal{M}_1^{s_{1},s_2}\left(s_{\Lambda_b}, s_{\Lambda}\right)=N_1\,
\sum_{m=t, +, -, 0}g_{mm}\Big\{H_m^1\left(s_{\Lambda_b}, s_{\Lambda}\right)\,L_{V,m}^{s_1,s_2}
+H_m^2\left(s_{\Lambda_b}, s_{\Lambda}\right)\,L_{A,m}^{s_1,s_2}\Big\},\label{Amp2ag}
\end{eqnarray}
where, $N_1\equiv\frac{G_{F}\alpha_{em}}{2\sqrt{2}\pi}V_{tb}V^{\ast}_{ts}$ and $s_{\Lambda_b}$ ($s_{\Lambda}$) corresponds to the spin of decaying (daughter) baryon. Also; $H_m^i\left(s_{\Lambda_b}, s_{\Lambda}\right)=\epsilon^{\ast\mu}(m)T^i_{\mu}\left(s_{\Lambda_b}, s_{\Lambda}\right)$, with $i=1,2$ and $L_{V(A),m}^{s_1,s_2}=\epsilon^{\mu^{\prime}}(m)L_{V(A),\mu^{\prime}}^{s_1,s_2}$ are the hadronic and leptonic helicity amplitudes, respectively. Using Eqs. (\ref{effH}) and (\ref{effH1}), with the definitions of hadronic matrix elements in Appendix {\ref{appendHME}}, the helicity amplitudes are gievn as
\begin{align}
H_{t}^{1}(s_{\Lambda_b},s_{\Lambda}) &= \left( C_9^{\text{eff}} + C_{9\ell}^{\text{NP}} +C_{9'\ell}^{\text{NP}}\right)  \mathcal{H}_t^V(s_{\Lambda_b},s_{\Lambda})    - \left( C_9^{\text{eff}} + C_{9\ell}^{\text{NP}} -C_{9'\ell}^{\text{NP}}\right)  \mathcal{H}^{A}_t(s_{\Lambda_b},s_{\Lambda}) ,\label{20N}
\\
    H_{t}^{2}(s_{\Lambda_b},s_{\Lambda}) &=\left( C_{10} + C_{10\ell}^{\text{NP}} +C_{10'\ell}^{\text{NP}}\right)  \mathcal{H}_t^V(s_{\Lambda_b},s_{\Lambda}) -\left( C_{10} + C_{10\ell}^{\text{NP}} -C_{10'\ell}^{\text{NP}}\right)  \mathcal{H}_{t}^{A}(s_{\Lambda_b},s_{\Lambda}),
\end{align}    
\begin{align}
 H_{0}^{1}(s_{\Lambda_b},s_{\Lambda}) &=  \left(C_{9}^{\text{eff}} + C_{9\ell}^{\text{NP}} + C_{9'\ell}^{\text{NP}}\right) \mathcal{H}_{0}^{V}(s_{\Lambda_b},s_{\Lambda})-\left(C_{9}^{\text{eff}} + C_{9\ell}^{\text{NP}} - C_{9'\ell}^{\text{NP}}\right) \mathcal{H}_{0}^{A} (s_{\Lambda_b},s_{\Lambda})\notag\\  
 \quad &- \frac{2m_b}{q^2}   C_{7}^{\text{eff}} \left[ \mathcal{H}_{0}^{T}(s_{\Lambda_b},s_{\Lambda}) +   \mathcal{H}_{0}^{T5}(s_{\Lambda_b},s_{\Lambda})\right],    
\\    
H_{0}^{2}(s_{\Lambda_b},s_{\Lambda}) &= \left(C_{10} + C_{10\ell}^{\text{NP}} + C_{10'\ell}^{\text{NP}}\right) \mathcal{H}^{V}_{0} (s_{\Lambda_b},s_{\Lambda}) -\left(C_{10} + C_{10\ell}^{\text{NP}} - C_{10'\ell}^{\text{NP}}\right)\mathcal{H}_{0}^{A}(s_{\Lambda_b},s_{\Lambda}),
\\ 
 H_{\pm}^{1}(s_{\Lambda_b},s_{\Lambda}) &= \left(C_{9}^{\text{eff}} + C_{9\ell}^{\text{NP}} + C_{9'\ell}^{\text{NP}}\right) \mathcal{H}_{\pm}^{V}(s_{\Lambda_b},s_{\Lambda})-\left(C_{9}^{\text{eff}} + C_{9\ell}^{\text{NP}} - C_{9'\ell}^{\text{NP}}\right) \mathcal{H}_{\pm}^{A}(s_{\Lambda_b},s_{\Lambda}) \notag\\  \quad &- \frac{2m_b}{q^2}   C_{7}^{\text{eff}} \left[ \mathcal{H}_{\pm}^{T}(s_{\Lambda_b},s_{\Lambda}) +   \mathcal{H}_{\pm}^{T5}(s_{\Lambda_b},s_{\Lambda})\right],
 \\
H_{\pm}^{2}(s_{\Lambda_b},s_{\Lambda}) &=\left(C_{10}+ C_{10\ell}^{\text{NP}} + C_{10'\ell}^{\text{NP}} \right) \mathcal{H}^{V}_{\pm}(s_{\Lambda_b},s_{\Lambda})   - \left(C_{10} + C_{10\ell}^{\text{NP}} - C_{10'\ell}^{\text{NP}} \right) \mathcal{H}^{A}_{\pm}(s_{\Lambda_b},s_{\Lambda}),\label{eq:25N}
\end{align}

where the helicity amplitudes with different polarizations of the virtual $j_{\text{eff}}^\mu$, and the vector quark current are
\begin{align}
\mathcal{H}^V_t(s_{\Lambda_b}, s_{\Lambda}) &= \epsilon^{*\mu}(t) \left\langle \Lambda(k, \text{s}_{\Lambda}) \left| \bar{s}\gamma_{\mu} b \right| \Lambda_b(p, \text{s}_{\Lambda_b}) \right\rangle = f^V_t(q^2) \frac{m_{\Lambda_b} - m_{\Lambda}}{\sqrt{q^2}} \left[\bar u_{\Lambda}(k, \text{s}_{\Lambda})\, u_{\Lambda_b}(p, \text{s}_{\Lambda_b})\right],
\\
\mathcal{H}^V_0(s_{\Lambda_b}, s_{\Lambda}) &=\epsilon^{*\mu}(0) \left\langle \Lambda(k, \text{s}_{\Lambda}) \left| \bar{s}\gamma_{\mu} b \right| \Lambda_b(p, \text{s}_{\Lambda_b}) \right\rangle = 2f^V_0(q^2) \frac{m_{\Lambda_b} + m_{\Lambda}}{s_{+}} (k \cdot \epsilon^*(0)) \left[\bar u_{\Lambda}(k, \text{s}_{\Lambda}) \,u_{\Lambda_b}(p, s_{\Lambda_b})\right],
\\
\mathcal{H}^V_{\pm}(s_{\Lambda_b}, s_{\Lambda}) &= \epsilon^{*\mu}(\pm) \left\langle \Lambda(k,\text{s}_{\Lambda}) \left| \bar{s}\gamma_{\mu} b \right| \Lambda_b(p,\text{s}_{\Lambda_b}) \right\rangle = f^V_{\perp}(q^2) \left[\bar u_{\Lambda}(k,\text{s}_{\Lambda})\slashed{\epsilon}^{*}(\pm)\,u_{\Lambda_b}(p,\text{s}_{\Lambda_b})\right].
\end{align}
The analogous expressions for the axial-vector quark current are
\begin{align}
\mathcal{H}^{A}_t(s_{\Lambda_b}, s_{\Lambda}) &=\epsilon^{*\mu}(t) \left\langle \Lambda(k, s_{\Lambda}) \left| \bar{s}\gamma_\mu \gamma_5 b \right| \Lambda_b(p, \text{s}_{\Lambda_b}) \right\rangle = -f^{A}_t(q^2) \frac{m_{\Lambda_b}+m_{\Lambda}}{\sqrt{q^2}} \left[\bar u_{\Lambda}(k, s_{\Lambda})\,\gamma_5\, u_{\Lambda_b}(p, \text{s}_{\Lambda_b})\right],
\\
\mathcal{H}^{A}_0(s_{\Lambda_b}, s_{\Lambda}) &= \epsilon^{*\mu}(0) \left\langle \Lambda(k, \text{s}_{\Lambda}) \left| \bar{s}\gamma_\mu \gamma_5 b \right| \Lambda_b(p, \text{s}_{\Lambda_b}) \right\rangle = -2f^{A}_0(q^2) \frac{m_{\Lambda_b} - m_{\Lambda}}{s_{-}} (k \cdot \epsilon^*(0)) \left[\bar{u}_{\Lambda}(k, \text{s}_{\Lambda})\,\gamma_5\, u_{\Lambda_b}(p, \text{s}_{\Lambda_b})\right],
\\
\mathcal{H}^{A}_{\pm}(s_{\Lambda_b}, s_{\Lambda}) &= \epsilon^{*\mu}(\pm) \left\langle \Lambda(k,\text{s}_{\Lambda}) \left| \bar{s}\gamma_\mu \gamma_5 b \right| \Lambda_b(p,\text{s}_{\Lambda_b}) \right\rangle = f^{A}_{\perp}(q^2) \left[\bar{u}_{\Lambda}(k,\text{s}_{\Lambda})\slashed{\epsilon}^{*}(\pm)\,\gamma_5\,u_{\Lambda_b}(p,\text{s}_{\Lambda_b})\right].
\end{align}
The other helicity amplitudes with $q^{\nu}$ projections of the tensor and pseudo-tensor currents are given as,
\begin{align}
\mathcal{H}^T_0(s_{\Lambda_b}, s_{\Lambda}) & = \epsilon^{*\mu}(0) \left\langle \Lambda(k, \text{s}_{\Lambda}) \left| \bar{s}\,i \sigma_{\mu\nu} q^\nu b \right| \Lambda_b(p, \text{s}_{\Lambda_b}) \right\rangle = -2f^T_0(q^2) \frac{q^2}{s_{+}} (k \cdot \epsilon^*(0))\left[\bar{u}_{\Lambda}(k, \text{s}_{\Lambda}) u_{\Lambda_b}(p, \text{s}_{\Lambda_b})\right], \\
\mathcal{H}^T_{\pm}(s_{\Lambda_b}, s_{\Lambda}) &= \epsilon^{*\mu}(\pm) \left\langle \Lambda(k, \text{s}_{\Lambda}) \left| \bar{s}\,i \sigma_{\mu\nu} q^\nu b \right| \Lambda_b(p, \text{s}_{\Lambda_b}) \right\rangle = -f^T_{\perp}(q^2) (m_{\Lambda_b} + m_{\Lambda}) \left[\bar{u}_{\Lambda}(k, s_{\Lambda}) \slashed{\epsilon}^* (\pm) u_{\Lambda_b}(p, \text{s}_{\Lambda_b})\right],
\end{align}
and
\begin{align}
\mathcal{H}^{T5}_0(s_{\Lambda_b}, s_{\Lambda}) &= \epsilon^{*\mu}(0) \left\langle \Lambda(k, \text{s}_{\Lambda}) \left| \bar{s}\,i \sigma_{\mu\nu} q^\nu \gamma_5 b \right| \Lambda_b(p, \text{s}_{\Lambda_b}) \right\rangle = -2f^{T5}_0(q^2) \frac{q^2}{s_{-}} (k \cdot \epsilon^*(0))\left[\bar{u}_{\Lambda}(k, \text{s}_{\Lambda}) \,\gamma_5\, u_{\Lambda_b}(p, s_{\Lambda_b})\right], \\
\mathcal{H}^{T5}_{\pm}(s_{\Lambda_b}, s_{\Lambda}) &= \epsilon^{*\mu}(\pm) \left\langle \Lambda(k, \text{s}_{\Lambda}) \left| \bar{s}\,i \sigma_{\mu\nu} q^\nu \gamma_5 b \right| \Lambda_b(p, \text{s}_{\Lambda_b}) \right\rangle = f^{T5}_{\perp}(q^2) (m_{\Lambda_b} - m_{\Lambda}) \left[\bar{u}_{\Lambda}(k, s_{\Lambda}) \slashed{\epsilon}^* (\pm) \gamma_5\, u_{\Lambda_b}(p, s_{\Lambda_b})\right],
\end{align}
where $s_\pm$ are defined as $s_\pm \equiv (m_{\Lambda_b} \pm m_{\Lambda})^2 - q^2$ and $q^2=(p-k)^2$ is square of the momentum transfer while $f^{(V,A,T,T5)}_{(t,0,\perp)}(q^2)$ are the different FFs.

We compute the helicity amplitudes defined in Eqs.~(\ref{20N})-(\ref{eq:25N}) for all possible spin configurations. By performing this calculation, together with the relevant kinematic conventions, we identify all non-vanishing contributions arising from both the SM and potential NP operators. These non-zero terms are summarized below \footnote{The $H_{t}^{1}\left(\pm \frac{1}{2}, \mp \frac{1}{2}\right)$ helicity amplitudes do not contribute as the corresponding vector lepton current is conserved for the time-like polarization of the virtual $j_{\text{eff}}^{\mu}$.}
\begin{align}
H_{t}^{2}\left(\pm \frac{1}{2}, \mp \frac{1}{2}\right)&=\mp\frac{m_{\Lambda_b}-m_{\Lambda}}{\sqrt{q^2}}\sqrt{s_{+}}
(C_{10}+C_{10\ell}^{\text{NP}}+C_{10^{\prime}\ell}^{\text{NP}})f^V_t(q^2)\notag
\\
&-\frac{m_{\Lambda_b}+m_{\Lambda}}{\sqrt{q^2}}\sqrt{s_{-}}
(C_{10}+C_{10\ell}^{\text{NP}}-C_{10^{\prime}\ell}^{\text{NP}})f^A_t(q^2),\label{Hamp002}
\\
H_{0}^{1}\left(\pm \frac{1}{2}, \mp \frac{1}{2}\right)&=\mp\sqrt{\frac{s_-}{q^2}}\Big[(C_{9}^{\text{eff}}+C_{9\ell}^{\text{NP}}+C_{9^{\prime}\ell}^{\text{NP}})
(m_{\Lambda_b}+m_{\Lambda})f^V_{0}(q^2)+2m_b C_{7}^{\text{eff}}f^T_{0}(q^2)\Big]\notag
\\
&-\sqrt{\frac{s_+}{q^2}}\Big[(C_{9}^{\text{eff}}+C_{9\ell}^{\text{NP}}-C_{9^{\prime}\ell}^{\text{NP}})
(m_{\Lambda_b}-m_{\Lambda})f^A_{0}(q^2)+2m_b C_{7}^{\text{eff}}f^{T5}_{0}(q^2)\Big],\label{Hamp003}
\end{align}
\begin{align}
H_{0}^{2}\left(\pm \frac{1}{2}, \mp \frac{1}{2}\right)&=\mp\sqrt{\frac{s_-}{q^2}}(C_{10}+C_{10\ell}^{\text{NP}}+C_{10^{\prime}\ell}^{\text{NP}})(m_{\Lambda_b}+m_{\Lambda})f^V_{0}(q^2)\notag\\
&-\sqrt{\frac{s_+}{q^2}}(C_{10}+C_{10\ell}^{\text{NP}}-C_{10^{\prime}\ell}^{\text{NP}})(m_{\Lambda_b}-m_{\Lambda})f^A_{0}(q^2),\label{Hamp004}
\\
H_{\pm}^{1}\left(\pm \frac{1}{2}, \pm\frac{1}{2}\right)&=\pm\sqrt{2s_-}\Big[(C_{9}^{\text{eff}}+C_{9\ell}^{\text{NP}}+C_{9^{\prime}\ell}^{\text{NP}})f^V_{\perp}(q^2)
+\frac{2m_b}{q^2} C_{7}^{\text{eff}}(m_{\Lambda_b}+m_{\Lambda})f^T_{\perp}(q^2)\Big]\notag
\\
&-\sqrt{2s_+}\Big[(C_{9}^{\text{eff}}+C_{9\ell}^{\text{NP}}-C_{9^{\prime}\ell}^{\text{NP}})f^A_{\perp}(q^2)
+\frac{2m_b}{q^2} C_{7}^{\text{eff}}(m_{\Lambda_b}-m_{\Lambda})f^{T5}_{\perp}(q^2)\Big],\label{Hamp005}
\\
H_{\pm}^{2}\left(\pm \frac{1}{2}, \pm\frac{1}{2}\right)&=\pm\sqrt{2s_-}(C_{10}+C_{10\ell}^{\text{NP}}+C_{10^{\prime}\ell}^{\text{NP}})f^V_{\perp}(q^2)
-\sqrt{2s_+}(C_{10}+C_{10\ell}^{\text{NP}}-C_{10^{\prime}\ell}^{\text{NP}})f^A_{\perp}(q^2).\label{Hamp006}
\end{align}
\subsection{$\Lambda \rightarrow N \pi$ decay amplitude}\label{cascade-amp1}
The $\Lambda(k,s_{\Lambda})\to N(p_3, s_N)\pi(p_4)$ decay, where $N\pi=\{p\pi^-,n\pi^0\}$, $p_3$ is the momentum of proton (or neutron) with spin $s_N$ and $p_4$ is the momentum of pion. This transition is described, within the SM, by the $\Delta S=1$ effective Hamiltonian \cite{Boer:2014kda}:
 \begin{equation}
 \mathcal{H}^{\rm eff}_{\Delta S=1} = \frac{4G_F}{\sqrt{2}} V_{ud}^\ast V_{us} \big[ \bar{d}\gamma_\mu P_L u \big] \big[ \bar{u} \gamma^\mu P_L s\big]\, .\label{eq:cascade1}
 \end{equation}  
The corresponding decay amplitude is given as 
\begin{eqnarray}
	\mathcal{M}_2(s_{\Lambda},s_N) &=& N_2 H_2(s_{\Lambda},s_N),\label{eq:cascade2}
\end{eqnarray}
where, $N_2=\frac{4G_F}{\sqrt{2}} V_{ud}^\ast V_{us}$, and
\begin{eqnarray}
	H_2(s_{\Lambda},s_N)
    &=&\langle N(p_3,s_N)\pi(p_4)\big| [ \bar{d}\gamma_\mu P_L u ][ \bar{u} \gamma^\mu P_L s ] \big| \Lambda(k,s_{\Lambda})\rangle,\notag\\
	&=& \bar{u}_{N}(p_3,s_N) ( \omega + \xi \gamma_5 ) u_{\Lambda}(k,s_{\Lambda}).\label{eq:cascade3}
\end{eqnarray}
Here, $\omega$ and $\xi$ represent the two independent hadronic parameters, which can be extracted from the decay width and polarization measurements of $\Lambda\to p\pi^-$ decay. Using the spinors according to the cascade decay kinematics defined in Appendix \ref{Kinematics A}, we obtain,
\begin{eqnarray}
H_2\bigg(\pm\frac{1}{2},\pm\frac{1}{2}\bigg) &=& \bigg( \sqrt{r_+}\omega \mp \sqrt{r_-} \xi \bigg)\cos\frac{\theta_\Lambda}{2}\, ,\label{eq:cascade4}\\
H_2\bigg(\pm\frac{1}{2},\mp\frac{1}{2}\bigg) &=& \bigg( \pm\sqrt{r_+}\omega + \sqrt{r_-} \xi \bigg)\sin\frac{\theta_\Lambda}{2} \, ,\label{eq:cascade5}
%
%
\label{eq:cascade7}
\end{eqnarray} 
where $r_\pm = (m_\Lambda \pm m_N)^2 - m_\pi^2$.

\subsection{Four-fold $\Lambda_b \rightarrow \Lambda(\rightarrow N \pi)  \ell^+\ell^-$ decay}\label{full-amp123}
The four-fold $\Lambda_b \rightarrow \Lambda(\rightarrow N \pi)  \ell^+\ell^-$ decay can be described by the full four-fold decay amplitude as~\footnote{We have only considered the spin structure of the intermediate spin $1/2$ $\Lambda$ state, in the full amplitude, and omitted its propagator as it is treated on-shell.}
\begin{eqnarray}
\mathcal{M}^{s_{1},s_2}\left(s_{\Lambda_b}, s_{N}\right)&=&
\sum_{s_{\Lambda}^{(a)}}\mathcal{M}_1^{s_{1},s_2}\left(s_{\Lambda_b}, s_{\Lambda}^{(a)}\right)\,\mathcal{M}_2\left(s_{\Lambda}^{(a)},s_N\right),\notag\\
&=&N_1\,
\sum_{s_{\Lambda}^{(a)}}\sum_{m=t, +, -, 0}g_{mm}\Big\{H_m^1\left(s_{\Lambda_b}, s_{\Lambda}^{(a)}\right)\,L_{V,m}^{s_1,s_2}
+H_m^2\left(s_{\Lambda_b}, s_{\Lambda}^{(a)}\right)\,L_{A,m}^{s_1,s_2}\Big\}
N_2\,H_2(s_{\Lambda}^{(a)},s_N).
\label{eq:FullFF1}
\end{eqnarray}
For the unpolarized four-fold $\Lambda_b \rightarrow \Lambda(\rightarrow N \pi)  \ell^+\ell^-$ decay, we get the squared amplitude, by averaging over the initial state spin and summing over the final states spins, as
\begin{eqnarray}
\Big|\mathcal{M}\Big|^2&=&\frac{1}{2\,s_{\Lambda_b}+1}\sum_{s_{\Lambda_b},s_N}\sum_{s_1,s_2}\Big|\mathcal{M}^{s_{1},\,s_2}\left(s_{\Lambda_b}, s_{N}\right)\Big|^2,\notag\\
&=&\frac{1}{2}\sum_{s_{\Lambda_b}}\sum_{s_{\Lambda}^{(a)},s_{\Lambda}^{(b)}}\sum_{s_1,s_2}\sum_{m,n=t, +, -, 0}|N_1|^2\,g_{mm}g_{nn}\Bigg[L_{V,m}^{s_1,s_2}L_{V,n}^{\ast s_1,s_2}H_m^1\left(s_{\Lambda_b}, s_{\Lambda}^{(a)}\right) H_n^{1\ast}\left(s_{\Lambda_b}, s_{\Lambda}^{(b)}\right)\notag\\
&+&L_{A,m}^{s_1,s_2}L_{A,n}^{\ast s_1,s_2}H_m^2\left(s_{\Lambda_b}, s_{\Lambda}^{(a)}\right) H_n^{2\ast}\left(s_{\Lambda_b}, s_{\Lambda}^{(b)}\right)+L_{V,m}^{s_1,s_2}L_{A,n}^{\ast s_1,s_2}H_m^1\left(s_{\Lambda_b}, s_{\Lambda}^{(a)}\right) H_n^{2\ast}\left(s_{\Lambda_b}, s_{\Lambda}^{(b)}\right)\notag\\
&+&L_{A,m}^{s_1,s_2}L_{V,n}^{\ast s_1,s_2}H_m^2\left(s_{\Lambda_b}, s_{\Lambda}^{(a)}\right) H_n^{1\ast}\left(s_{\Lambda_b}, s_{\Lambda}^{(b)}\right)\Bigg]\sum_{s_{N}}|N_2|^2{H}_2\left(s^{(a)}_{\Lambda},s_N\right) H_2^*\left(s^{(b)}_{\Lambda},s_N\right).\label{eq:FullFF2}
\end{eqnarray}
\section{Four-fold angular distributions}
\subsection{Unpolarized lepton case}\label{unpol-Rate}
Considering four-body phase space with the squared amplitude, and combining all the elements, the final unpolarized four-fold differential decay distribution is obtained as \cite{Boer:2014kda, Das:2018iap}
\begin{align}
    \frac{d^4\Gamma}{dq^2 d\cos{\theta_{\ell}} d\cos{\theta_{\Lambda}} d\phi} =& \frac{3}{8 \pi}\mathcal{B}(\Lambda\to  N\pi)\Big[
    \Big( K_{\mathrm{1ss}} \sin^2{\theta_{\ell}} + K_{\mathrm{1cc}} \cos^2{\theta_{\ell}} + K_{\mathrm{1c}} \cos{\theta_{\ell}} \Big) \notag \\
    &+ \Big( K_{\mathrm{2ss}} \sin^2{\theta_{\ell}} + K_{\mathrm{2cc}} \cos^2{\theta_{\ell}} + K_{\mathrm{2c}} \cos{\theta_{\ell}} \Big) \cos{\theta_{\Lambda}} \notag \\
    &+ \Big( K_{\mathrm{3sc}} \sin{\theta_{\ell}} \cos{\theta_{\ell}} + K_{\mathrm{3s}} \sin{\theta_{\ell}} \Big) \sin{\phi} \sin{\theta_{\Lambda}} \notag \\
    &+ \Big( K_{\mathrm{4sc}} \sin{\theta_{\ell}} \cos{\theta_{\ell}} + K_{\mathrm{4s}} \sin{\theta_{\ell}} \Big) \cos{\phi} \sin{\theta_{\Lambda}}\Big],
    \label{eq:finaltrig}
\end{align}
where the angular coefficients, including the lepton mass effects, are given in both helicity and transversity basis in Appendix \ref{AppE} and Appendix \ref{AppF}, respectively.

\subsection{Polarized lepton case}\label{ANGO}
We define the spin four-vector of the lepton $\ell^-$, in its rest frame, by using the three orthogonal unit vectors, $\mathbf{\hat{e}}_i$, where $i=$ L, N, and T correspond to the longitudinal, normal, and transverse polarizations of the lepton, respectively.  
\begin{equation}
s_{1L}^{\mu} \equiv (0, \,\hat{e}_L) = \left(0, \frac{\vec{{p}}_{\ell^-}}{|\vec{{p}}_{\ell^-}|}\right),\qquad 
    s_{1N}^{\mu} \equiv (0, \hat{e}_N) = \left(0, \frac{\vec{{k}} \times \vec{{p}}_{\ell^-}}{|\vec{{k}} \times \vec{{p}}_{\ell^-}|}\right),\qquad 
    s_{1T}^{\mu} \equiv (0, \hat{e}_T) = \left(0, \hat{e}_N \times \hat{e}_L\right).\label{eq:polariza}
\end{equation}
Here, $\vec{p}_{\ell^-}$, and $\vec{k}$ denote the three-momentum vectors of $\ell^-$, and $\Lambda$, respectively. When boosted to $\ell\bar\ell$ center of mass frame, the longitudinal polarization four vector gets transformed, while the other two remain unchanged.
\begin{equation}
    s_{1L,{\text{CM}}}^{\mu} = \left(\frac{|\vec{p}_{\ell^-}|}{m_\ell}, \frac{E_{\ell} \vec{{p}}_{\ell^-}}{m_\ell |\vec{p}_{\ell^-}|}\right).\label{Slongitudinal}
\end{equation}
To get the four-fold differential decay distribution, with L, N and T polarized lepton, we employ the spin projector $\frac{1}{2}(1+\gamma_5\slashed{s_1})$ in the leptonic tensors. In particular, we make the replacement $(\slashed{p}_{\ell^-}+m_{\ell})\rightarrow(\slashed{p}_{\ell^-}+m_{\ell})\frac{1}{2}(1+\gamma_5\slashed{s}_1)$.

\subsubsection{Longitudinal four-fold angular distribution}\label{Lsection}
For the case of longitudinal lepton polarization, L polarized four-fold angular decay distribution is obtained as
\begin{align}  \frac{d^4\Gamma_{\text{L}}\left(\vec{s}_{\ell^-}=\pm\mathbf{\hat{e}}_L\right)}{dq^2 d\cos{\theta_{\ell}} d\cos{\theta_{\Lambda}} d\phi} =& \frac{3}{8 \pi}\mathcal{B}(\Lambda\to  N\pi)\Big[
    \Big( \mathcal{K}_{\mathrm{1ss \text{L}}} \sin^2{\theta_{\ell}} +  \mathcal{K}_{\mathrm{1cc \text{L}}} \cos^2{\theta_{\ell}} + \mathcal{K}_{\mathrm{1c \text{L}}} \cos{\theta_{\ell}} \Big) \notag \\
    &+ \Big(\mathcal{K}_{\mathrm{2ss \text{L}}} \sin^2{\theta_{\ell}} +  \mathcal{K}_{\mathrm{2cc \text{L}}} \cos^2{\theta_{\ell}} +  \mathcal{K}_{\mathrm{2c \text{L}}} \cos{\theta_{\ell}} \Big) \cos{\theta_{\Lambda}} \notag \\
    &+ \Big(\mathcal{K}_{\mathrm{3sc \text{L}}} \sin{\theta_{\ell}} \cos{\theta_{\ell}} +  \mathcal{K}_{\mathrm{3s \text{L}}}\sin{\theta_{\ell}} \Big) \sin{\phi} \sin{\theta_{\Lambda}} \notag \\
    &+ \Big(\mathcal{K}_{\mathrm{4sc \text{L}}} \sin{\theta_{\ell}} \cos{\theta_{\ell}} +  \mathcal{K}_{\mathrm{4s \text{L}}}\sin{\theta_{\ell}} \Big) \cos{\phi} \sin{\theta_{\Lambda}}\Big].
    \label{eq:lon}
\end{align}
To keep track of the spin orientation of the L polarized lepton, we use the parameter $\xi_L$ and decompose each angular coefficient as follows
\begin{equation}
 \mathcal{K}_{\{\cdots\}\text{L}} = \frac{K_{\{\cdots\}}}{2}+ \xi_L\,\mathcal{K}_{\{\cdots\}\text{L}}^\prime\, ,
 \label{L new form}
\end{equation}
where $\{\cdots\}$ corresponds to the suffixes $1ss, 1cc, 1c, 2ss,2cc,2c,3sc,3s,4sc,4s$. The two terms of $\mathcal{K}_{\{\cdots\}\text{L}}$ correspond to spin orientation independent (unpolarized) and dependent (polarized) parts of each angular coefficient, respectively. The parameter $\xi_L=\pm1$ for $\vec{s}_{\ell^-}=\pm\mathbf{\hat{e}}_L$, respectively.
With longitudinally polarized lepton, the $q^2$ dependent analytical expressions of $\mathcal{K}^{\prime}_{\{\cdots\}}$ in terms of helicity and transversity amplitudes are given in Appendix \ref{AppE} and Appendix \ref{AppF}, respectively.
\subsubsection{Normal four-fold angular distribution}
\label{Nonsection}
In contrast to the case of longitudinal polarization, for the case of a normally polarized lepton, all polarization-dependent terms appear as new angular structures with additional angular coefficients, identified with parameter $\xi_N=\pm1$, and they do not merge with the previously known unpolarized angular structures. The corresponding differential decay distribution becomes
\begin{align}
\frac{d^4\Gamma_{\text{N}}\left(\vec{s}_{\ell^-}=\pm\mathbf{\hat{e}}_N\right)}{dq^2d\cos{\theta_{\ell}}d\cos{\theta_{\Lambda}}d\phi} =&\frac{3}{8 \pi}\mathcal{B}(\Lambda\to  N\pi)\Big[ 
\Big(\mathcal{K}_{\mathrm{1ss\text{N}}} \sin^2{\theta_{\ell}} + \mathcal{K}_{\mathrm{1cc\text{N}}} \cos^2{\theta_{\ell}} +\mathcal{K}_{\mathrm{1c\text{N}}} \cos{\theta_{\ell}}+\xi_N\,\mathcal{K}_{\mathrm{1s\text{N}}}\sin{\theta_{\ell}}\Big) \notag \\
&+
\Big(\mathcal{K}_{\mathrm{2ss\text{N}}} \sin^2{\theta_{\ell}} + \mathcal{K}_{\mathrm{2cc\text{N}}} \cos^2{\theta_{\ell}} + \mathcal{K}_{\mathrm{2c\text{N}}} \cos{\theta_{\ell}} +\xi_N\, \mathcal{K}_{\mathrm{2s\text{N}}} \sin{\theta_{\ell}}\Big)\cos{\theta_{\Lambda}} \notag \\
&+ \Big(\mathcal{K}_{\mathrm{3sc\text{N}}} \sin{\theta_{\ell}} \cos{\theta_{\ell}} + \mathcal{K}_{\mathrm{3s\text{N}}}\sin{\theta_{\ell}} +\xi_N\,\big(\mathcal{K}_{\mathrm{3c\text{N}}}\cos{\theta_{\ell}}+\mathcal{K}_{\mathrm{3\text{N}}}\big)\Big) \sin{\phi} \sin{\theta_{\Lambda}} \notag \\
&+ \Big(\mathcal{K}_{\mathrm{4sc\text{N}}} \sin{\theta_{\ell}} \cos{\theta_{\ell}} + \mathcal{K}_{\mathrm{4s\text{N}}}\sin{\theta_{\ell}} +\xi_N\,\big(\mathcal{K}_{\mathrm{4c\text{N}}}\cos{\theta_{\ell}}+\mathcal{K}_{\mathrm{4 \text{N}}}\big)\Big) \cos{\phi} \sin{\theta_{\Lambda}}\Big].
\label{eq:nor}
\end{align}
The angular coefficient $\mathcal{K}_{\{\cdots\}\text{N}}$ with suffixes $1ss, 1cc, 1c, 2ss,2cc,2c,3sc,3s,4sc,4s$ are related to the corresponding unpolarized angular coefficients by $\mathcal{K}_{\{\cdots\}\text{N}}=\frac{K_{\{\cdots\}}}{2}$. The additional six polarization dependent angular coefficients, are $\mathcal{K}_{\mathrm{1s \text{N}}}$, $\mathcal{K}_{\mathrm{2s\text{N}}}$, $\mathcal{K}_{\mathrm{3c\text{N}}}$, $\mathcal{K}_{\mathrm{3\text{N}}}$, $\mathcal{K}_{\mathrm{4c\text{N}}}$, and $\mathcal{K}_{\mathrm{4\text{N}}}$. Among these six coefficients, two of them are real, i.e., $\mathcal{K}_{\mathrm{3c\text{N}}}$ and $\mathcal{K}_{\mathrm{3\text{N}}}$, while the remaining are imaginary. Their analytical expressions in terms of helicity and transversity amplitudes are given in Appendix \ref{AppE} and Appendix \ref{AppF}, respectively. 
\subsubsection{Transverse four-fold angular distribution}
For the transversely polarized lepton case, twelve additional angular coefficients, identified with parameter $\xi_T=\pm1$, are contributing to the T polarized four-fold differential decay distribution, which is obtained as
\begin{align}
    \frac{d^4\Gamma_{\text{T}}\left(\vec{s}_{\ell^-}=\pm\mathbf{\hat{e}}_T\right)}{dq^2d\cos{\theta_{\ell}}d\cos{\theta_{\Lambda}}d\phi}  = \frac{3}{8 \pi}\mathcal{B}(\Lambda\to  N\pi)&\Big[\Big(
    \mathcal{K}_{\mathrm{1ss\text{T}}} \sin^2{\theta_{\ell}} + \mathcal{K}_{\mathrm{1cc\text{T}}} \cos^2{\theta_{\ell}}   +\mathcal{K}_{\mathrm{1c\text{T}}} \cos{\theta_{\ell}}\notag\\
&+\xi_T\,\big(\mathcal{K}_{\mathrm{1sc\text{T}}}\sin{\theta_{\ell}}\cos{\theta_{\ell}}+\mathcal{K}_{\mathrm{1s\text{T}}} \sin{\theta_{\ell}} \big)\Big)\notag\\
&+\Big(\mathcal{K}_{\mathrm{2ss\text{T}}} \sin^2{\theta_{\ell}} + \mathcal{K}_{\mathrm{2cc\text{T}}} \cos^2{\theta_{\ell}} +  \mathcal{K}_{\mathrm{2c\text{T}}} \cos{\theta_{\ell}}\notag\\
&+\xi_T\,\big( \mathcal{K}_{\mathrm{2sc\text{T}}} \cos{\theta_{\ell}} \sin{\theta_{\ell}}+\mathcal{K}_{\mathrm{2s\text{T}}} \sin{\theta_{\ell}}\big)  \Big) \cos{\theta_{\Lambda}}\notag\\
&+ \sin{\phi} \sin{\theta_{\Lambda}}\Big(\mathcal{K}_{\mathrm{3sc\text{T}}} \sin{\theta_{\ell}} \cos{\theta_{\ell}}+\mathcal{K}_{\mathrm{3s\text{T}}}\sin{\theta_{\ell}}\notag\\
&+\xi_T\,\big(\mathcal{K}_{\mathrm{3ss\text{T}}}\sin^2{\theta_{\ell}}+\mathcal{K}_{\mathrm{3cc\text{T}}}\cos^2{\theta_{\ell}}+\mathcal{K}_{\mathrm{3c\text{T}}}\cos{\theta_{\ell}} +\mathcal{K}_{\mathrm{3 \text{T}}}\big)\Big)\notag\\
&+\cos{\phi} \sin{\theta_{\Lambda}}\Big(\mathcal{K}_{\mathrm{4sc\text{T}}} \sin{\theta_{\ell}} \cos{\theta_{\ell}}  +\mathcal{K}_{\mathrm{4s\text{T}}}\sin{\theta_{\ell}}\notag\\
    &+\xi_T\,\big(\mathcal{K}_{\mathrm{4ss \text{T}}}\sin^2{\theta_{\ell}}+\mathcal{K}_{\mathrm{4cc\text{T}}}\cos^2{\theta_{\ell}}+ \mathcal{K}_{\mathrm{4c\text{T}}}\cos{\theta_{\ell}}+\mathcal{K}_{\mathrm{4 \text{T}}}\big)\Big) \Big].
    \label{eq:trans}
\end{align}
 The angular coefficient $\mathcal{K}_{\{\cdots\}\text{T}}$ with suffixes $1ss, 1cc, 1c, 2ss,2cc,2c,3sc,3s,4sc,4s$ are related to the corresponding unpolarized angular coefficients by $\mathcal{K}_{\{\cdots\}\text{T}}=\frac{K_{\{\cdots\}}}{2}$. We found twelve new angular coefficients in this case. Among them, the four coefficients 
$\mathcal{K}_{\mathrm{1sT}}$, $\mathcal{K}_{\mathrm{2sT}}$, $\mathcal{K}_{\mathrm{1scT}}$, 
and $\mathcal{K}_{\mathrm{2scT}}$ appear in the expressions of the well-known polarized observables, namely the differential branching ratio, hadron forward backward asymmetry, lepton 
forward--backward asymmetry, and lepton hadron forward backward asymmetry, as provided in 
Eqs.~(\ref{DBRpolT}) and (\ref{LHLHApolT}). In addition, four of the newly obtained coefficients are purely imaginary, i.e.,
$\mathcal{K}_{\mathrm{3ssT}}$,
$\mathcal{K}_{\mathrm{3ccT}}$,  
$\mathcal{K}_{\mathrm{3cT}}$, 
$\mathcal{K}_{\mathrm{3T}}$,
while the remaining four are real:
$\mathcal{K}_{\mathrm{4ssT}}$,
$\mathcal{K}_{\mathrm{4ccT}}$,
$\mathcal{K}_{\mathrm{4cT}}$,  
$\mathcal{K}_{\mathrm{4T}}$.
The polarization asymmetries of these real angular coefficients within the SM are illustrated in FIG.~\ref{Tspin}. The analytical expressions of all the angular coefficients are provided in terms of helicity and transversity amplitudes in Appendix \ref{AppE} and Appendix \ref{AppF}, respectively.
 

\section{Physical observables}\label{physObs}
In this section, we present the extraction of various unpolarized and polarized physical observables from the corresponding four-fold differential decay distributions. In addition, we construct several polarization asymmetries from the polarized angular distributions, which constitute one of the main objectives of this study.
 \subsection{Unpolarized lepton case}
Starting from the unpolarized four-fold angular decay distribution given in Eq.~(\ref{eq:finaltrig}), various decay observables can be extracted using the appropriate weight functions, as described in Ref.~\cite{Boer:2014kda}. For completeness, we present below the expressions for the relevant physical observables and the other possible normalized angular coefficients.

\begin{itemize}
\item Integrating over angles, the differential decay rate in terms of the angular coefficient 
 becomes
\begin{eqnarray}
\frac{d\Gamma\left(\Lambda_b \rightarrow \Lambda(\rightarrow  N \pi)\ell^{+}\ell^{-}\right)}{dq^{2}} &=& \mathcal{B}(\Lambda\to  N\pi)\big(2K_{1ss} + K_{1cc}\big)\;.\label{DDRunpol}
\label{DBRunpol}    
\end{eqnarray}
where $\mathcal{B}(\Lambda\to  N\pi)$ is the contribution from the cascade decay $\Lambda\to N\pi$. If we set aside the cascade decay, the decay rate for  $\Lambda_b \rightarrow \Lambda\ell^{+}\ell^{-}$ will then take the form
\begin{eqnarray}
\frac{d\Gamma\left(\Lambda_b \rightarrow \Lambda\ell^{+}\ell^{-}\right)}{dq^{2}} &=& 2K_{1ss} + K_{1cc}\;.\label{DDRunpol2fold}
\end{eqnarray}
The differential branching ratios for $\Lambda_b \rightarrow \Lambda(\rightarrow  N \pi)\ell^{+}\ell^{-}$ and $\Lambda_b \rightarrow \Lambda\ell^{+}\ell^{-}$ decays are given by
\begin{align}
    \frac{d\mathcal{B} \left(\Lambda_b \rightarrow \Lambda(\rightarrow  N \pi)\ell^{+}\ell^{-}\right)}{dq^2} = \tau_{B}  \frac{d\Gamma \left(\Lambda_b \rightarrow \Lambda(\rightarrow  N \pi)\ell^{+}\ell^{-}\right)}{dq^2}, \quad
     \frac{d\mathcal{B} (\Lambda_b \rightarrow \Lambda\ell^{+}\ell^{-})}{dq^2} = \tau_{B} \frac{d\Gamma (\Lambda_b \rightarrow \Lambda\ell^{+}\ell^{-})}{dq^2}.\label{brratioAB}
\end{align}
The binned differential branching ratio can be defined as
\begin{equation}
\left\langle d\mathcal{B}/dq^2\right\rangle_{\left[q^{2}_{\text{min}},\, q^{2}_{\text{max}}\right]}=\frac{\int^{q^{2}_{\text{max}}}_{q^{2}_{\text{min}}} \left(d\mathcal{B}/dq^2\right)\,dq^2}{q^{2}_{\text{max}}-q^{2}_{\text{min}}}. \label{Binned-BR}
\end{equation}
\item The forward-backward asymmetry of leptons, hadron and that of combined lepton-hadron can be calculated as: 
\begin{eqnarray}
\mathcal{A}_{\text{FB}}^{\ell} = \frac{3}{2}\frac{K_{1c}}{(2K_{1ss} + K_{1cc})},\quad 
\mathcal{A}_{\text{FB}}^{\Lambda} = \frac{1}{2}\frac{(2K_{2ss}+K_{2cc})}{(2K_{1ss} + K_{1cc})},\quad
\mathcal{A}_{\text{FB}}^{\ell\Lambda} =\frac{3}{4} \frac{K_{2c}}{(2K_{1ss} + K_{1cc})}.\label{LHLHAunpol}    
\end{eqnarray}
\item The other possible normalized angular coefficients are
\begin{eqnarray}
\hat{K}_{2ss} &=& \frac{K_{2ss}}{2K_{1ss} + K_{1cc}},  \quad 
\hat{K}_{2cc} = \frac{K_{2cc}}{2K_{1ss} + K_{1cc}}, \quad \hat{K}_{3sc} = \frac{K_{3sc}}{2K_{1ss} + K_{1cc}},  \notag\\
\hat{K}_{3s}&=& \frac{K_{3s}}{2K_{1ss} + K_{1cc}},\quad\hat{K}_{4sc} = \frac{K_{4sc}}{2K_{1ss} + K_{1cc}},\quad\hat{K}_{4s}= \frac{K_{4s}}{2K_{1ss} + K_{1cc}}.\label{ACoefunpol}    
\end{eqnarray}
The corresponding binned angular coefficients for the unpolarized lepton case are defined as
\begin{equation}
\hat{K}_{{\{\cdots\}}_{\left[q^{2}_{\text{min}},\, q^{2}_{\text{max}}\right]}}=\frac{\int^{q^{2}_{\text{max}}}_{q^{2}_{\text{min}}}\,{K}_{\{\cdots\}}\,dq^2}{\int^{q^{2}_{\text{max}}}_{q^{2}_{\text{min}}}(d\Gamma/dq^2)\,dq^2}, \label{Binned-Acoefficients}
\end{equation}
where $\{\cdots\}$ corresponds to the suffixes $2ss, 2cc, 3sc, 3s, 4sc, 4s$.

\end{itemize}

\subsection{Polarized longitudinal lepton case}
For the Longitudinally (L) polarized lepton, the spin-dependent observables can be extracted from Eq. (\ref{eq:lon}) by using similar weight functions as used for the unpolarized lepton case. The expressions are given as:
\begin{itemize}
\item The differential decay rate and the differential branching ratio of $\Lambda_b \rightarrow \Lambda(\rightarrow  N \pi)\ell^{+}\ell^{-}$ decay, with longitudinally polarized lepton become
\begin{eqnarray}
\frac{d\Gamma_{\text{L}}\left(\vec{s}_{\ell^-}=\pm\mathbf{\hat{e}}_L\right)}{dq^{2}} &=& \mathcal{B}(\Lambda\to  N\pi)\big(2\mathcal{K}_{1ss\text{L}} + \mathcal{K}_{1cc\text{L}}\big),\notag\\
&=&\mathcal{B}(\Lambda\to  N\pi)\left(K_{1ss} + \frac{K_{1cc}}{2} +\xi_L \left(2\,\mathcal{K}_{1ss\text{L}}^\prime+ \mathcal{K}_{1cc\text{L}}^\prime\right)\right),
\label{DDRpolL}\\
\frac{d\mathcal{B}_{\,\text{L}}\left(\vec{s}_{\ell^-}=\pm\mathbf{\hat{e}}_L\right)}{dq^{2}} &=&\tau_{\Lambda_b}\frac{d\Gamma_{\text{L}}\left(\vec{s}_{\ell^-}=\pm\mathbf{\hat{e}}_L\right)}{dq^{2}}.
\label{DBRpolL}    
\end{eqnarray}

\item The unpolarized differential decay rate can be obtained as
\begin{eqnarray}
\frac{d\Gamma\left(\Lambda_b \rightarrow \Lambda(\rightarrow  N \pi)\ell^{+}\ell^{-}\right)}{dq^{2}}=\frac{d\Gamma_{\text{L}}\left(\vec{s}_{\ell^-}=+\mathbf{\hat{e}}_L\right)}{dq^{2}}+\frac{d\Gamma_{\text{L}}\left(\vec{s}_{\ell^-}=-\mathbf{\hat{e}}_L\right)}{dq^{2}}.
\label{NormpolL}    
\end{eqnarray}

\item The different forward-backward asymmetries for the longitudinally polarized lepton case become
\begin{eqnarray}
\mathcal{A}_{\text{FBL}}^{\ell}\left(\vec{s}_{\ell^-}=\pm\mathbf{\hat{e}}_L\right) &=& \frac{3}{2}\frac{\mathcal{K}_{1c\text{L}}}{(2K_{1ss} + K_{1cc})},\quad 
\mathcal{A}_{\text{FBL}}^{\Lambda}\left(\vec{s}_{\ell^-}=\pm\mathbf{\hat{e}}_L\right) = \frac{1}{2}\frac{(2\mathcal{K}_{2ss\text{L}}+\mathcal{K}_{2cc\text{L}})}{(2K_{1ss} + K_{1cc})},\notag\\
\mathcal{A}_{\text{FBL}}^{\ell\Lambda}\left(\vec{s}_{\ell^-}=\pm\mathbf{\hat{e}}_L\right) &=&\frac{3}{4} \frac{\mathcal{K}_{2c\text{L}}}{(2K_{1ss} + K_{1cc})}.\label{LHLHApolL}    
\end{eqnarray}
\item The other normalized angular coefficients as:
\begin{eqnarray}
\hat{\mathcal{K}}_{2ss\text{L}}\left(\vec{s}_{\ell^-}=\pm\mathbf{\hat{e}}_L\right) &=& \frac{\mathcal{K}_{2ss\text{L}}}{2K_{1ss} + K_{1cc}},  \; 
\hat{\mathcal{K}}_{2cc\text{L}}\left(\vec{s}_{\ell^-}=\pm\mathbf{\hat{e}}_L\right) = \frac{\mathcal{K}_{2cc\text{L}}}{2K_{1ss} + K_{1cc}}, \; \hat{\mathcal{K}}_{3sc\text{L}}\left(\vec{s}_{\ell^-}=\pm\mathbf{\hat{e}}_L\right) = \frac{\mathcal{K}_{3sc\text{L}}}{2K_{1ss} + K_{1cc}}, \notag \\
\hat{\mathcal{K}}_{3s\text{L}}\left(\vec{s}_{\ell^-}=\pm\mathbf{\hat{e}}_L\right)&=& \frac{\mathcal{K}_{3s\text{L}}}{2K_{1ss} + K_{1cc}},\;\hat{\mathcal{K}}_{4sc\text{L}}\left(\vec{s}_{\ell^-}=\pm\mathbf{\hat{e}}_L\right) = \frac{\mathcal{K}_{4sc\text{L}}}{2K_{1ss} + K_{1cc}},\;\hat{\mathcal{K}}_{4s\text{L}}\left(\vec{s}_{\ell^-}=\pm\mathbf{\hat{e}}_L\right)= \frac{\mathcal{K}_{4s\text{L}}}{2K_{1ss} + K_{1cc}}.\notag\\
\label{ACoefpolL}    
\end{eqnarray}
\end{itemize}
\subsection{Polarized normal lepton case}
Starting from four-fold angular decay distribution, with the Normal (N) polarized lepton, given in Eq. (\ref{eq:nor}), different decay observables are constructed as
\begin{itemize}
\item The differential decay rate and the corresponding differential branching ratio will read as:
\begin{eqnarray}
\frac{d\Gamma_{\text{N}}\left(\vec{s}_{\ell^-}=\pm\mathbf{\hat{e}}_N\right)}{dq^{2}} &=& \mathcal{B}(\Lambda\to  N\pi)\big(2\mathcal{K}_{1ss\text{N}} + \mathcal{K}_{1cc\text{N}}+\frac{3\pi}{4}\xi_N\,\mathcal{K}_{1s\text{N}}\big).\label{DDRpolN}\\
\frac{d\mathcal{B}_{\,\text{N}}\left(\vec{s}_{\ell^-}=\pm\mathbf{\hat{e}}_N\right)}{dq^{2}} &=&\tau_{\Lambda_b}\frac{d\Gamma_{\text{N}}\left(\vec{s}_{\ell^-}=\pm\mathbf{\hat{e}}_N\right)}{dq^{2}}.
\label{DBRpolN}    
\end{eqnarray}


\item For normally polarized lepton, the different forward-backward asymmetries become
\begin{eqnarray}
\mathcal{A}_{\text{FBN}}^{\ell}\left(\vec{s}_{\ell^-}=\pm\mathbf{\hat{e}}_N\right) &=& \frac{3}{2}\frac{\mathcal{K}_{1c\text{N}}}{(2K_{1ss} + K_{1cc})},\quad 
\mathcal{A}_{\text{FBN}}^{\Lambda}\left(\vec{s}_{\ell^-}=\pm\mathbf{\hat{e}}_N\right) = \frac{(8\mathcal{K}_{2ss\text{N}}+4\mathcal{K}_{2cc\text{N}}+3\pi \xi_N\,\mathcal{K}_{2s\text{N}})}{8\,(2K_{1ss} + K_{1cc})},\notag\\
\mathcal{A}_{\text{FBN}}^{\ell\Lambda}\left(\vec{s}_{\ell^-}=\pm\mathbf{\hat{e}}_N\right) &=&\frac{3}{4} \frac{\mathcal{K}_{2c\text{N}}}{(2K_{1ss} + K_{1cc})}.\label{LHLHApolN}    
\end{eqnarray}

\item The additional spin-dependent normalized angular coefficients
\begin{eqnarray}
\hat{\mathcal{K}}_{3c\text{N}}\left(\vec{s}_{\ell^-}=\pm\mathbf{\hat{e}}_N\right) &=& \frac{\xi_N\,\mathcal{K}_{3c\text{N}}}{2K_{1ss} + K_{1cc}},  \quad\; 
\hat{\mathcal{K}}_{3\text{N}}\left(\vec{s}_{\ell^-}=\pm\mathbf{\hat{e}}_N\right) = \frac{\xi_N\,\mathcal{K}_{3\text{N}}}{2K_{1ss} + K_{1cc}},\notag \\
\hat{\mathcal{K}}_{4c\text{N}}\left(\vec{s}_{\ell^-}=\pm\mathbf{\hat{e}}_N\right) &=& \frac{\xi_N\,\mathcal{K}_{4c\text{N}}}{2K_{1ss} + K_{1cc}},  \quad\; 
\hat{\mathcal{K}}_{4\text{N}}\left(\vec{s}_{\ell^-}=\pm\mathbf{\hat{e}}_N\right) = \frac{\xi_N\,\mathcal{K}_{4\text{N}}}{2K_{1ss} + K_{1cc}}.
\label{ACoefpolN}    
\end{eqnarray}
\end{itemize}
\subsection{Polarized transverse lepton case}
Similarly, for transversely polarized lepton, the expressions of observables are given as follows:
\begin{eqnarray}
\frac{d\Gamma_{\text{T}}\left(\vec{s}_{\ell^-}=\pm\mathbf{\hat{e}}_T\right)}{dq^{2}} &=& \mathcal{B}(\Lambda\to  N\pi)\big(2\mathcal{K}_{1ss\text{T}} + \mathcal{K}_{1cc\text{T}}+\frac{3\pi}{4}\xi_T\,\mathcal{K}_{1s\text{T}}\big),\label{DDRpolT}\\
\frac{d\mathcal{B}_{\,\text{T}}\left(\vec{s}_{\ell^-}=\pm\mathbf{\hat{e}}_T\right)}{dq^{2}} &=&\tau_{\Lambda_b}\frac{d\Gamma_{\text{T}}\left(\vec{s}_{\ell^-}=\pm\mathbf{\hat{e}}_T\right)}{dq^{2}}.
\label{DBRpolT}    
\end{eqnarray}


\begin{eqnarray}
\mathcal{A}_{\text{FBT}}^{\ell}\left(\vec{s}_{\ell^-}=\pm\mathbf{\hat{e}}_T\right) &=& \frac{1}{2}\frac{(3\mathcal{K}_{1c\text{T}}+2\xi_T\,\mathcal{K}_{1sc\text{T}})}{(2K_{1ss} + K_{1cc})},\quad 
\mathcal{A}_{\text{FBT}}^{\Lambda}\left(\vec{s}_{\ell^-}=\pm\mathbf{\hat{e}}_T\right) = \frac{(8\mathcal{K}_{2ss\text{T}}+4\mathcal{K}_{2cc\text{T}}+3\pi \xi_T\,\mathcal{K}_{2s\text{T}})}{8\,(2K_{1ss} + K_{1cc})},\notag\\
\mathcal{A}_{\text{FBT}}^{\ell\Lambda}\left(\vec{s}_{\ell^-}=\pm\mathbf{\hat{e}}_T\right) &=&\frac{1}{4} \frac{(3\mathcal{K}_{2c\text{T}}+2 \xi_T\,\mathcal{K}_{2sc\text{T}})}{(2K_{1ss} + K_{1cc})}.\label{LHLHApolT}    
\end{eqnarray}
Similarly, the additional spin-dependent normalized angular coefficients
\begin{eqnarray}
\hat{\mathcal{K}}_{\{\cdots\}\text{T}}\left(\vec{s}_{\ell^-}=\pm\mathbf{\hat{e}}_T\right) &=& \frac{\xi_T\,\mathcal{K}_{\{\cdots\}\text{T}}}{2K_{1ss} + K_{1cc}},
\label{ACoefpolTTT}    
\end{eqnarray}
where $\{\cdots\}$ corresponds to the suffixes $3ss, 3cc, 3c, 3, 4ss, 4cc, 4c, 4$. Furthermore, the binned differential branching ratios $\left\langle{d\mathcal{B}_{\,\text{i}}\left(\vec{s}_{\ell^-}=\pm\mathbf{\hat{e}}_i\right)}/{dq^{2}} \right\rangle_{\left[q^{2}_{\text{min}},\, q^{2}_{\text{max}}\right]}$ and the binned angular coefficients $\hat{\mathcal{K}}_{\{\cdots\}\text{i}}\left(\vec{s}_{\ell^-}=\pm\mathbf{\hat{e}}_i\right)_{\left[q^{2}_{\text{min}},\, q^{2}_{\text{max}}\right]}$, for $i=\text{L}$, $\text{N}$, and $\text{T}$ polarized lepton can be defined in analogy to Eqs. (\ref{Binned-BR}) and (\ref{Binned-Acoefficients}).


\subsection{Polarization asymmetries observables}
\label{Polarizationasymmetriesobservables}
\begin{itemize}

\item The longitudinal, normal and transverse single lepton polarization asymmetry can be defined as
\begin{equation}
    P_{i}(q^2)=\frac{\frac{d\Gamma_{i}\left(\vec{s}_{\ell^-}=+\mathbf{\hat{e}}_i\right)}{dq^2}-\frac{d\Gamma_{i}\left(\vec{s}_{\ell^-}=-\mathbf{\hat{e}}_i\right)}{dq^2}}{\frac{d\Gamma_{i}\left(\vec{s}_{\ell^-}=+\mathbf{\hat{e}}_i\right)}{dq^2}+\frac{d\Gamma_{i}\left(\vec{s}_{\ell^-}=-\mathbf{\hat{e}}_i\right)}{dq^2}},
    \label{34}
\end{equation}
where the polarization $i=\text{L}$, $\text{N}$, and $\text{T}$, dependent decay widths are given in Eqs. (\ref{DDRpolL}), (\ref{DDRpolN}), and (\ref{DDRpolT}), respectively. Further, unit vector $\hat{e}_i$, with plus and minus designate the spin of the lepton along and opposite to the corresponding polarized unit vector, respectively. The corresponding expressions are given as
\begin{eqnarray}
P_{L} = \frac{2\left(2\,\mathcal{K}_{1ss\text{L}}^\prime+ \mathcal{K}_{1cc\text{L}}^\prime\right)}{(2K_{1ss} + K_{1cc})},\quad 
P_{N} = \frac{3\pi}{2}\frac{\mathcal{K}_{1s\text{N}}}{(2K_{1ss} + K_{1cc})},\quad
P_{T} =\frac{3\pi}{2} \frac{\mathcal{K}_{1s\text{T}}}{(2K_{1ss} + K_{1cc})}.\label{SLpolA}    
\end{eqnarray}
Also, the binned single lepton polarization asymmetry is defined as
\begin{equation}
    \left\langle P_{i}\right\rangle_{\left[q^{2}_{\text{min}},\, q^{2}_{\text{max}}\right]}=\frac{\int^{q^{2}_{\text{max}}}_{q^{2}_{\text{min}}}\left[\frac{d\Gamma_{i}\left(\vec{s}_{\ell^-}=+\mathbf{\hat{e}}_i\right)}{dq^2}-\frac{d\Gamma_{i}\left(\vec{s}_{\ell^-}=-\mathbf{\hat{e}}_i\right)}{dq^2}\right]\,dq^2}{\int^{q^{2}_{\text{max}}}_{q^{2}_{\text{min}}}\left[\frac{d\Gamma_{i}\left(\vec{s}_{\ell^-}=+\mathbf{\hat{e}}_i\right)}{dq^2}+\frac{d\Gamma_{i}\left(\vec{s}_{\ell^-}=-\mathbf{\hat{e}}_i\right)}{dq^2}\right]\,dq^2}.
    \label{345binned}
\end{equation}

\item The singly polarized forward-backward asymmetry of leptons, hadron and that of the combined lepton-hadron are defined as
\begin{eqnarray}
\mathcal{\overline{A}}_{\text{FBi}}^{j}(q^2) = \mathcal{A}_{\text{FBi}}^{j}\left(\vec{s}_{\ell^-}=+\mathbf{\hat{e}}_i\right)-\mathcal{A}_{\text{FBi}}^{j}\left(\vec{s}_{\ell^-}=-\mathbf{\hat{e}}_i\right),\label{SLpolLHLHA}    
\end{eqnarray}
where, $j$ belongs to the type of forward-backward asymmetry, i.e., $j=\ell$, $\Lambda$, and $\ell\Lambda$, while $i=\text{L}$, $\text{N}$, and $\text{T}$. The expressions of different forward-backward asymmetries for the L, N and T polarized lepton case become
\begin{align}
&\mathcal{\overline{A}}_{\text{FBL}}^{\ell} = \frac{3\,\mathcal{K}_{1c\text{L}}^{\prime}}{(2K_{1ss} + K_{1cc})},
&\mathcal{\overline{A}}_{\text{FBL}}^{\Lambda} = \frac{(2\mathcal{K}_{2ss\text{L}}^{\prime}+\mathcal{K}_{2cc\text{L}}^{\prime})}{(2K_{1ss} + K_{1cc})},\quad\quad\,\,\,
&\mathcal{\overline{A}}_{\text{FBL}}^{\ell\Lambda} =\frac{3}{2} \frac{\mathcal{K}_{2c\text{L}}^{\prime}}{(2K_{1ss} + K_{1cc})}.\label{SLpolLHLHAexpL}\\    
&\mathcal{\overline{A}}_{\text{FBN}}^{\ell} = 0, 
& \mathcal{\overline{A}}_{\text{FBN}}^{\Lambda} =\frac{3\pi}{4} \frac{\mathcal{K}_{2s\text{N}}}{(2K_{1ss} + K_{1cc})},\quad\quad
&\mathcal{\overline{A}}_{\text{FBN}}^{\ell\Lambda} =0.\label{SLpolLHLHAexpN}\\    
&\mathcal{\overline{A}}_{\text{FBT}}^{\ell} = \frac{2\,\mathcal{K}_{1sc\text{T}}}{(2K_{1ss} + K_{1cc})},
&\mathcal{\overline{A}}_{\text{FBT}}^{\Lambda} =\frac{3\pi}{4} \frac{\mathcal{K}_{2s\text{T}}}{(2K_{1ss} + K_{1cc})},\quad\quad
&\mathcal{\overline{A}}_{\text{FBT}}^{\ell\Lambda} = \frac{\mathcal{K}_{2sc\text{T}}}{(2K_{1ss} + K_{1cc})}.\label{SLpolLHLHAexpT} 
\end{align}
\item The other polarization asymmetries in the normalized individual angular coefficients are defined as
\begin{eqnarray}
\mathcal{\overline{K}}_{\{\cdots\}\,i} = \hat{\mathcal{K}}_{\{\cdots\}\,i}\left(\vec{s}_{\ell^-}=+\mathbf{\hat{e}}_i\right)-\hat{\mathcal{K}}_{\{\cdots\}\,i}\left(\vec{s}_{\ell^-}=-\mathbf{\hat{e}}_i\right),\label{SLpolangobs}
\end{eqnarray}
where $i=\text{L}$, $\text{N}$, and $\text{T}$, and $\{\cdots\}$ refers only to the suffixes of the spin-dependent angular coefficients in each case.
\end{itemize}

\section{Phenomenological Analysis}
\label{Phenom}
In this section, we analyze the various physical observables discussed above in the SM and also in different NP scenarios. To do so, the input parameters that we used are given in TABLE \ref{numericalvalues}. 
\begin{table}[H]
\caption{Values of the used input parameters.}\label{numericalvalues}
\centering
\begin{tabular}{|cccc|}
\hline \hline
\( \text{Constant} \) & \( \text{Values} \)  & \( \text{Constant} \) & \( \text{Values} \) \\
\hline
\( G_F \) & \( 1.166378 \times 10^{-5} \) GeV\(^{-2}\) & \( |V_{tb}V^*_{ts}| \) & \( 0.0397^{+0.0008}_{-0.0006} \) \\
\( m_b \) & \( 4.18^{+0.03}_{-0.02} \) GeV & \( \alpha_{em}(m_b) \) & 1/133.28 \\
\( \alpha_s(m_b) \) & 0.2233 & \( \alpha \) & \( 0.642\pm 0.013 \)  \\
\( m_{\mu} \) & 0.106 GeV & \(m_b^{\text{pole}} \) & \( 4.91 \pm 0.12 \) GeV \\
$\mathcal{B}\left(\Lambda\to p \pi^-\right)$ & \( 0.64 \,\% \)  & 
\( m_c^{\text{pole}} \) & \( 1.77\pm 0.14\) GeV \\
$\mathcal{B}\left(\Lambda\to n \pi^0\right)$ & $0.36 \,\% $ & \( \mu \) & 5 GeV \\
\( m_{\Lambda_b} \) & 5.619 GeV  &  \( \tau_{\Lambda_b} \) & \( (1.471 \pm 0.009) \times 10^{-12} \) s \\
\( m_{\Lambda} \) &1.116 GeV & & \\
\hline \hline
\end{tabular}
\end{table}
Within our simplified factorization approach, the essential hadronic input functions are the $\Lambda_b\rightarrow \Lambda$ FFs \cite{Detmold:2016pkz}. To this end, we have used high precision Lattice QCD calculation of the relevant form
factors. The FFs used in \cite{Detmold:2016pkz} are related to our notation of the FFs as $f^V_{t, 0, \perp}=f_{0, +, \perp}$,
$f^A_{t, 0, \perp}=g_{0, +, \perp}$, $f^T_{0, \perp}=h_{+, \perp}$, and $f^{T5}_{0, \perp}=\tilde h_{+, \perp}$. To obtain the $q^2$ dependence and to estimate the uncertainties, the lattice calculations are
fitted to two $z-$parameterizations, the so-called “nominal” fit and the “higher-order” fit. For the sake of completeness, the values of the fit parameters for the two cases, taken from Ref. \cite{Detmold:2016pkz}, are presented in Tables \ref{NFFFTable} and \ref{HOFFFTable}, respectively. However, the lengthy correlation matrices of the form factors can be seen in Ref. \cite{Detmold:2016pkz}. 

\begin{table}[t!]
\caption{Central values along with uncertainties of the parameters for the nominal fit form factors \cite{Detmold:2016pkz}.}\label{NFFFTable}
\centering
\begin{tabular}{|c c c c|}
\hline\hline
Parameter & Value & Parameter & Value \\
\hline

$a^{f^V_{0}}_{0}$  & $0.4221 \pm 0.0188$ & $a^{f^A_{t}}_{1}$  & $-1.0290 \pm 0.1614$ \\

$a^{f^V_{0}}_{1}$  & $-1.1386 \pm 0.1683$ & $a^{f^A_{\perp}}_{1}$  & $-1.1357 \pm 0.1911$ \\

$a^{f^V_{t}}_{0}$  & $0.3725 \pm 0.0213$ & $a^{f^T_{0}}_{0}$  & $0.4960 \pm 0.0258$ \\

$a^{f^V_{t}}_{1}$  & $-0.9389 \pm 0.2250$ & $a^{f^T_{0}}_{1}$  & $-1.1275 \pm 0.2537$ \\

$a^{f^V_{\perp}}_{0}$  & $0.5182 \pm 0.0251$ & $a^{f^{T}_{\perp}}_{0}$  & $0.3876 \pm 0.0172$ \\

$a^{f^V_{\perp}}_{1}$  & $-1.3495 \pm 0.2413$ & $a^{f^{T}_{\perp}}_{1}$  & $-0.9623 \pm 0.1550$ \\

$a^{f^A_{0}}_{0}$, $a^{f^A_{\perp}}_{0}$ & $0.3563 \pm 0.0142$ & $a^{f^{T5}_{0}}_{0}, a^{f^{T5}_{\perp}}_{0}$  & $0.3403 \pm 0.0133$ \\

$a^{f^A_{0}}_{1}$  & $-1.0612 \pm 0.1678$ & $a^{f^{T5}_{0}}_{1}$  & $-0.7697 \pm 0.1612$ \\

$a^{f^A_{t}}_{0}$  & $0.4028 \pm 0.0182$ & $a^{f^{T5}_{\perp}}_{1}$  & $-0.8008 \pm 0.1537$ \\

\hline\hline
\end{tabular}
\end{table}

\begin{table}[h!]
\caption{Central values along with uncertainties of the parameters for higher-order fit form factors \cite{Detmold:2016pkz}.}\label{HOFFFTable}
\centering
\begin{tabular}{|c c c c|}
\hline\hline
Parameter & Value & Parameter & Value \\
\hline

$a^{f^V_{0}}_{0}$  & $0.4229 \pm 0.0274$ & $a^{f^A_{t}}_{2}$  & $1.1490 \pm 1.0327$ \\

$a^{f^V_{0}}_{1}$  & $-1.3728 \pm 0.3068$ & $a^{f^A_{\perp}}_{1}$  & $-1.3607 \pm 0.2949$ \\

$a^{f^V_{0}}_{2}$  & $1.7972 \pm 1.1506$ & $a^{f^A_{\perp}}_{2}$  & $2.4621 \pm 1.3711$ \\

$a^{f^V_{t}}_{0}$  & $0.3604 \pm 0.0277$ & $a^{f^T_{0}}_{0}$  & $0.4753 \pm 0.0423$ \\

$a^{f^V_{t}}_{1}$  & $-0.9248 \pm 0.3453$ & $a^{f^T_{0}}_{1}$  & $-0.8840 \pm 0.3997$ \\

$a^{f^V_{t}}_{2}$  & $0.9861 \pm 1.1988$ & $a^{f^T_{0}}_{2}$  & $-0.8190 \pm 1.6760$ \\

$a^{f^V_{\perp}}_{0}$  & $0.5148 \pm 0.0353$ & $a^{f^{T}_{\perp}}_{0}$  & $0.3745 \pm 0.0313$ \\

$a^{f^V_{\perp}}_{1}$  & $-1.4781 \pm 0.4030$ & $a^{f^{T}_{\perp}}_{1}$  & $-0.9439 \pm 0.2766$ \\

$a^{f^V_{\perp}}_{2}$  & $1.2496 \pm 1.6396$ & $a^{f^{T}_{\perp}}_{2}$  & $1.1606 \pm 1.0757$ \\

$a^{f^A_{0}}_{0}, a^{f^A_{\perp}}_{0}$  & $0.3522 \pm 0.0205$ & $a^{f^{T5}_{0}}_{0}, a^{f^{T5}_{\perp}}_{0}$  & $0.3256 \pm 0.0248$ \\

$a^{f^A_{0}}_{1}$  & $-1.2968 \pm 0.2732$ & $a^{f^{T5}_{0}}_{1}$  & $-0.9603 \pm 0.2303$ \\

$a^{f^A_{0}}_{2}$  & $2.7106 \pm 1.0665$ & $a^{f^{T5}_{0}}_{2}$  & $2.9780 \pm 1.0041$ \\

$a^{f^A_{t}}_{0}$  & $0.4059 \pm 0.0267$ & $a^{f^{T5}_{\perp}}_{1}$  & $-0.9634 \pm 0.2268$ \\

$a^{f^A_{t}}_{1}$  & $-1.1622 \pm 0.2929$ & $a^{f^{T5}_{\perp}}_{2}$  & $2.4782 \pm 0.9549$ \\

\hline\hline
\end{tabular}
\end{table}

To see the impact of the vector and axial vector NP WCs given in Eq. (\ref{Heff main}), we use the global fit data from Table 9 of Ref. \cite{Alguero:2023jeh}, and the best-fit values with $1\sigma$ intervals are summarized in Table \ref{tab:bestfitWC}. Here, we have selected only those NP scenarios that exhibit relatively greater pull \cite{Alguero:2023jeh}, are distinguishable in the observables we considered, and can be easily realized within specific NP models. For simplicity, we have renamed them as scenarios S1, S2, S3, and S4 in Table \ref{tab:bestfitWC}. The NP WCs in Eq. (\ref{Heff main}) are related to those in Table \ref{tab:bestfitWC} within the NP hypothesis that includes both LFU contributions (universal across lepton flavors) and LFUV contributions affecting muons only, i.e.,
\begin{eqnarray}
C_{i^{(\prime)}e, \,\tau}^{\text{NP}}=C_{i^{(\prime)}}^{\text{U}},\qquad\quad C_{i^{(\prime)}\mu}^{\text{NP}}=C_{i^{(\prime)}}^{\text{U}}
+C_{i^{(\prime)}\mu}^{\text{V}},\label{WC910}
\end{eqnarray}
where the superscript ``U'' and ``V'' refer to the LFU and LFUV contribution, respectively, and $i=9, 10$. The selected NP scenarios can be realized mainly in $Z^{\prime}$ and leptoquark models. For instance, in the $Z^{\prime}$ extensions, S1 can emerge with both vector and axial couplings to muons, while S2 can be realized with pure vector NP couplings to muons. Additionally, S3 can be generated via off-shell photon penguins \cite{Crivellin:2018yvo} in a LQ model, while S4 can be generated in $Z^{\prime}$ model with vector couplings
to muons and additional vector-like quarks with the quantum numbers of left-handed quarks doublets \cite{Bobeth:2016llm}.

\begin{table*}[t!]
\begin{center}
 \caption{\small Best-fit values of the WCs and the $1\sigma$ ranges of the different NP scenarios, as presented in \cite{Alguero:2023jeh}.}\label{tab:bestfitWC}
			\begin{tabular}{|clcc|}
				\hline\hline
				Scenario & & Best-fit value & $1\sigma$ \\
				\hline
                S1 & $C_{9\mu}^{\text{V}}$                          &$-1.02$       & $[-1.43, -0.61]$  \\
                 & $C_{10\mu}^{\text{V}}$   &$-0.35$       & $[-0.75, -0.00]$  \\
                 & $C_{9}^{\text{U}}=C_{10}^{\text{U}}$   &$+0.19$       & $[-0.16, +0.58]$  \\
                S2 & $C_{9\mu}^{\text{V}}$   &$-0.21$       & $[-0.39, -0.02]$  \\
                   & $C_{9}^{\text{U}}$                            &$-0.97$       & $[-1.21, -0.72]$  \\
                S3 & $C_{9\mu}^{\text{V}}=-C_{10\mu}^{\text{V}}$     &$-0.08$       & $[-0.14, -0.02]$  \\
                   & $C_{9}^{\text{U}}$                              &$-1.10$       & $[-1.27, -0.91]$  \\
                S4 & $C_{9\mu}^{\text{V}}$                           &$-0.68$       & $[-0.84, -0.52]$  \\
                   & $C_{10^{\prime}}^{\text{U}}$                    &$-0.03$       & $[-0.15, +0.09]$  \\
                \hline\hline
	\end{tabular}
	\end{center}
\end{table*}
In the next step, with these input parameters, the task is to perform the phenomenological analysis of the above mentioned physical observables which correspond to the L, N and T polarization of a single lepton in the decay process \( \Lambda_b \to  \Lambda(\to N\pi) \mu^+ \mu^- \). For all results, lepton is taken as the muon. Furthermore, we omit predictions in the intermediate $q^2$ region due to significant non-local hadronic effects, particularly from charmonium resonances such as $J/\psi$ and $\psi(2S)$.
\subsection{Lepton polarization effects in various angular observables within SM}
\subsubsection{Longitudinal polarization}\label{L PDO}
In FIG.~\ref{Longitudinal}, we present the differential branching ratios and other angular observables as functions of $q^2$, comparing the unpolarized case with the scenario where the final-state lepton is longitudinally polarized. For predicting the differential branching ratios, we consider the values of the branching ratios of the cascade decays, $\mathcal{B}\left(\Lambda\to p \pi^-\right)$, and $\mathcal{B}\left(\Lambda\to n \pi^0\right)$, from the particle data group (PDG) \cite{ParticleDataGroup:2024cfk}, which are also reported in TABLE \ref{numericalvalues}. The bands in these plots originate from the errors in the input parameters, mainly the form factors. From the plots, we can read that for the longitudinally polarized lepton case in the SM:
\begin{itemize}
    \item The differential branching ratios  for the spin $-\tfrac{1}{2}$ case of the polarized lepton and the unpolarized lepton cases nearly coincide in the low, intermediate, and the high-$q^2$ range. However, the contribution from the spin $+\tfrac{1}{2}$ component remains negligibly small across almost the entire $q^2$ domain. Consequently, the differential branching ratios with the longitudinally polarized lepton case do not offer a very promising opportunity to probe the effects of lepton polarization and the different spin possibilities.
\begin{figure}[t]
\centering
\scalebox{1}{
\begin{tabular}{ccc} 
\includegraphics[width=2.2in,height=2.2in]{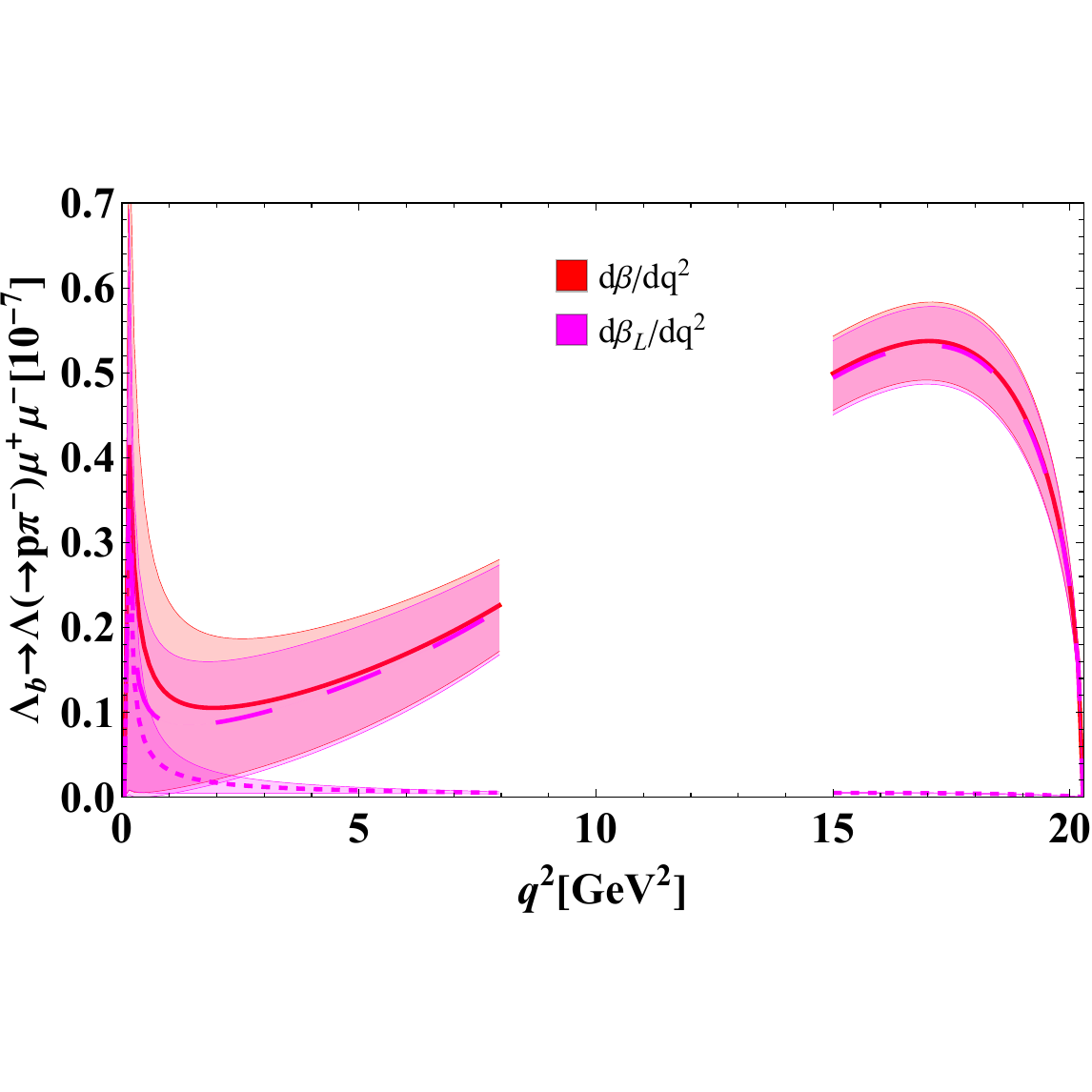} & 
\includegraphics[width=2.2in,height=2.2in]{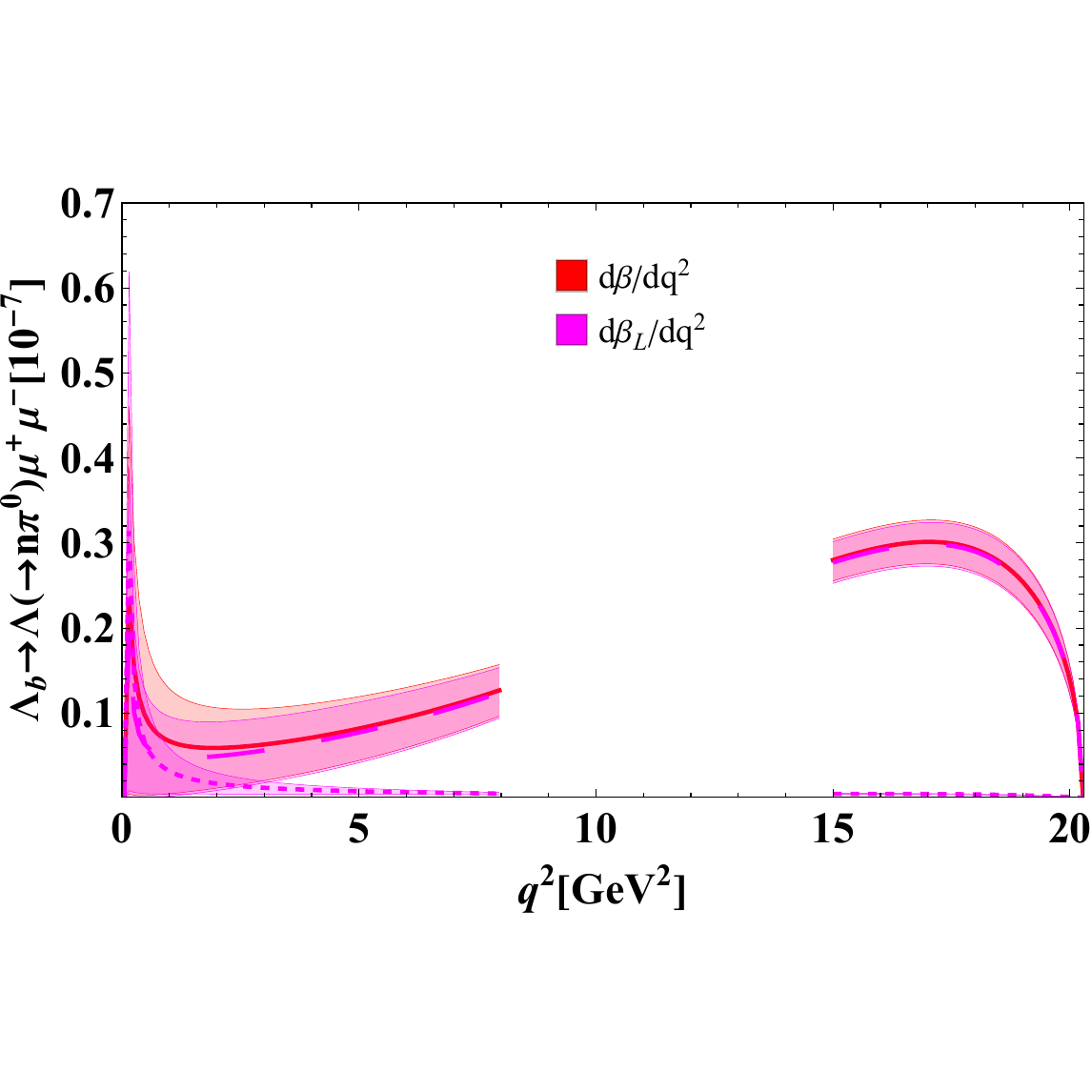 } & 
\includegraphics[width=2.2in,height=2.2in]{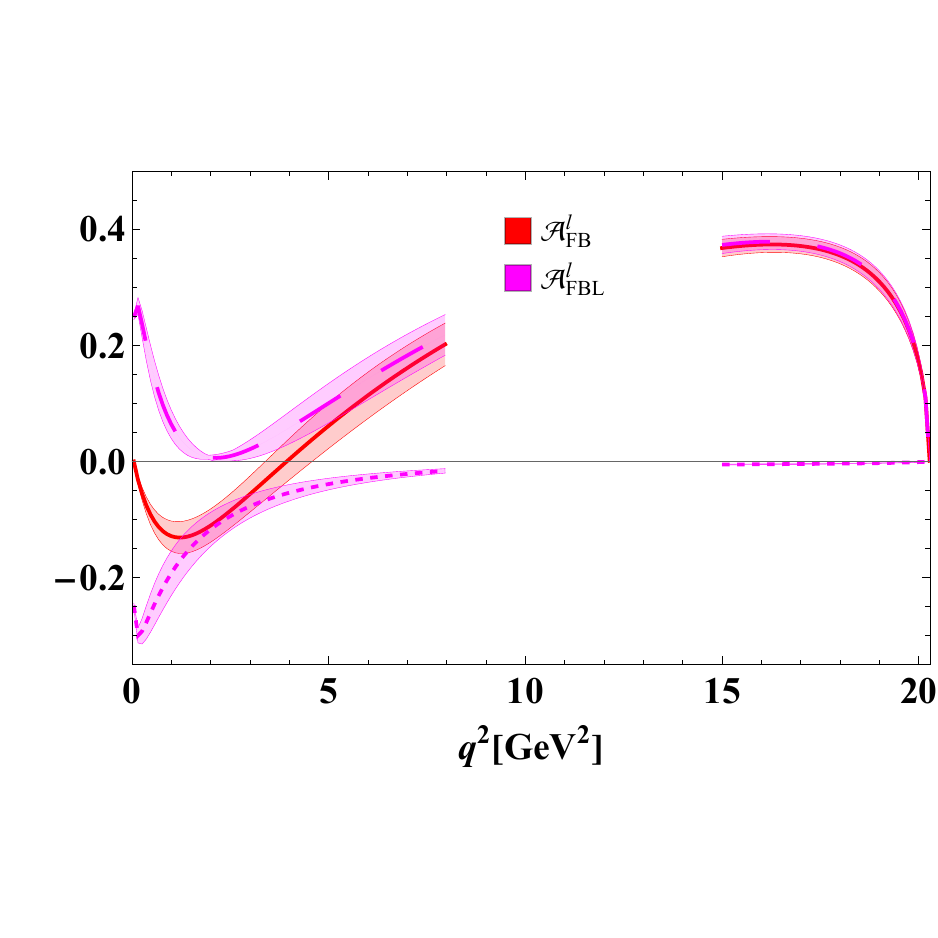}  \vspace{-2cm} \\ 
\includegraphics[width=2.2in,height=2.2in]{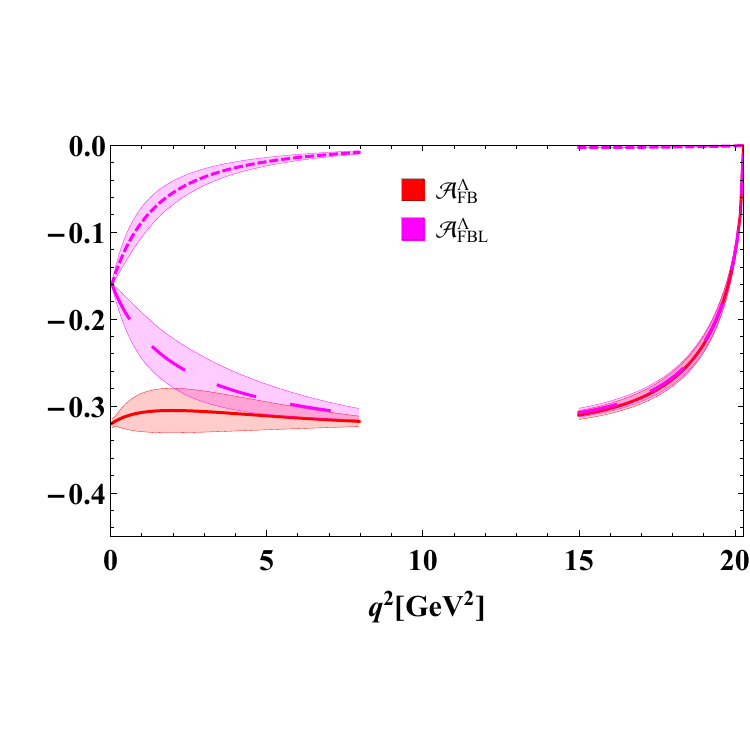} & 
\includegraphics[width=2.2in,height=2.2in]{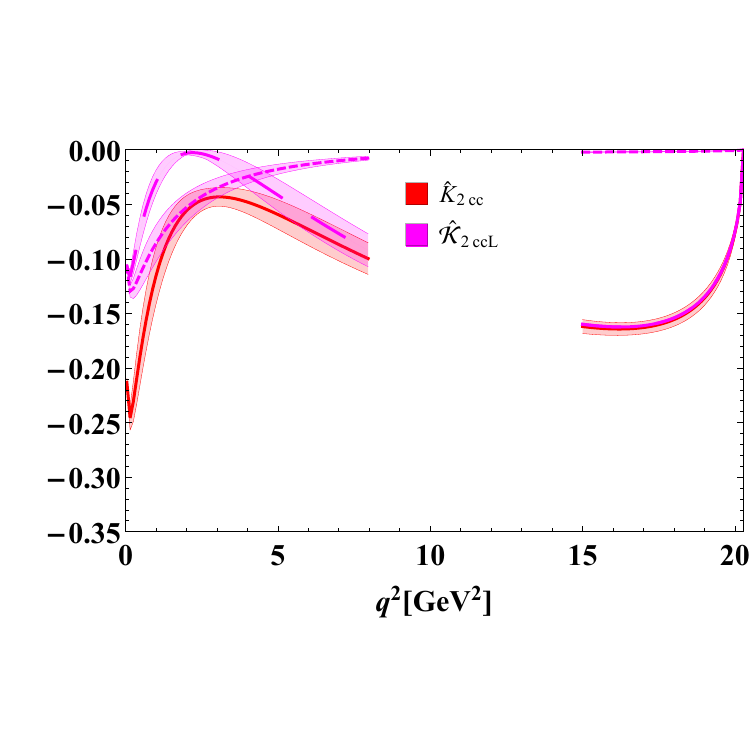 } & 
\includegraphics[width=2.2in,height=2.2in]{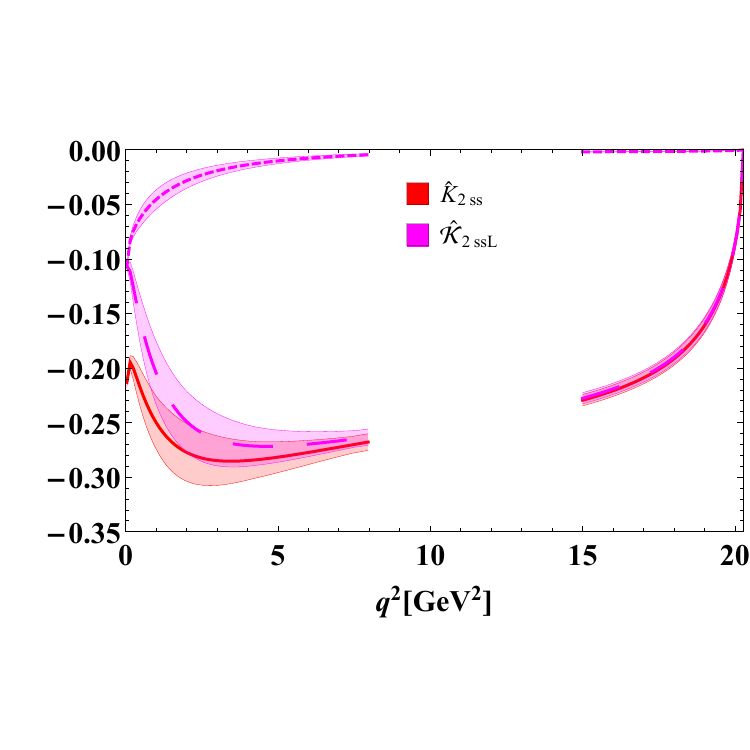 }\vspace{-2cm} \\ \includegraphics[width=2.2in,height=2.2in]{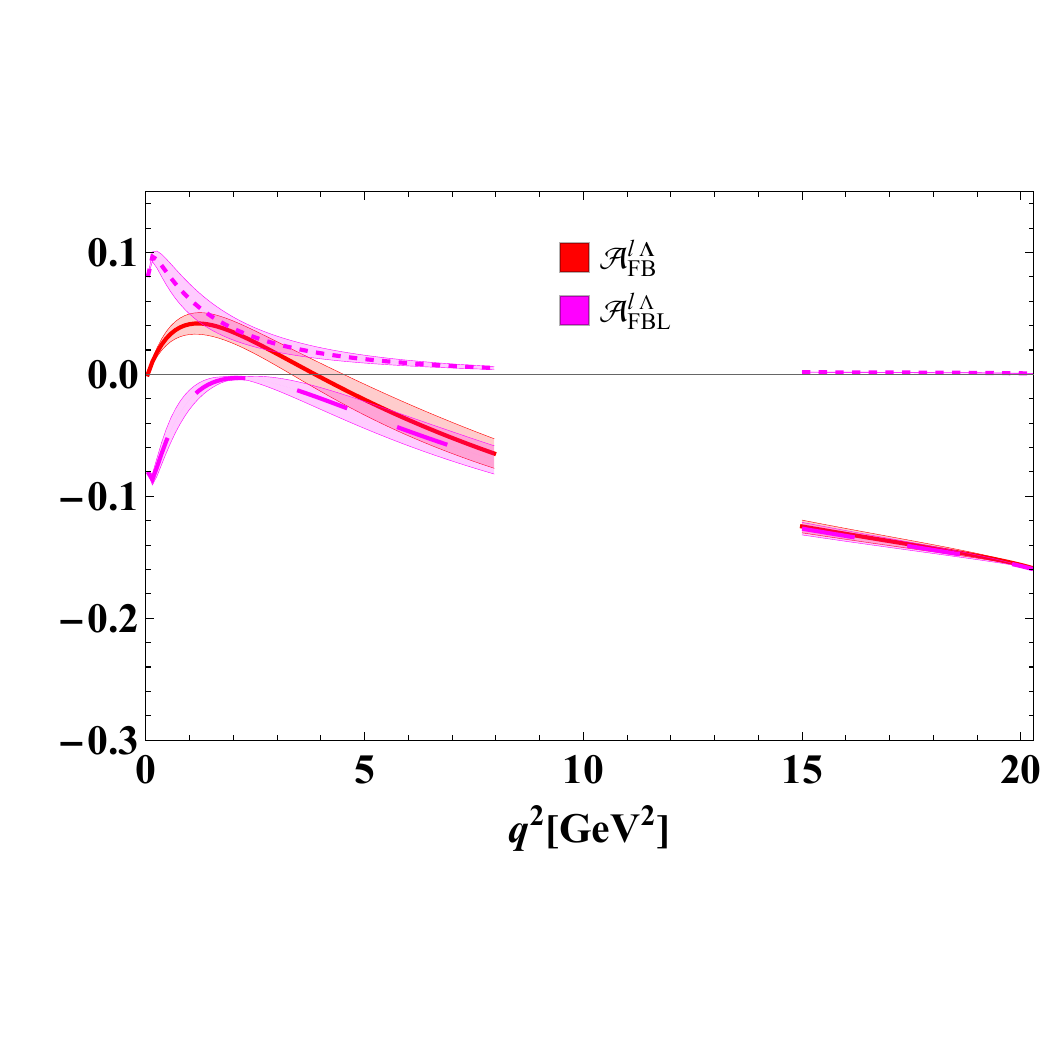} & 
\includegraphics[width=2.2in,height=2.2in]{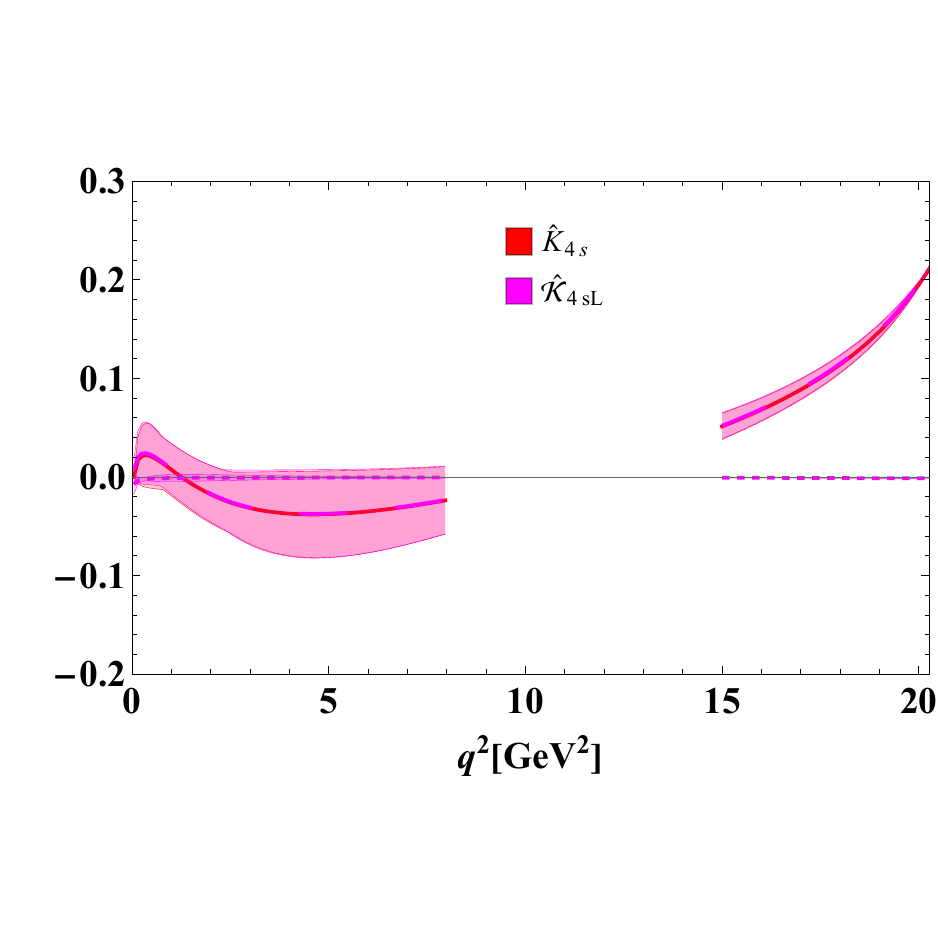}  
& 
\includegraphics[width=2.2in,height=2.2in]{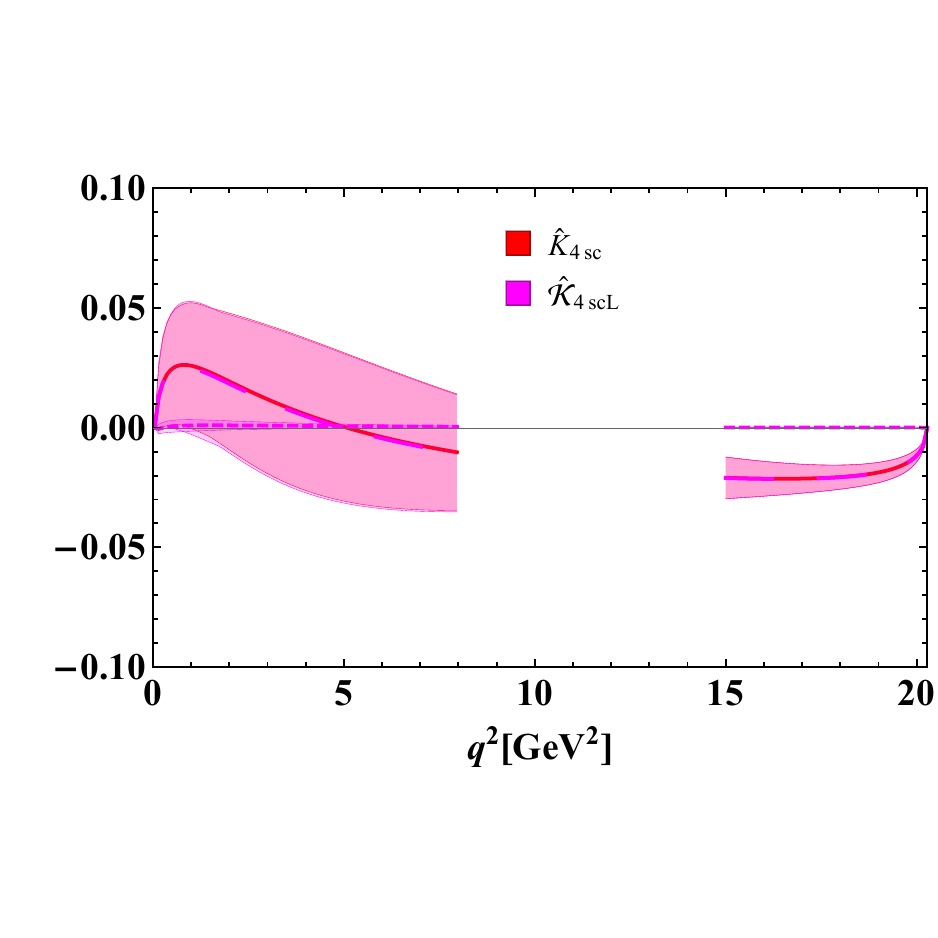}\vspace{-1.0cm}
\end{tabular}}
\caption{Differential branching ratios and various observables vs. $q^{2}$ for SM unpolarized and longitudinally polarized final state lepton. Dotted line represent the spin $+\tfrac{1}{2}$ contributions, while the dashed line represent the spin $-\tfrac{1}{2}$ contributions. For all results $\ell=\mu$.} \label{Longitudinal}
\end{figure}   
\item The zero-crossing point of the lepton forward-backward asymmetry, \( \mathcal{A}_{\text{FB}}^\ell \), is a key observable for probing the SM. As illustrated in FIG.~\ref{Longitudinal}, the polarized case allows for a clear separation of the contributions from different helicity states. Notably, there is no zero-crossing observed for either spin configuration. The spin \( -\tfrac{1}{2} \) contribution remains positive throughout the entire \( q^2 \) range. At high \( q^2 \), the total contribution arises solely from the spin \( -\tfrac{1}{2} \) state, whereas in the low \( q^2 \) region, both spin components contributed.
\item Just like \( \mathcal{A}_{\text{FB}}^\ell \), the unpolarized and polarized cases in \( \mathcal{A}_{\text{FB}}^{\ell\Lambda} \) are clearly distinguishable, with the contributions from the spin \( -\tfrac{1}{2} \) and \( +\tfrac{1}{2} \) helicity states clearly separated in the whole $q^2$ region. As in the previous case, no zero-crossing is observed for the polarized lepton, nor for either helicity component.
\item In the case of the forward-backward asymmetry associated with the final-state \( \Lambda \), i.e., \( \mathcal{A}_{\text{FB}}^\Lambda \), the distribution in the polarized lepton scenario differs significantly from the unpolarized case across the entire \( q^2 \) range. The trend for the polarized lepton with spin \( +\tfrac{1}{2} \) approaches zero from the negative side, in contrast to both the unpolarized case and the polarized case with spin \( -\tfrac{1}{2} \). 
\item In FIG.~\ref{Longitudinal}, we observe that, except for \( \hat{\mathcal{K}}_{2ccL} \) and \( \hat{\mathcal{K}}_{2ssL} \), the results for polarized leptons and individual lepton helicity states are largely masked by the theoretical uncertainties within the SM. In the case of \( \hat{\mathcal{K}}_{2ccL} \), the contribution from the polarized lepton with spin \( +\tfrac{1}{2} \) increases in the low to intermediate \( q^2 \) region, which is in clear contrast to both the spin \( -\tfrac{1}{2} \) and unpolarized cases. On the other hand, for \( \hat{\mathcal{K}}_{2ssL} \), despite the value slightly higher than the unpolarized case, the behavior of the polarized lepton with spin \( -\tfrac{1}{2} \) closely follows that of the unpolarized distribution, while the contribution from spin \( +\tfrac{1}{2} \) remains nearly zero across the entire \( q^2 \) range. These contrasting behaviors of \( \hat{\mathcal{K}}_{2ccL} \) and \( \hat{\mathcal{K}}_{2ssL} \) also stand out relative to the other angular observables, where such helicity-dependent distinctions are not pronounced.

\end{itemize}

\subsubsection{Transverse polarization}
\begin{figure}[t!]
\centering
\scalebox{1}{
\begin{tabular}{ccc} 
\includegraphics[width=2.2in,height=2.2in]{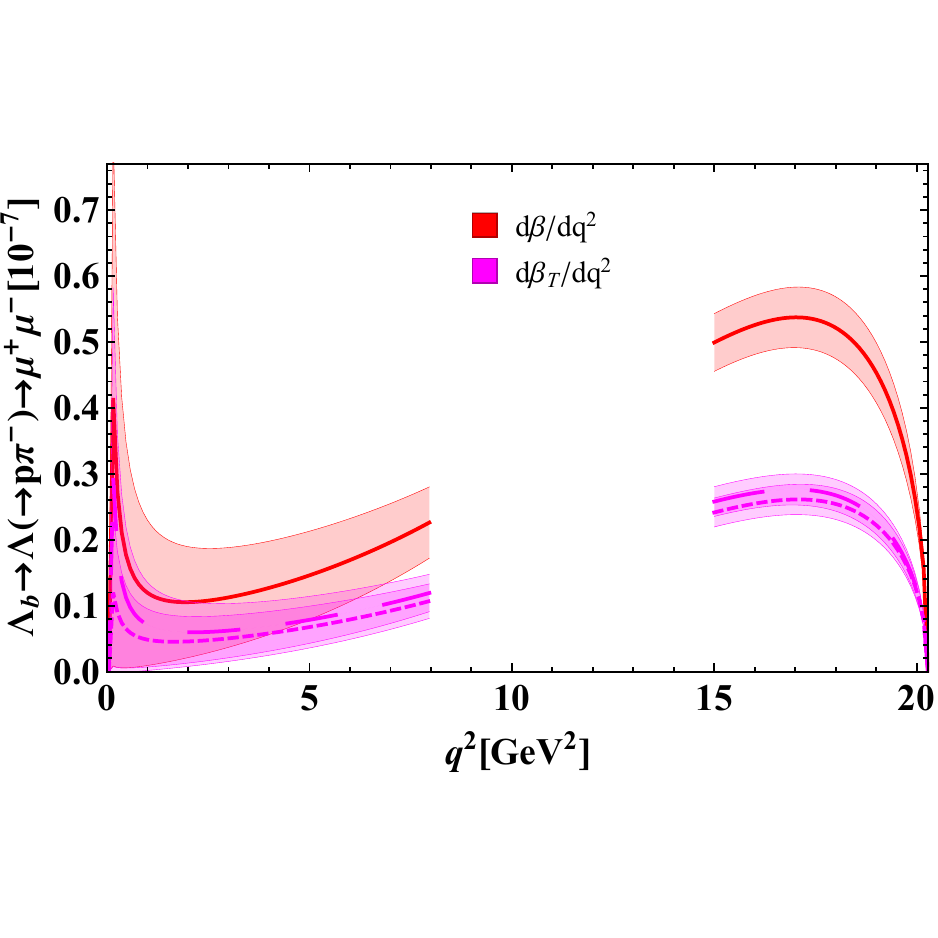} & 
\includegraphics[width=2.2in,height=2.2in]{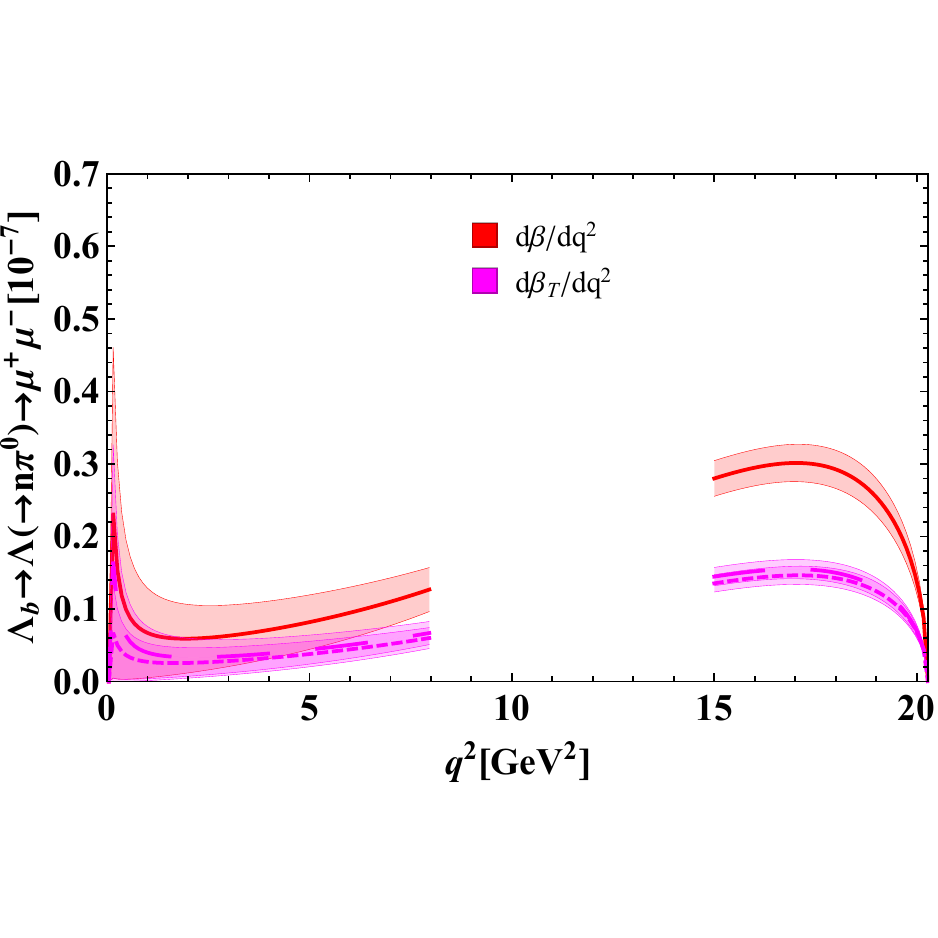} & 
\includegraphics[width=2.2in,height=2.2in]{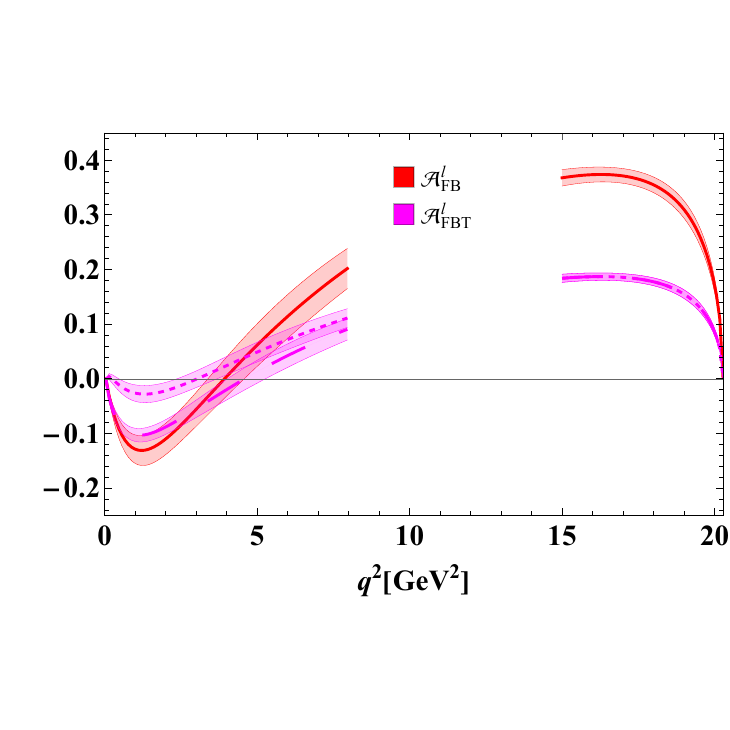}  \vspace{-2cm} \\ 
\includegraphics[width=2.2in,height=2.2in]{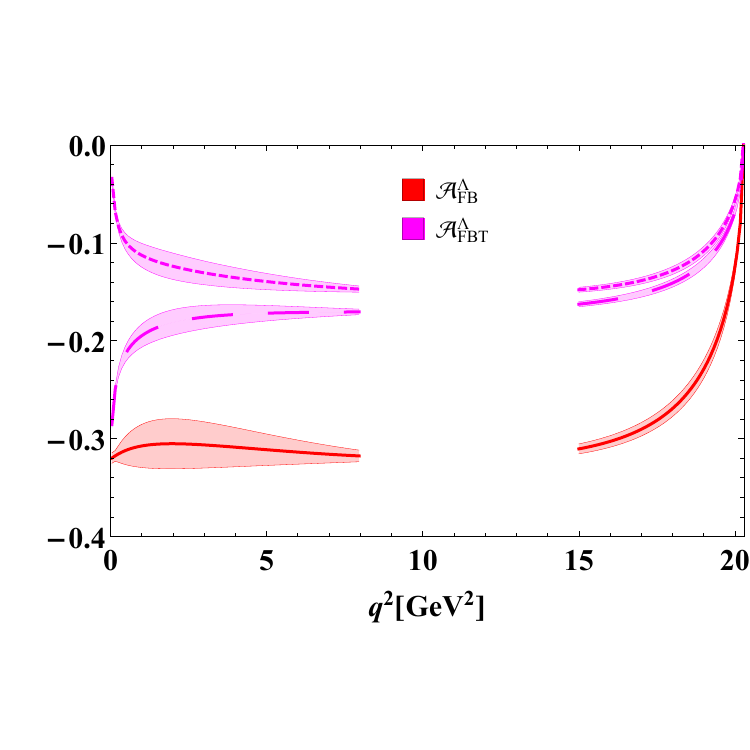} &
\includegraphics[width=2.2in,height=2.2in]{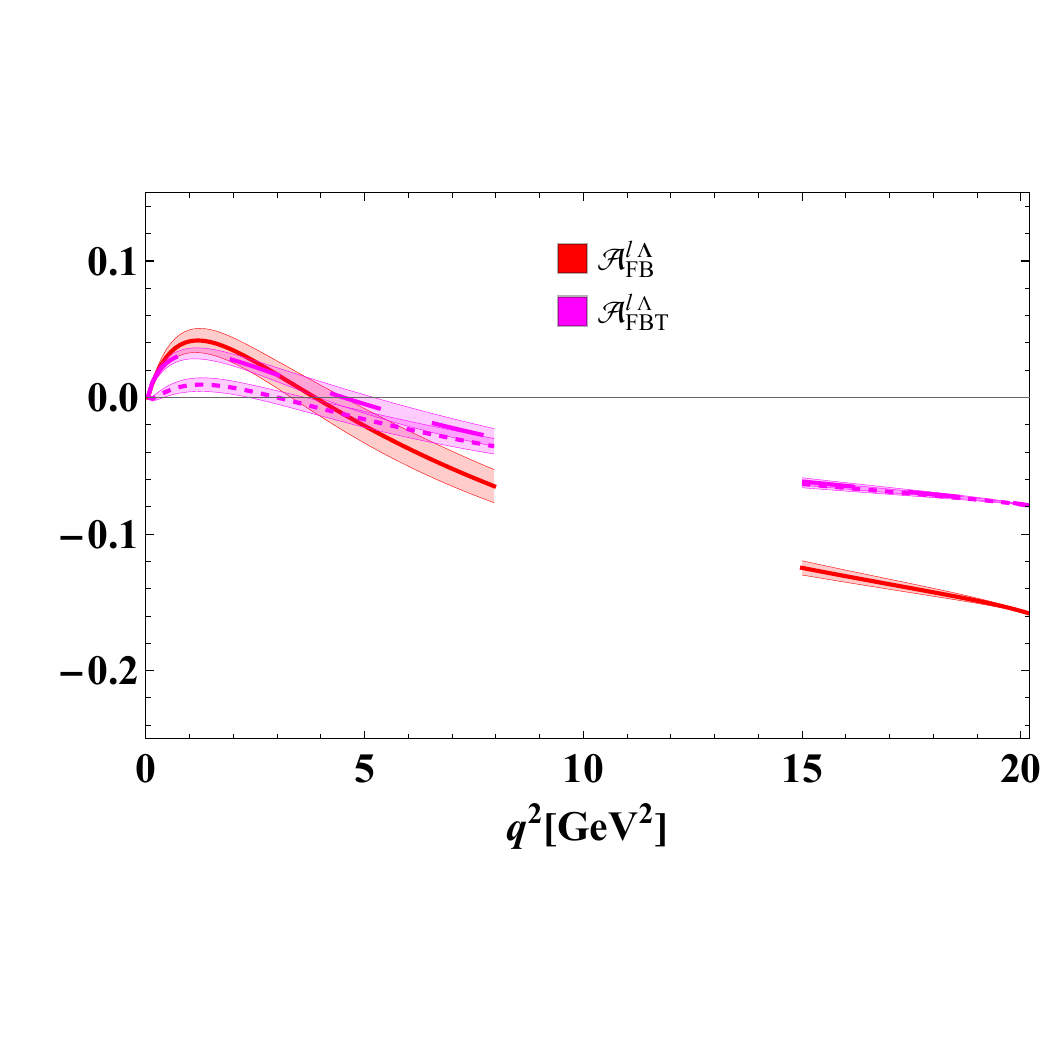}\vspace{-1.0cm}
\end{tabular}}
\caption{Differential branching ratios and various observables vs. $q^{2}$ for SM unpolarized and transversely polarized final state lepton. Dotted line represent the spin $+\tfrac{1}{2}$ contributions, while the dashed line represent the spin $-\tfrac{1}{2}$ contributions. For all results $\ell=\mu$.}\label{Trans}
\end{figure}

In FIG.~\ref{Trans}, we present the differential branching ratios and the forward--backward asymmetries as functions of $q^2$, comparing the unpolarized case with the scenario where the final-state lepton is transversely polarized. For clarity, the contributions from the two lepton spin states are also shown separately in order to examine their relative impact across different $q^2$ regions. The main observations can be summarized as follows:
\begin{itemize}
    \item The transversely polarized branching ratio is decreased compared to the unpolarized case in the high $q^2$ region, where both spin states contribute almost the same.
    \item In contrast to the longitudinal polarization lepton case, the forward-backward asymmetries $\mathcal{A}_{\text{FBT}}^{\ell}$ and $\mathcal{A}_{\text{FBT}}^{\ell\Lambda}$ for the transversely polarized lepton also crosses zero, similar to the unpolarized case. However, the position of the zero is slightly shifted when the spin of the final-state polarized lepton is taken to be $-\tfrac{1}{2}$ and $+\tfrac{1}{2}$ for $\mathcal{A}_{\text{FBT}}^{\ell}$ and $\mathcal{A}_{\text{FBT}}^{\ell\Lambda}$, respectively. In the high-$q^2$ region, the magnitude of these asymmetries decreases compared to the unpolarized scenario. Moreover, the uncertainties arising from the FFs are relatively small in this region, which makes them a particularly sensitive observable to study the impact of lepton polarization.
\item In the case of transversely polarized lepton and with spin $+\tfrac{1}{2}$ and $-\tfrac{1}{2}$ the magnitude of $\mathcal{A}_{\text{FBT}}^\Lambda$ remains higher in the whole $q^2$ range from the corresponding unpolarized $\mathcal{A}_{\text{FB}}^\Lambda$, while the two contributions remain distinct from each other in the low to intermediate $q^2$ region. 
\end{itemize}

\subsection{Polarization asymmetry observables}\label{NP in L}
In this section, we show the asymmetry, the difference between the spin $+\tfrac{1}{2}$ and $-\tfrac{1}{2}$, for the observables defined in Eqs. (\ref{SLpolA})--(\ref{SLpolangobs}). The techniques for the extraction of these asymmetries are discussed in section (\ref{Polarizationasymmetriesobservables}). The expressions of these polarization asymmetries containing spin-dependent angular coefficients can be written in terms of helicity or transversity amplitudes, which encode the short-distance SM and NP Wilson coefficients (see for instance Eqs. (\ref{Hamp002})--(\ref{Hamp006})). To obtain the SM predictions of these asymmetries, we set the contributions of all NP WCs to zero, while in the other case, we also analyze NP impact on these asymmetries by taking different values of the NP WCs along with the SM WCs, and compare the results with their corresponding SM predictions.  Specifically, we consider four representative NP scenarios, selected to be consistent with current experimental observations of the anomalies observed in rare \( b \)-hadron decays. The values of the associated WCs for each scenario are listed in TABLE~\ref{tab:bestfitWC}. For the NP predictions of these observables, we take the lower limit of the $1\sigma$ range. As in the SM case, the uncertainty bands shown in the plots reflect the propagation of uncertainties from hadronic FFs and other input parameters.
 \begin{figure}[t!]
\centering
\scalebox{1}{
\begin{tabular}{ccc} 
\includegraphics[width=2.2in,height=2.2in]{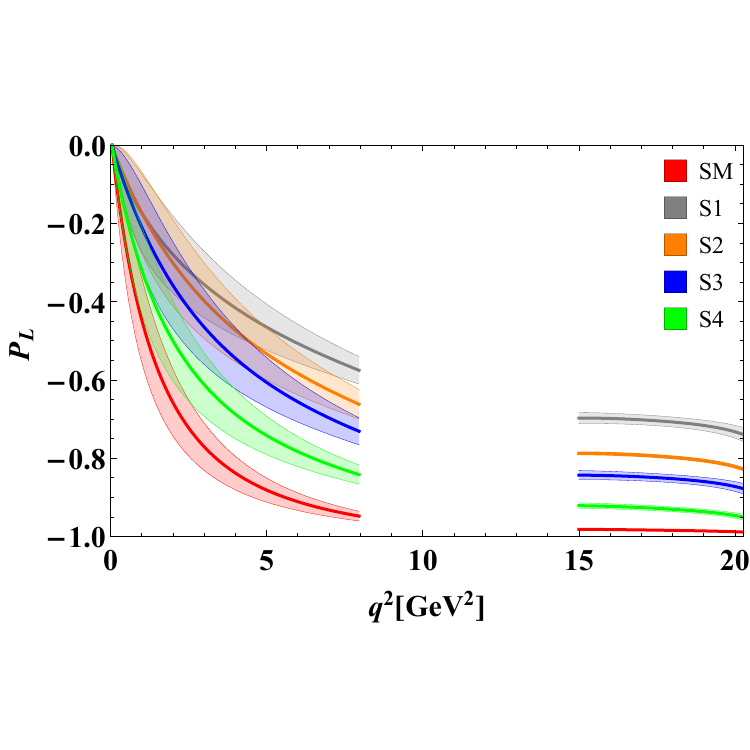} & 
\includegraphics[width=2.2in,height=2.2in]{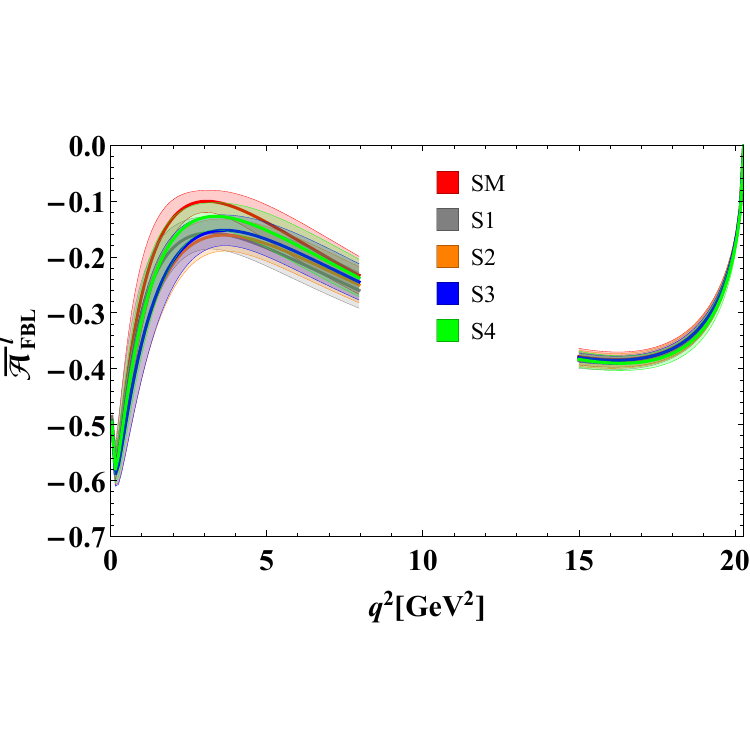} & 
\includegraphics[width=2.2in,height=2.2in]{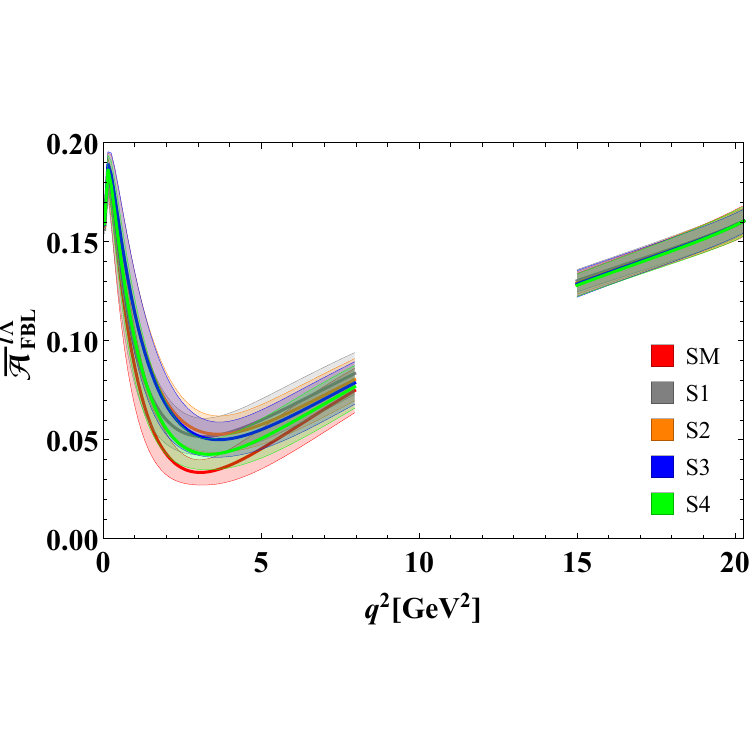}  \vspace{-2cm} \\ 
\includegraphics[width=2.2in,height=2.2in]{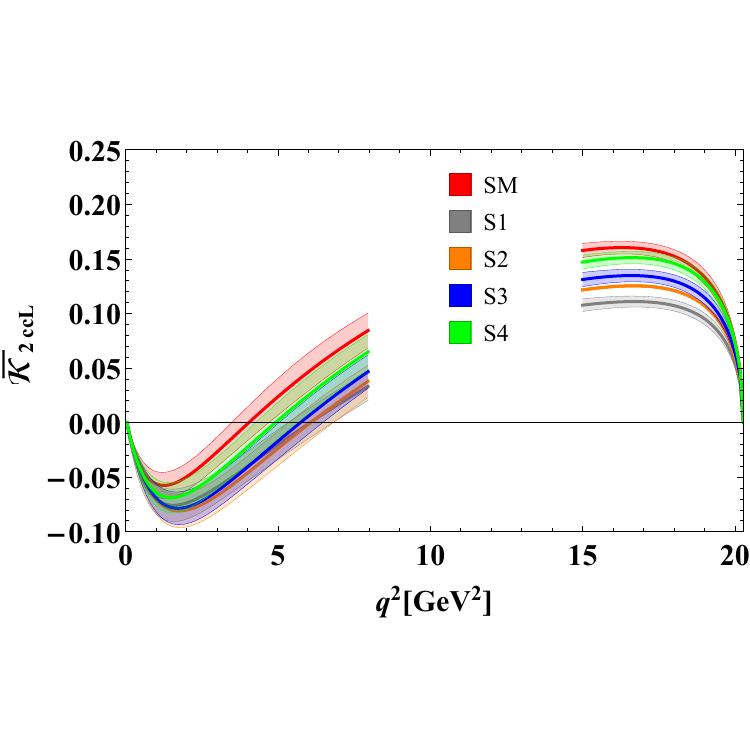} & 
\includegraphics[width=2.2in,height=2.2in]{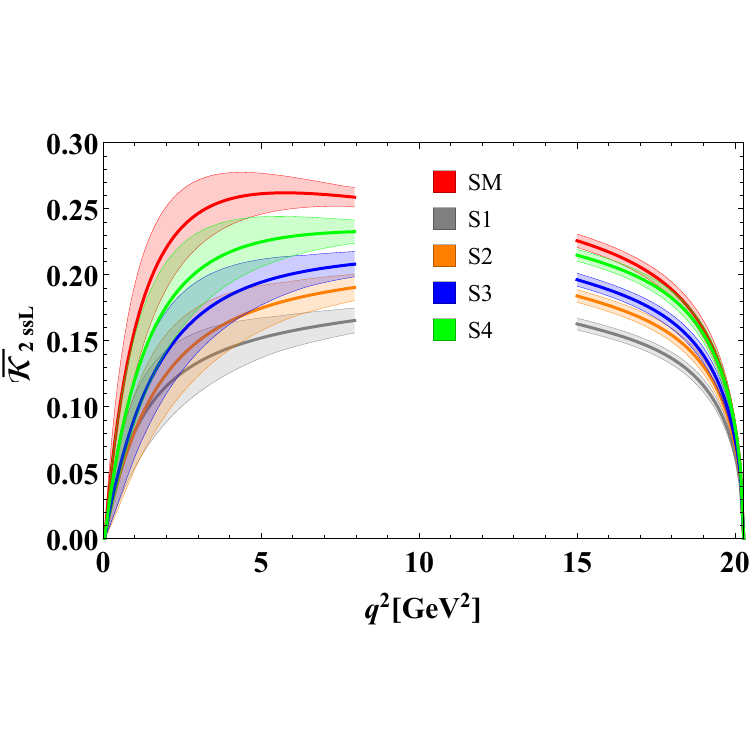} & 
\includegraphics[width=2.2in,height=2.2in]{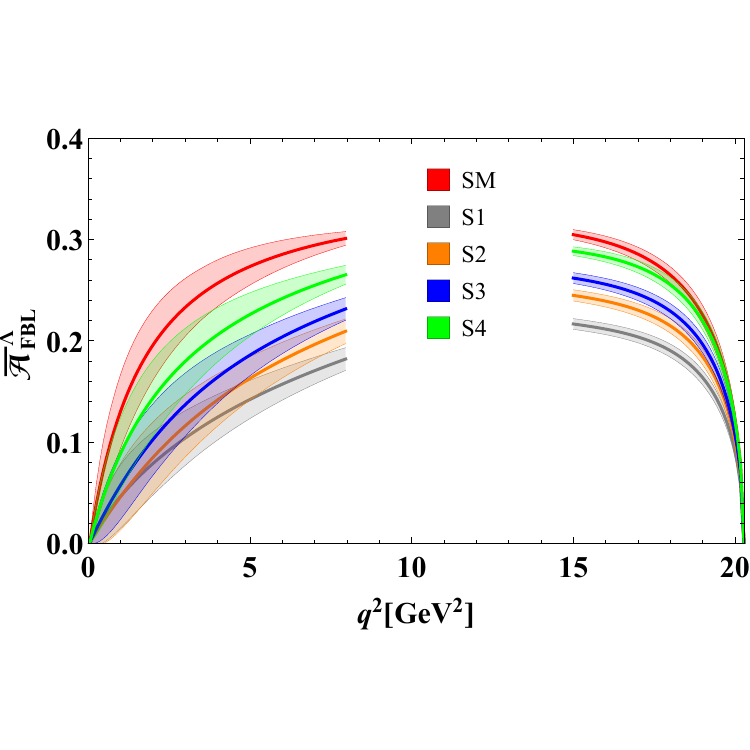}\vspace{-2cm} \\ \includegraphics[width=2.2in,height=2.2in]{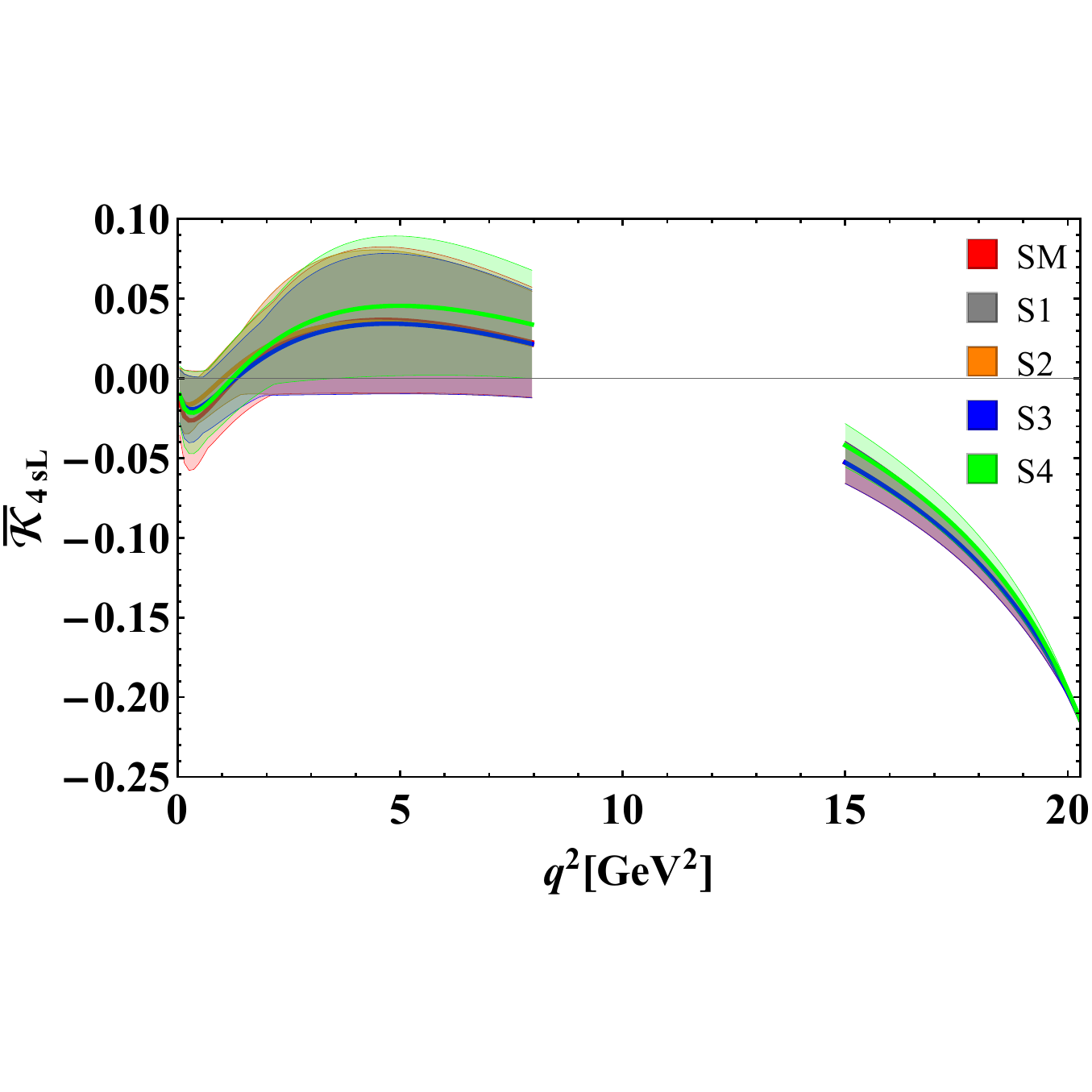} & 
\includegraphics[width=2.2in,height=2.2in]{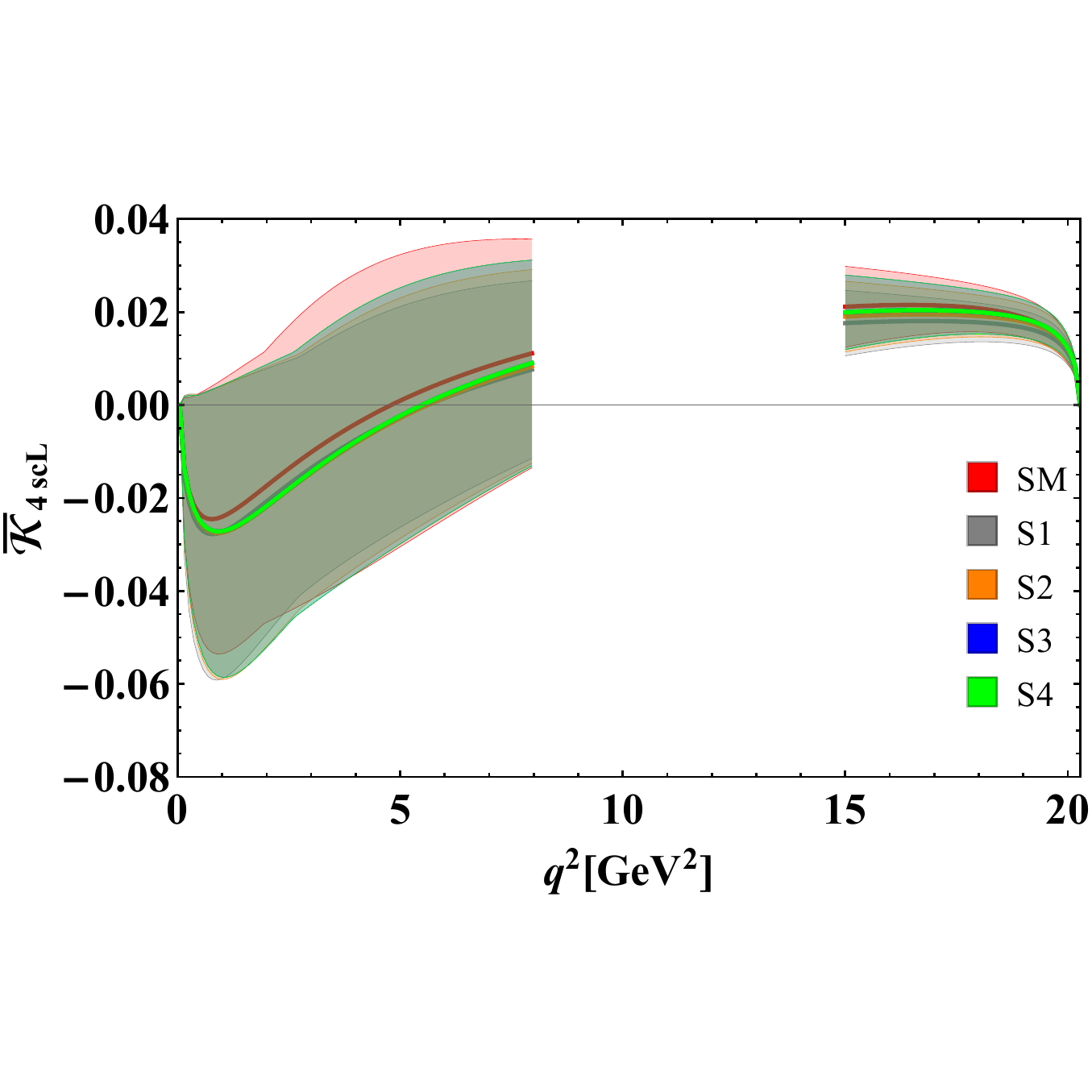}\vspace{-1.0cm} 
\end{tabular}}
\caption{Predictions for NP sensitivity in asymmetry of the longitudinal polarized observables as a function of $q^2$. Red curves represent the SM, while gray, orange, blue and green curves correspond to NP-scenario 1, 2, 3 and 4, respectively. For all results $\ell=\mu$.}\label{PrimeK}
\end{figure}
\subsubsection{Longitudinal polarization case}
\begin{itemize}
\item The longitudinal lepton polarization \( P_L \) as a function of \( q^2 \) is shown in FIG.~\ref{PrimeK} for the SM and various NP scenarios. It is evident that the NP effects deviate significantly from the SM predictions, lying well outside the SM uncertainty band. Moreover, the different NP scenarios are clearly distinguishable from both the SM and from each other. The largest deviation from the SM occurs in scenario S1, where the magnitude of \( P_L \) decreases noticeably. This behavior arises due to the negative contributions to the effective WCs \( C_9^{eff} \) and \( C_{10} \), with \( C_{10} \) becoming more negative relative to the SM. A similar trend is observed in scenarios S2 and S3. In contrast, scenario S4 does not modify the observable largely, resulting in relatively minor deviations from the SM prediction.
\item Another set of interesting observables are the spin-dependent asymmetries related to the lepton and the lepton-hadron forward-backward asymmetries, denoted by \( \overline{\mathcal{A}}_{\text{FBL}}^\ell \) and \( \overline{\mathcal{A}}_{\text{FBL}}^{\ell\Lambda} \), respectively, as shown in FIG. \ref{PrimeK}. As discussed in Eq.~(\ref{SLpolLHLHA}), these observables correspond to the difference between the two lepton spin configurations for a given polarization state. In the case of \( \overline{\mathcal{A}}_{\text{FBL}}^\ell \), there is no zero-crossing in either the SM or any of the NP scenarios. This is consistent with the SM result shown in FIG.~\ref{Longitudinal}, where the zero-crossing in \( \mathcal{A}_{FB}^\ell \) arises from the sum of the spin \( +\tfrac{1}{2} \) and \( -\tfrac{1}{2} \) contributions. Therefore, taking their difference—i.e., forming \( \overline{\mathcal{A}}_{FB}^\ell \)—eliminates the possibility of a zero-crossing. A similar behavior is observed for \( \overline{\mathcal{A}}_{FB}^{\ell\Lambda} \).
These spin-dependent observables are particularly important, as they allow for a clean separation of the two helicity configurations in the SM for polarized leptons. Among the NP scenarios considered, scenario S1 shows the most pronounced deviation from the SM, making its effects more readily observable in these asymmetries compared to the other scenarios.
\begin{figure}[b!]
\begin{tabular}{cc}
\includegraphics[scale=0.62]{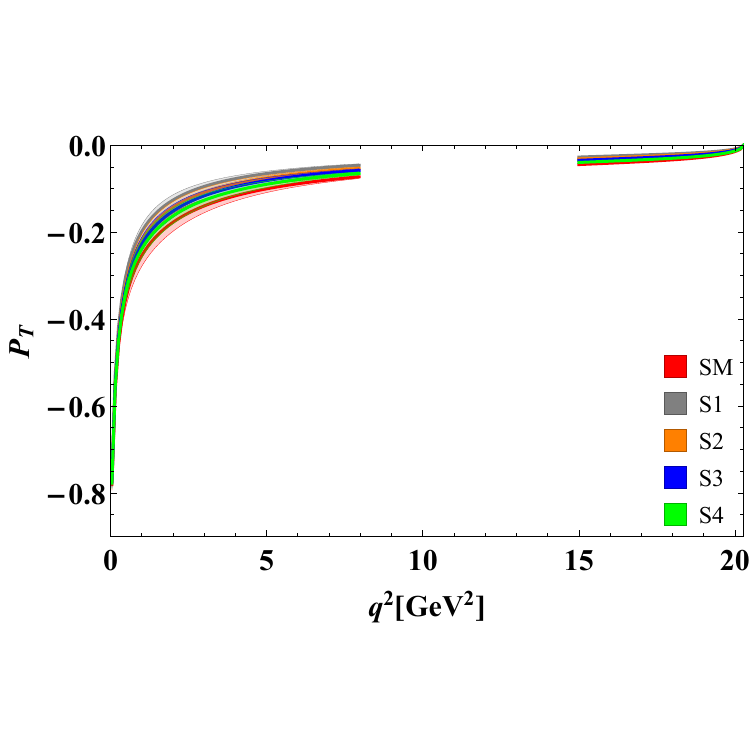} 
&\includegraphics[scale=0.45]{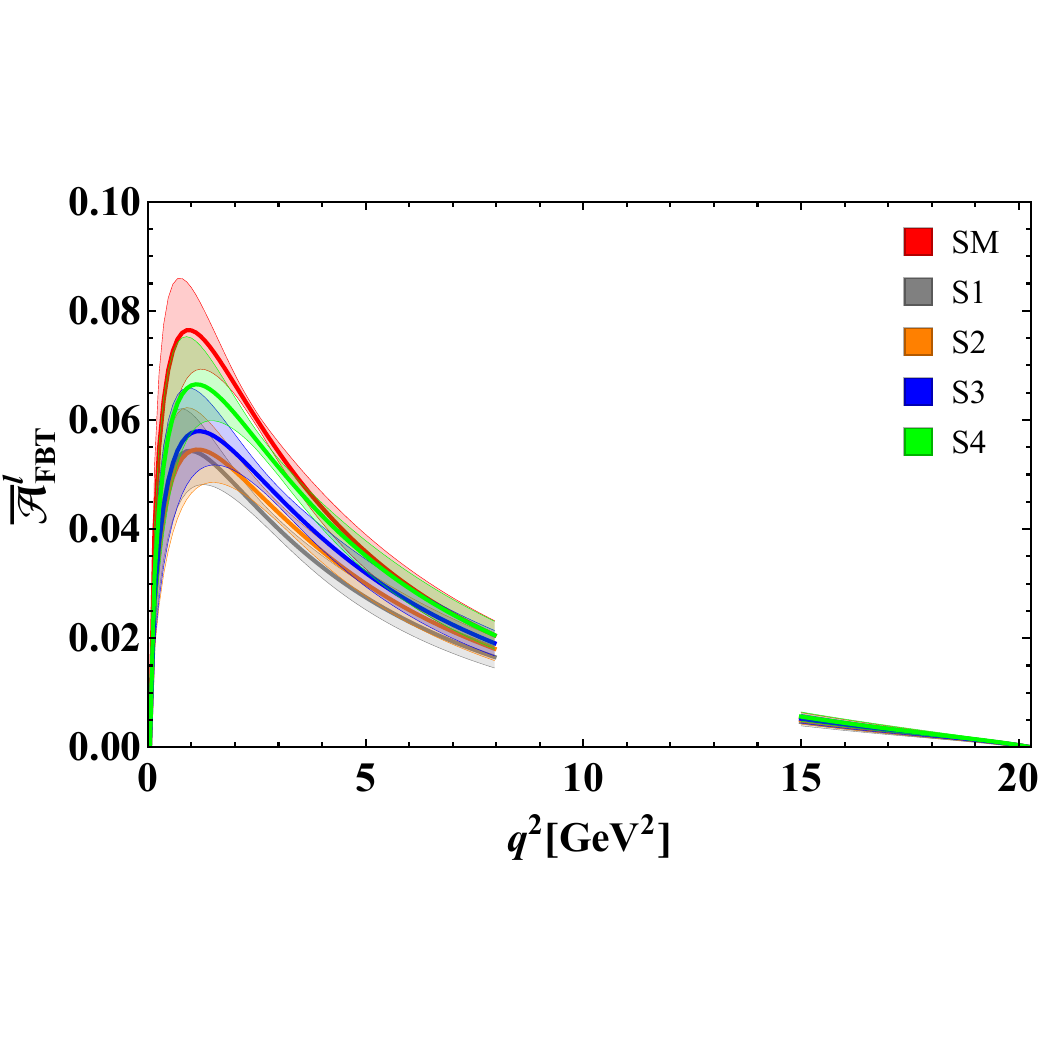}\vspace{-2.5cm}\\
\includegraphics[scale=0.62]{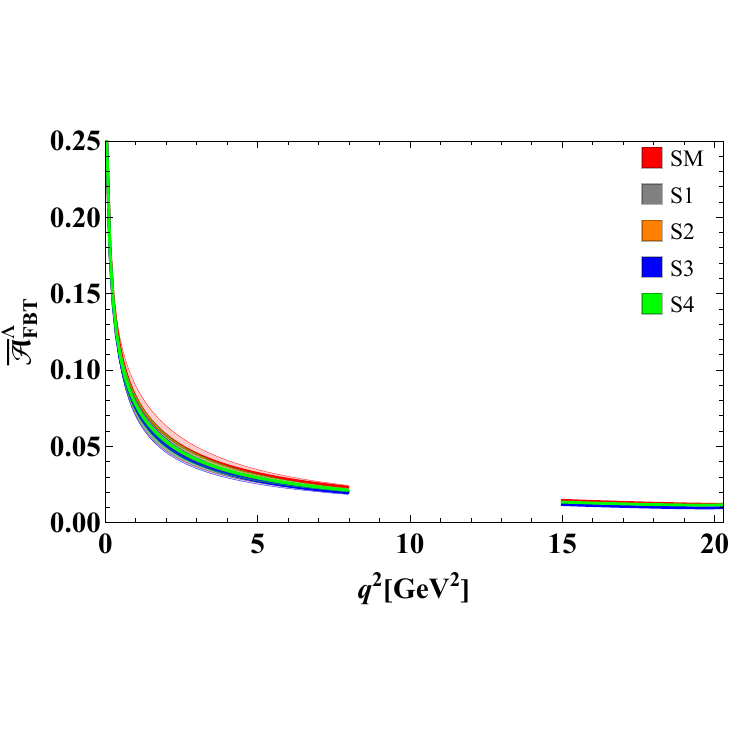} 
&\includegraphics[scale=0.45]{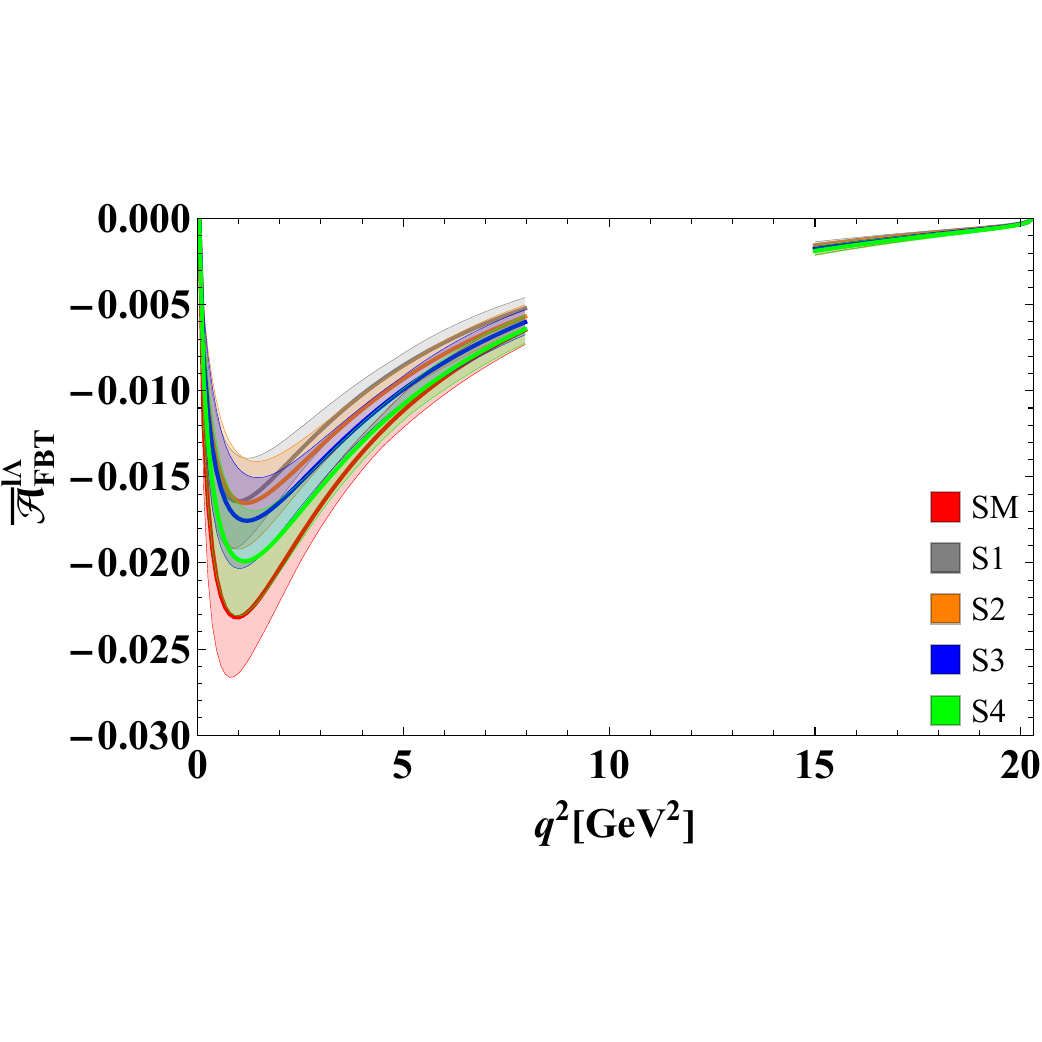}\vspace{-1.5cm}
\end{tabular}
\caption{Predictions for NP sensitivity in asymmetry of the transverse polarized observables as a function of $q^2$. Red curves represent the SM, while gray, orange, blue and green curves correspond to NP-scenario 1, 2, 3 and 4, respectively. For all results $\ell=\mu$.}\label{TNPplot}
\end{figure}
\item In contrast to \( \hat{\mathcal{K}}_{2ccL} \), as shown in FIG.~\ref{Longitudinal}, the observable \( \overline{\mathcal{K}}_{2ccL} \) exhibits a zero-crossing. This occurs because, in this case, the contributions from both spin helicity states enter with the same sign, leading to a cancellation at a specific value of \( q^2 \) within the SM. The position of this zero is particularly sensitive to NP contributions that modify the effective WCs \( C_9^{\text{eff}} \) and \( C_{10} \). Consequently, a precise measurement of the zero-crossing point in \( \overline{\mathcal{K}}_{2ccL} \) can serve as a powerful probe to distinguish between different NP scenarios.
\item Significant NP effects are also observed in the angular coefficient \( \overline{\mathcal{K}}_{2ssL} \) and the hadron forward-backward asymmetry \( \overline{\mathcal{A}}_{\text{FBL}}^\Lambda \) within the low-\( q^2 \) region (\( 0.045 \leq q^2 \leq 8~\text{GeV}^2 \)). Once again, the largest deviation from the SM prediction occurs in scenario S1, where the observable decreases substantially when evaluated as the difference between spin helicity states for polarized leptons.
\item In contrast to the aforementioned spin asymmetries, although NP effects are present in the angular coefficients \( \overline{\mathcal{K}}_{4sL} \) and \( \overline{\mathcal{K}}_{4scL} \), their magnitudes are suppressed by roughly an order compared to the leading observables. As a result, observing these effects experimentally would require significantly larger datasets, making their detection more challenging.
\end{itemize}

\subsubsection{Transverse polarization case}
For the case of transverse lepton polarization, the analytical expressions for the various angular coefficients, dependent on $\xi_{T}$ parameter, are provided in Eqs.~(\ref{ZiTexpF})--(\ref{ziTexpL}). It follows from these expressions that all angular coefficients are proportional to the lepton mass \( m_\ell \) and are therefore suppressed relative to their counterparts in the longitudinal polarization case. This behavior is clearly visible in the plots shown in FIG.~\ref{TNPplot} for $P_T$, $\overline{\mathcal{A}}_{\text{FBT}}^\ell$, $\overline{\mathcal{A}}_{\text{FBT}}^{\Lambda}$, and $\overline{\mathcal{A}}_{\text{FBT}}^{\ell\Lambda}$, where the magnitudes are approximately an order of magnitude smaller than those obtained for the longitudinally polarized case. Furthermore, the effects of NP are significant only in Scenario~S1, and predominantly in $\overline{\mathcal{A}}_{\text{FBT}}^\ell$ and $\overline{\mathcal{A}}_{\text{FBT}}^{\ell\Lambda}$, within the low- to intermediate-$q^2$ region.

\begin{figure}[b!]
\centering
\begin{tabular}{cc} 
        \includegraphics[scale=0.45]{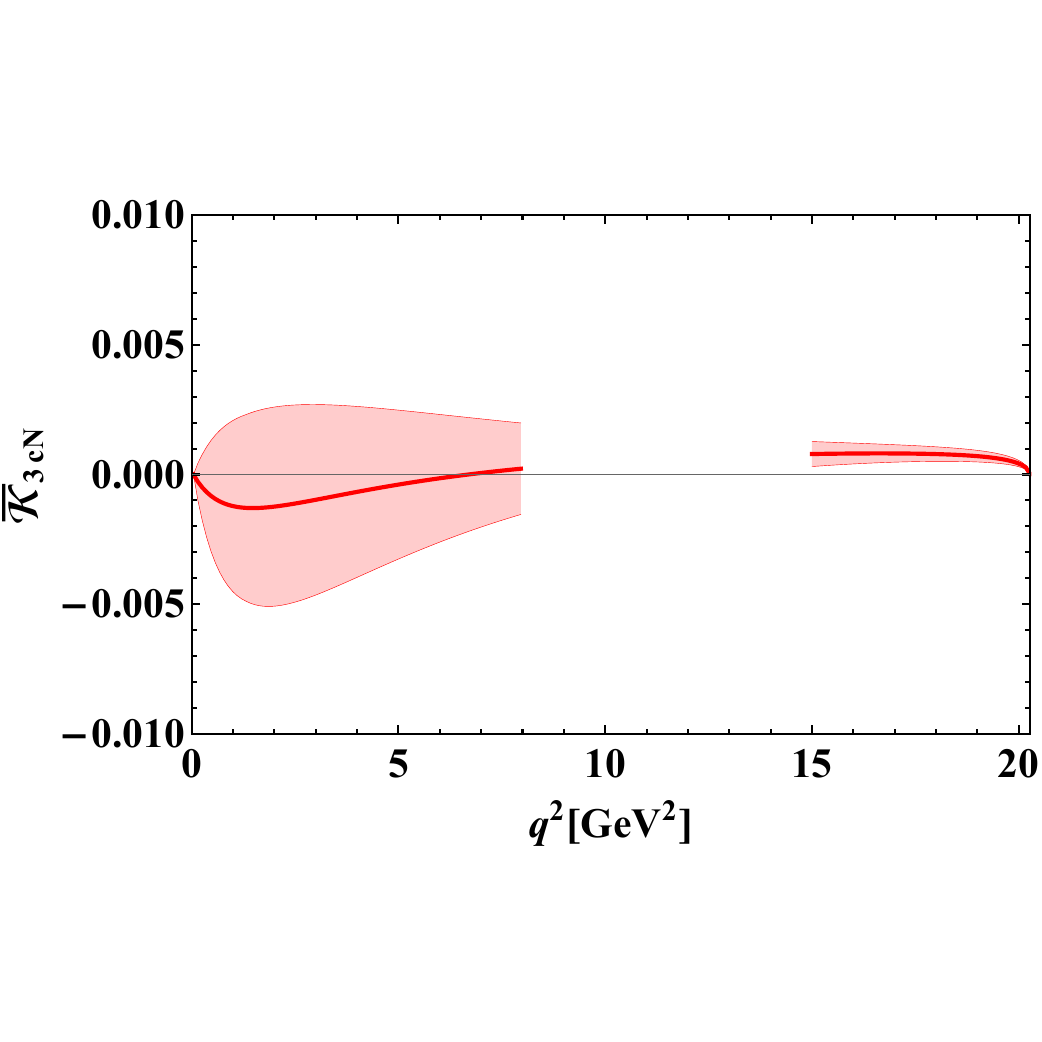}  
        &\includegraphics[scale=0.45]{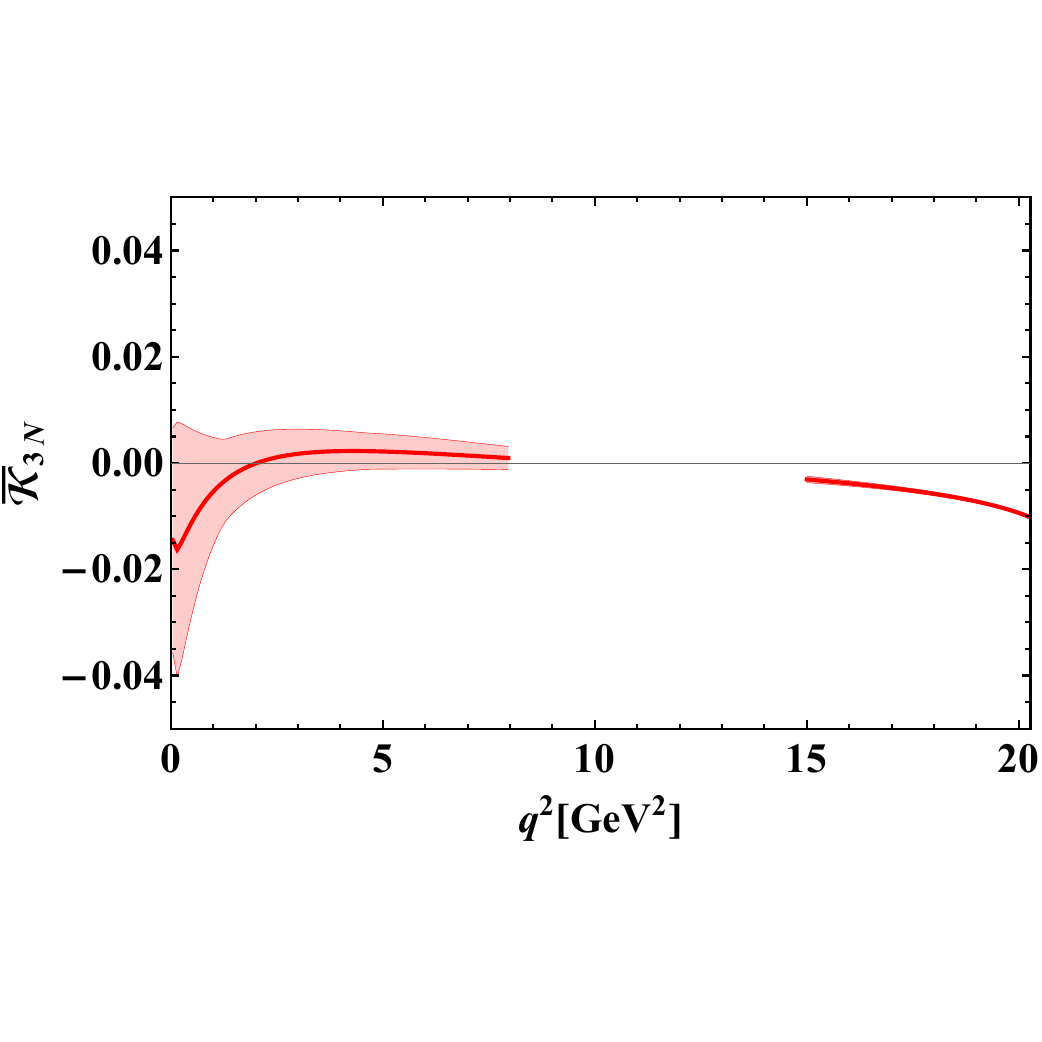}\vspace{-1.5cm}
\end{tabular}
\caption{\textcolor{black}{SM predictions for polarization asymmetries of the new real angular coefficients, $\overline{\mathcal{K}}_{\text{3cN}}$ and $\overline{\mathcal{K}}_{\text{3N}}$, for the normally polarized muon.}}\label{Nspin}
\end{figure}
\begin{figure}[t!]
\begin{tabular}{cc}
\includegraphics[scale=0.45]{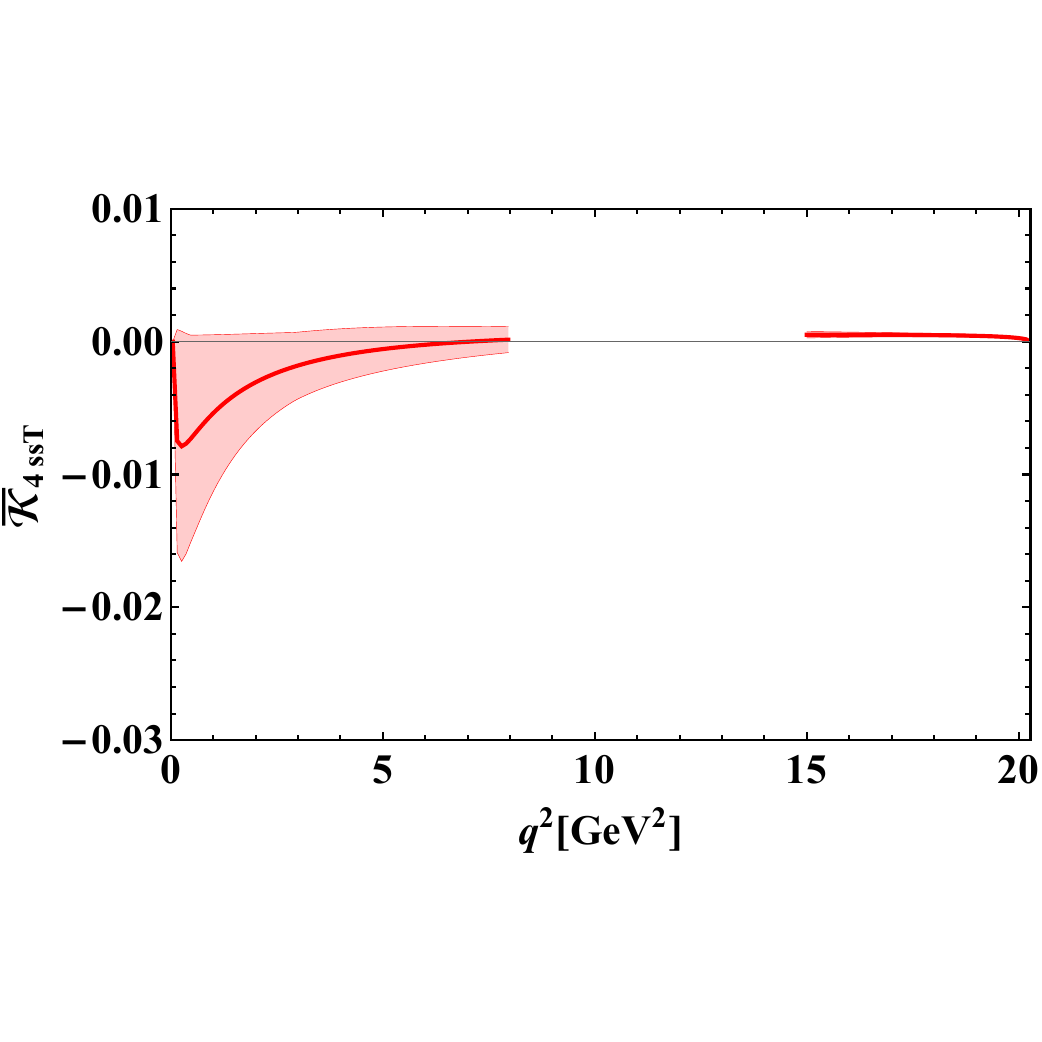} 
&\includegraphics[scale=0.45]{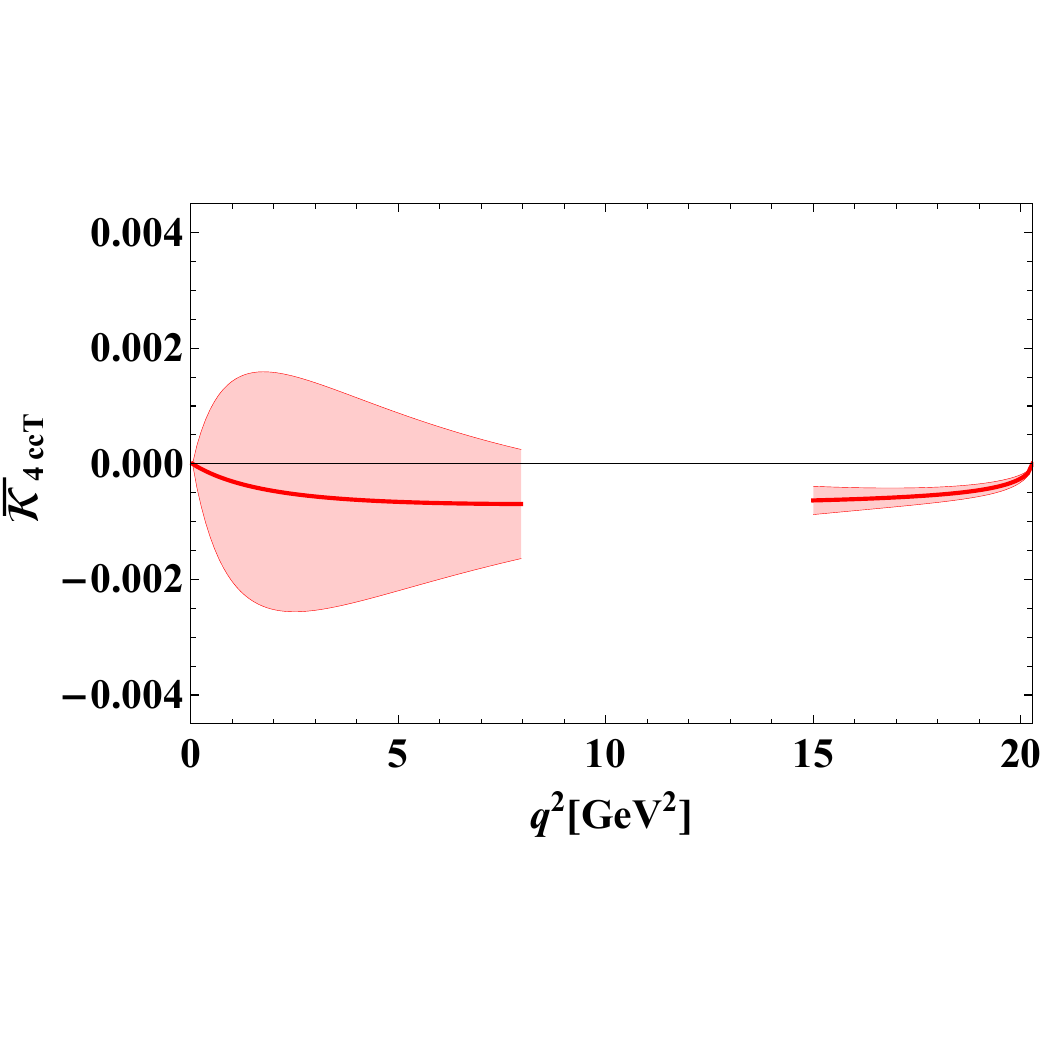}\vspace{-2.5cm}\\
\includegraphics[scale=0.45]{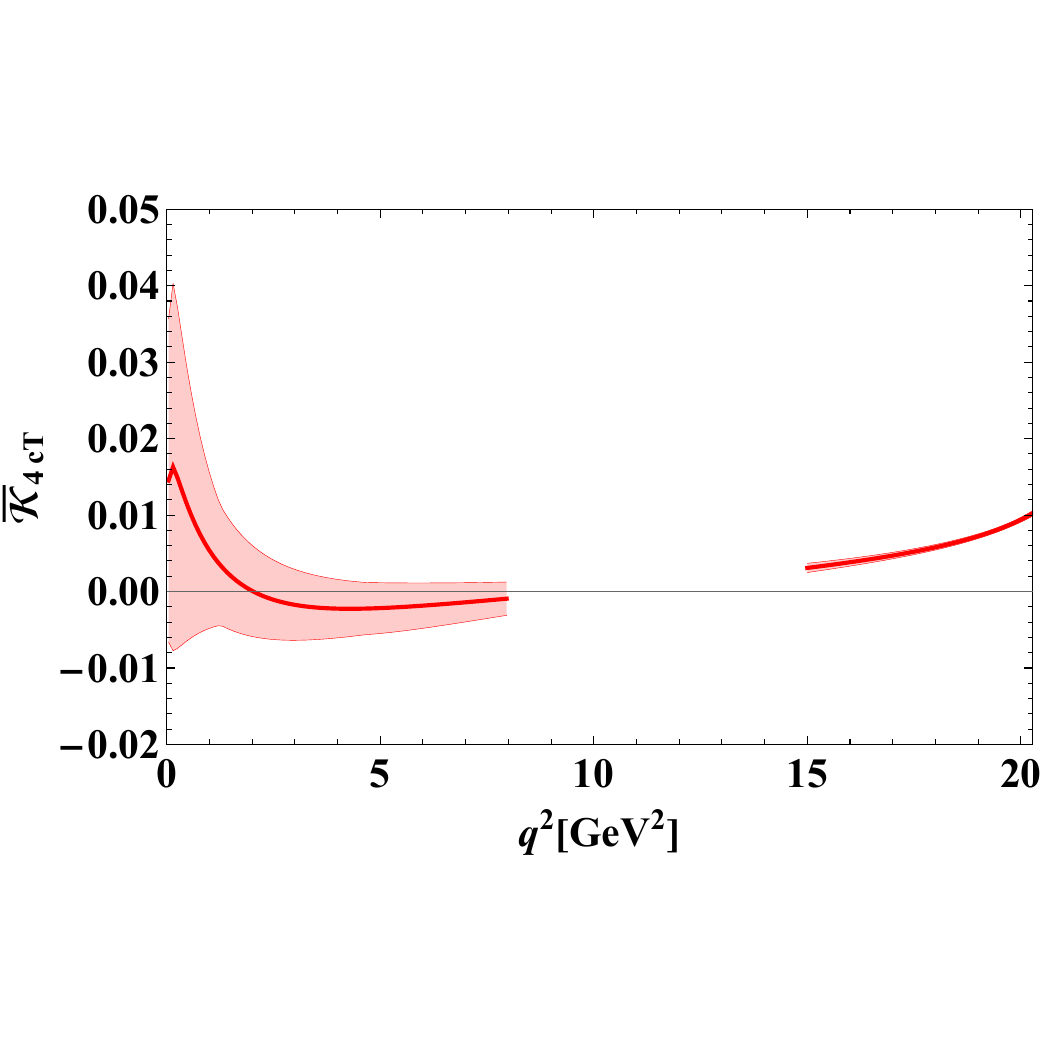} 
&\includegraphics[scale=0.45]{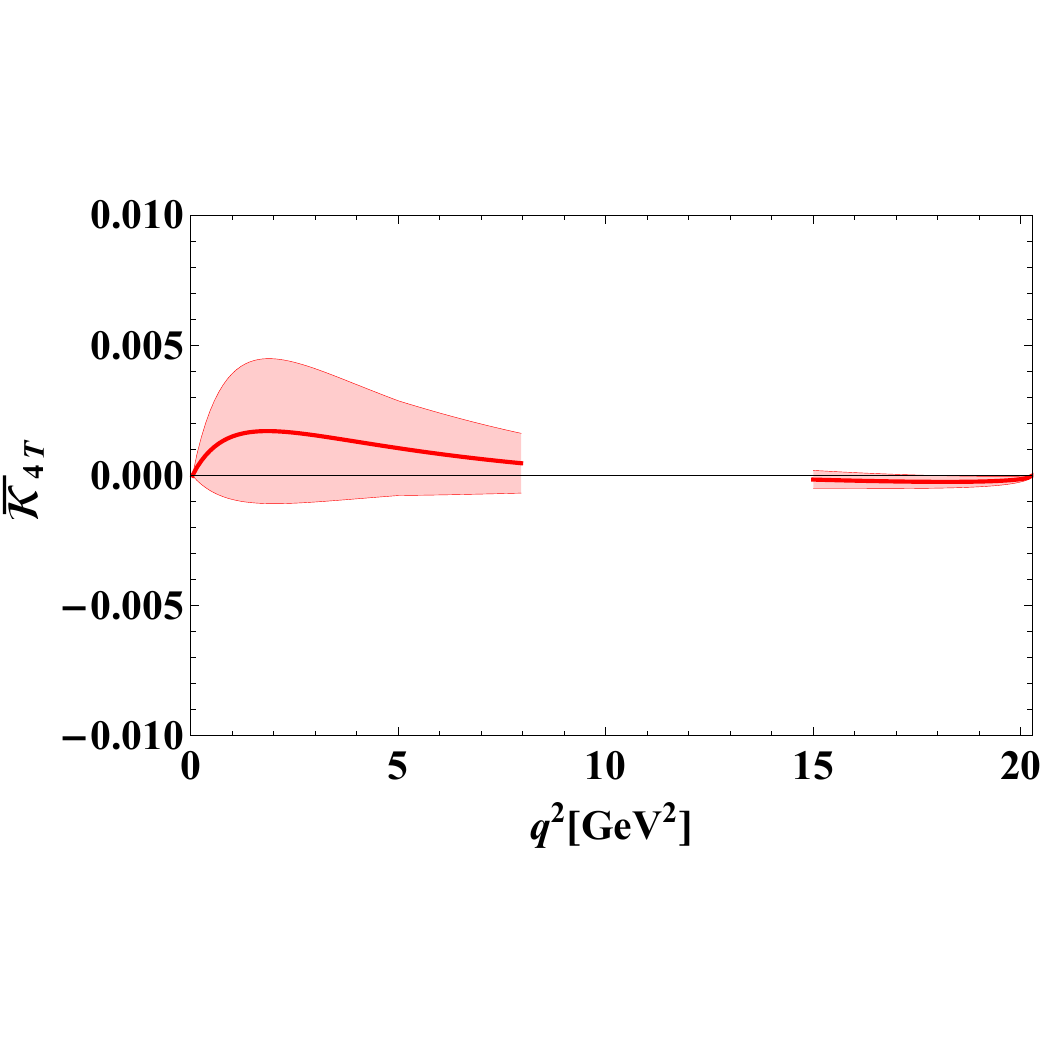}\vspace{-1.5cm}
\end{tabular}
\caption{SM predictions for polarization asymmetries of the new real angular coefficients, $\overline{\mathcal{K}}_{\text{4ssT}}$, $\overline{\mathcal{K}}_{\text{4ccT}}$, $\overline{\mathcal{K}}_{\text{4cT}}$, and $\overline{\mathcal{K}}_{\text{4T}}$, for the transversely polarized muon.}\label{Tspin}
\end{figure}
\subsection{Polarization asymmetries for new angular coefficients in N and T cases}
In FIG. \ref{Nspin} and \ref{Tspin}, we present the SM predictions of the polarization asymmetries of the additional individual real angular coefficients with N and T polarized lepton case, respectively. The bands in the polarization asymmetries vs $q^2$ plots originate mainly from the form-factors uncertainties. In contrast to the longitudinal L case, the polarization asymmetries of the presented angular coefficients associated with N and T polarization cases are suppressed within the SM, and the potential NP contributions are indistinguishable from the SM expectations. Consequently, for the N and T polarization cases, only the SM results are shown.
\subsection{Binned predictions for the physical observables}
The binned numerical values of the SM and NP predictions (when considered) for all the observables presented in Figs.~\ref{Longitudinal}-\ref{Tspin}, in two different $q^2$ bins, are given in Tables \ref{Bin1-analysis001}-\ref{Bin1-analysis005}, of appendix \ref{AppG}. One is called as the low-central $q^2=[0.045-8.0]$ GeV$^2$ bin while the other as high $q^2=[15.0-20.27]$ GeV$^2$ bin. As the intermediate $q^2$ region is dominated by the charmonium resonances, we refrain from giving predictions of observables in that region. We find that most observables exhibit deviations at the level of 
$\sim (1 - 2)\sigma$, indicating moderate sensitivity to NP scenarios. 
In particular, from Tables~\ref{Bin1-analysis001}-\ref{Bin1-analysis005}, the differential and integrated branching ratios show modest deviations, 
typically within $\mathcal{O}(10\%-20\%)$, reflecting partial NP contributions. 
Among these, low-central $q^2$ bins exhibit slightly enhanced sensitivity, 
while high-$q^2$ regions remain closer to the SM predictions due to smaller 
hadronic uncertainties.
The longitudinal polarization $\langle P_L \rangle$ displays sizable deviations 
from the SM, especially in the low-central $q^2$ region where shifts of 
$\mathcal{O}(30\%-40\%)$ are observed across different NP scenarios. 
Similarly, the forward-backward asymmetries 
$\overline{\mathcal{A}}^{\ell}_{\mathrm{FBL}}$ and 
$\overline{\mathcal{A}}^{\Lambda}_{\mathrm{FBL}}$ exhibit noticeable 
NP-induced shifts, particularly at low-central $q^2$, while remaining relatively stable 
in the high-$q^2$ region.
Among the angular observables, $\overline{\mathcal{K}}_{\mathrm{2ccL}}$ and 
$\overline{\mathcal{K}}_{\mathrm{2ssL}}$ show enhanced sensitivity in the 
high-$q^2$ bin, where theoretical uncertainties are smaller and deviations 
from the SM reach a visible level. In contrast, higher-order coefficients such as 
$\overline{\mathcal{K}}_{\mathrm{4sL}}$ and $\overline{\mathcal{K}}_{\mathrm{4scL}}$ 
remain largely consistent with the SM within uncertainties, indicating limited 
discriminating power with current precision.
For transversely polarized observables, 
$\langle P_T \rangle$ and $\overline{\mathcal{A}}^{\ell}_{\mathrm{FBT}}$ show 
small but systematic deviations from the SM, whereas 
$\overline{\mathcal{A}}^{\Lambda}_{\mathrm{FBT}}$ exhibits mild sensitivity. 
The newly introduced angular coefficients such as 
$\overline{\mathcal{K}}_{3\mathrm{cN}}$, 
$\overline{\mathcal{K}}_{3\mathrm{N}}$, and 
$\overline{\mathcal{K}}_{4\mathrm{T}}$ are strongly suppressed in the SM 
and affected by relatively large uncertainties, making their experimental 
extraction particularly challenging.


Lastly, rare $b\to s \ell \ell$ decays can be explored at high luminosity flavor facilities, such as LHCb \cite{Cerri:2018ypt}, and Belle II \cite{Belle-II:2018jsg}. A rough estimate suggests that establishing an 
observable at the $n\sigma$ level typically requires $\mathcal{O}(n^2)$ signal 
events after selection. For branching ratios, this implies roughly $\mathcal{O}(10^7 - 10^8)$ 
reconstructed $\Lambda_b$ decays for a $1\sigma$ measurement per $q^2$ bin.

The decay $\Lambda_b \to \Lambda \mu^+ \mu^-$ provides a clean experimental 
environment with high reconstruction efficiency, making it well suited for 
detailed angular and branching ratio analyses. With the full Run~1 and Run~2 
datasets corresponding to an integrated luminosity of about $9\,\mathrm{fb}^{-1}$, 
LHCb has already performed measurements of differential branching fractions 
and angular observables~\cite{LHCb:2015tgy}. The ongoing Run~3 is 
expected to significantly increase the dataset, while the LHCb Upgrade aims to 
collect up to $\sim 50\,\mathrm{fb}^{-1}$, and Upgrade~II targets 
$\mathcal{O}(300\,\mathrm{fb}^{-1})$ at the HL-LHC~\cite{Cerri:2018ypt}. 
These datasets correspond to an enormous increase in the available 
$b$-hadron sample, reaching $\mathcal{O}(10^{13})$ $b\bar{b}$ pairs within 
the detector acceptance.

This substantial increase in statistics implies that key observables such as 
$\langle P_L \rangle$, forward-backward asymmetries, and branching ratios, 
which already show $\sim (1 - 2)\sigma$ deviations, could become experimentally 
accessible with improved precision. However, higher-order angular coefficients 
and suppressed observables, particularly those involving transverse and normal 
polarization, will likely require the full Upgrade dataset and careful control of 
systematic uncertainties. A realistic assessment of the achievable sensitivity 
therefore necessitates dedicated experimental studies including detector 
acceptance, reconstruction efficiencies, and background contributions.

\section{Conclusions}\label{concl}
In this work, we have presented a detailed analysis of the polarization-dependent 
angular observables in the decay $\Lambda_b \to \Lambda(\to N \pi)\,\mu^+\mu^-$. 
By considering the final-state lepton polarized along the longitudinal (L), normal (N), 
and transverse (T) directions, we derived the corresponding four-fold angular 
distributions and identified the additional angular structures that arise beyond 
the unpolarized case.

For the longitudinal polarization, the spin-dependent terms appear naturally within 
the already established angular framework, leading to polarized counterparts of 
the well-known observables. In the case of normal polarization, we identified 
six additional angular coefficients, of which two are real, and four are purely imaginary. The numerical results for the case of a normally polarized lepton are not presented for physical observables such as the lepton and lepton--hadron forward--backward asymmetries, as well as the corresponding polarization asymmetries ($\mathcal{\overline{A}}_{\text{FBN}}^{\ell}$, $\mathcal{\overline{A}}_{\text{FBN}}^{\ell\Lambda}$), because they are unaffected by the normal polarization effects. Also, in other cases, such as the differential branching ratio, the hadron forward--backward asymmetry, and the related polarization asymmetries ($P_{N}$, $\mathcal{\overline{A}}_{\text{FBN}}^{\Lambda}$), the effects are negligibly small, and not shown, since these observables depend on additional angular coefficients containing purely imaginary contributions from the helicity or transversity amplitudes. However, the results for the polarization asymmetries $\overline{\mathcal{K}}_{3\mathrm{cN}}$, and $\overline{\mathcal{K}}_{3\mathrm{N}}$, originating from the two additional real angular coefficients, although suppressed, are presented in FIG. \ref{Nspin}.
For the transverse polarization, twelve new 
angular coefficients are obtained, consisting of eight real and four imaginary 
terms. Some of these angular coefficients modify different observables, including the differential branching ratio, and forward--backward asymmetries. The polarization asymmetries in the remaining real angular coefficients are analyzed within the SM framework.

We further investigated the sensitivity of these polarization-dependent 
observables to potential NP by employing global fit results for vector 
and axial-vector WCs. Our findings show that while the asymmetries 
associated with longitudinal and transverse polarizations exhibit some sensitivity 
to such modifications, the additional spin-dependent structures in the normal and 
transverse cases remain indistinguishable from SM predictions. 

In conclusion, our study demonstrates that incorporating lepton polarization 
significantly enriches the angular analysis, yielding new observables that can serve 
as potential probes of physics beyond the SM. The results indicate moderate to sizeable deviations from the SM expectations and the framework developed here provides a basis for future precision measurements. For future experimental probes, we also predict the binned SM and NP predictions for the observables, in two different $q^2$ bins. In particular, improved determinations of hadronic FFs and high-statistics data from experiments such as LHCb and Belle~II may enhance the discovery 
potential of polarization-dependent angular observables in $\Lambda_b$ decays.
\clearpage
\appendix
\section{Kinematics of the decay}
\label{Kinematics A}
\begin{figure}[H]
    \centering    
\includegraphics[width=0.8 \textwidth]{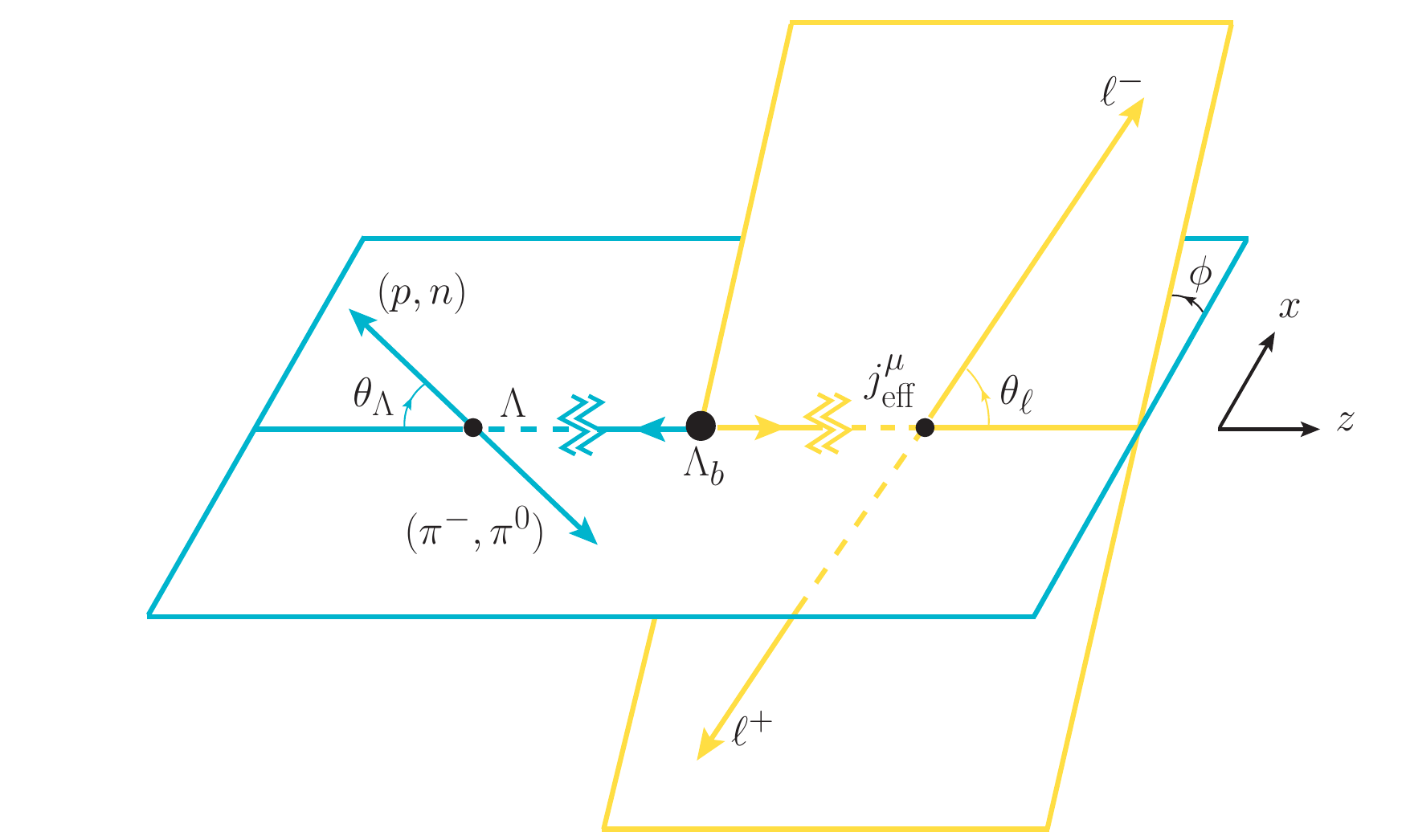}    \caption{Kinematics of $\Lambda_{b} \to \Lambda (\rightarrow N \pi) \ell^{+}\ell^{-}$ decays, where $N\pi=\{p\pi^-,n\pi^0\}$.}
\label{fig:kinematica}
\end{figure}

We consider the momenta and spins assignments of the particles using the cascade nature of the decay as follows
\begin{eqnarray}
\Lambda_b (p,s_{\Lambda_b}) &\to& \Lambda(k,s_{\Lambda})\left(j_{\text{eff}}(q)\to \ell^+\left(p_{\ell^+},s_2\right)\ell^-\left(p_{\ell^-},s_1\right)\right),\notag\\
\Lambda(k,s_{\Lambda})&\to& N(p_3, s_N)\pi(p_4), \quad\quad N\pi=\{p\pi^-,n\pi^0\}.\label{eqkinA01}
\end{eqnarray}
We consider the decay in the rest frame of the \(\Lambda_b\) baryon, where the final-state \(\Lambda\) baryon and the virtual vector boson—denoted as \(j_{\mathrm{eff}}^{\mu}\)—are assumed to move back-to-back, with the \(\Lambda\) traveling along the \(-z\)-axis. In their respective rest frames, the \(\Lambda\) subsequently decays to \(N\pi\), while the virtual boson decays to the lepton pair \(\ell^{+}\ell^{-}\). The relevant kinematic configuration is illustrated in FIG.~\ref{fig:kinematica}. We follow the same conventions for the definitions of the four independent variables, dilepton invariant mass squared $q^2$, the polar angles $\theta_{\ell}$, $\theta_{\Lambda}$, and the azimuthal angle $\phi$, as described in \cite{Gutsche:2013pp,Faessler:2002ut}.
\subsection{Kinematics in $\Lambda_b$ rest frame}
\label{Kinematics b}
The four momentum in the rest frame of $\Lambda_b$ are defined as follows \cite{Gutsche:2013pp, Faessler:2002ut}
\begin{eqnarray}
p^{\mu}=(m_{\Lambda_{b}}, 0, 0, 0), \qquad k^{\mu}=(E_{\Lambda}, 0, 0, -|\vec{k}|), \qquad q^{\mu}=(q^0, 0, 0, +|\vec{k}|),\label{eqkinA02}
\end{eqnarray}
where
\begin{eqnarray}
q^0=\frac{m_{\Lambda_{b}}^2-m_{\Lambda}^{2}+q^2}{2m_{\Lambda_{b}}}, \qquad E_{\Lambda}=\frac{m_{\Lambda_{b}}^2+m_{\Lambda}^{2}-q^2}{2m_{\Lambda_{b}}},
\qquad |\vec{k}|=\frac{\sqrt{\lambda(m^2_{\Lambda_{b}}, m^2_{\Lambda}, q^2)}}{2 m_{\Lambda_{b}}},\label{eqkinA03}
\end{eqnarray}
with $\lambda(m^2_{\Lambda_{b}}, m^2_{\Lambda}, q^2)$ being the K{\"a}ll{\'e}n function.
The polarization four-vectors of $j^{\mu}_{\text{eff}}$, in $\Lambda_b $ rest frame, are given as
\begin{eqnarray}
\epsilon^{\mu}(t)=\frac{1}{\sqrt{q^2}}(q^0, 0, 0, |\vec{k}|),\;\;
\epsilon^{\mu}(\pm)=\frac{1}{\sqrt{2}}(0, \mp1, -i, 0),\quad \epsilon^{\mu}(0)=\frac{1}{\sqrt{q^2}}(|\vec{k}|, 0, 0, q^0).\label{eqkinA04}
\end{eqnarray}

\subsection{Kinematics in $\ell\bar{\ell}$ rest frame}
\label{Kinematics c}

The four momenta in the $\ell\overline{\ell}$-CM frame are
\begin{align}
q^{\mu} &= \left( 
\sqrt{q^2},\, 
0,\, 
0,\, 
0
\right),\label{eqkinA05}
\\
p_{\ell^-}^{\mu} &= \left( 
E_{\ell},\, 
|\vec{p_{\ell}}| \sin\theta_{\ell} \cos\phi,\, 
|\vec{p_{\ell}}| \sin\theta_{\ell} \sin\phi,\, 
|\vec{p_{\ell}}| \cos\theta_{\ell}
\right),\label{eqkinA06}
\\
p_{\ell^+}^{\mu} &=\left( 
E_{\ell},\, 
-|\vec{p_{\ell}}| \sin\theta_{\ell} \cos\phi,\, 
-|\vec{p_{\ell}}| \sin\theta_{\ell} \sin\phi,\, 
-|\vec{p_{\ell}}| \cos\theta_{\ell}
\right),\label{eqkinA07}
\end{align}
with $ E_{\ell} = \frac{\sqrt{q^2}}{2}$, $|\vec{p_{\ell}}| = \frac{\sqrt{q^2}}{2} \beta_{\ell}$, and $\beta_{\ell} = \sqrt{1 - \frac{4m_{\ell}^2}{q^2}}$. In the di-lepton pair rest frame ($\ell\overline{\ell}$-CM), the transverse polarizations of $j_{\text{eff}}$ remain same, while the other two polarizations
read
\begin{eqnarray}
\varepsilon^{\mu}(t)=(1, 0, 0, 0),\qquad  \varepsilon^{\mu}(0)=(0, 0, 0, 1).\label{eqkinA08}
\end{eqnarray}

\subsection{Kinematics in $\Lambda$ rest frame}
\label{Kinematics d}
The $N\pi$ system is characterized by its invariant mass $k^2=m_\Lambda^2$. The four momenta in the rest frame of $\Lambda$ are taken as
\begin{align}
k^{\mu} &= \left( 
m_{\Lambda},\, 
0,\, 
0,\, 
0
\right),\label{eqkinA10}
\\
p_{3}^{\mu} &= \left( 
E_{N},\, 
|\vec{p_{3}}| \sin\theta_{\Lambda} ,\, 
0,\, 
-|\vec{p_{3}}| \cos\theta_{\Lambda}
\right),\label{eqkinA11}
\\
p_{4}^{\mu} &=\left( 
E_{\pi},\, 
-|\vec{p_{3}}| \sin\theta_{\Lambda} ,\, 
0,\, 
|\vec{p_{3}}| \cos\theta_{\Lambda}
\right),\label{eqkinA12}
\end{align}
where, $|\vec{p_{3}}|=\sqrt{\lambda(m^2_{\Lambda}, m^2_{N}, m^2_{\pi})}/2 m_{\Lambda}$.
\section{The expressions of WCs in the SM}\label{WCsC7C9}
The explicit expressions used for the WCs are given as follows:

\begin{align}
    C_7^{\text{eff}}(q^2) &= C_7 
    - \frac{1}{3} \left( C_3 + \frac{4}{3} C_4 + 20 C_5 + \frac{80}{3} C_6 \right) \nonumber - \frac{\alpha_s}{4 \pi} \left[ \left(C_1 - 6 C_2\right) F_{1,c}^{(7)}(q^2) + C_8 F_8^{(7)}(q^2) \right], \\
    C_9^{\text{eff}}(q^2) &= C_9 + \frac{4}{3} \left( C_3 + \frac{16}{3} C_5 + \frac{16}{9} C_6 \right) \nonumber - h(m_c^{\text{pole}}, q^2) \left( \frac{7}{2} C_3 + \frac{2}{3} C_4 + 38 C_5 + \frac{32}{3} C_6 \right) \nonumber \\
    &\quad + h(m_b^{\text{pole}}, q^2) \left( \frac{4}{3} C_1 + C_2 + 6 C_3 + 60 C_5 \right)  - \frac{\alpha_s}{4 \pi} \left[ C_1 F_{1,c}^{(9)}(q^2) + C_2 F_{2,c}^{(9)}(q^2) + C_8 F_8^{(9)}(q^2) \right],
\end{align}

where the functions \( h(m_q^{\text{pole}}, q^2) \) with \( q = c, b \) and \( F_{i}^{(7,9)}(q^2) \) are defined in \cite{Beneke:2001at}, while \( F_{1,c}^{(7,9)}(q^2) \) and \( F_{2,c}^{(7,9)}(q^2) \) are given for low \( q^2 \) in \cite{Asatryan:2001zw} and for high \( q^2 \) in \cite{Greub:2008cy}. The quark masses, $m_q^{\text{pole}}$ with \( q = c, b \), used in all of these functions are defined in the pole scheme.

\section{Transition matrix elements}\label{appendHME}
The matrix elements for the $\Lambda_b\to \Lambda$ transition can be written in the helicity basis \cite{Feldmann:2011xf}
\begin{align}
\left\langle \Lambda(k, s_{\Lambda})\left\vert \bar{s}\gamma_{\mu}b\right\vert \Lambda_b(p, s_{\Lambda_b})\right\rangle
&=\bar u_{\Lambda}(k, s_{\Lambda})\Big[f_t^V(q^{2})(m_{\Lambda_b}-m_{\Lambda})\frac{q_{\mu}}{q^2}\notag\\
&+f_0^V(q^2)\frac{m_{\Lambda_b}+m_{\Lambda}}{s_+}\big\{p_{\mu}+k_{\mu}-\frac{q_{\mu}}
{q^2}(m^2_{\Lambda_b}-m^2_{\Lambda})\big\}\notag
\\
&+f_{\perp}^V(q^2)\big\{\gamma_{\mu}-\frac{2m_{\Lambda}}{s_+}p_{\mu}-
\frac{2m_{\Lambda_b}}{s_+}k_{\mu}\big\}\Big]u_{\Lambda_b}(p, s_{\Lambda_b}).\label{HM1}
\end{align}
\begin{align}
\left\langle \Lambda(k, s_{\Lambda})\left\vert \bar{s}\gamma_{\mu}\gamma_5 b\right\vert \Lambda_b(p, s_{\Lambda_b})\right\rangle
&=-\bar u_{\Lambda}(k, s_{\Lambda})\gamma_5\Big[f_t^A(q^{2})(m_{\Lambda_b}+m_{\Lambda})\frac{q_{\mu}}{q^2}\notag\\
&+f_0^A(q^2)\frac{m_{\Lambda_b}-m_{\Lambda}}{s_-}\big\{p_{\mu}+k_{\mu}-\frac{q_{\mu}}
{q^2}(m^2_{\Lambda_b}-m^2_{\Lambda})\big\}\notag
\\
&+f_{\perp}^A(q^2)\big\{\gamma_{\mu}+\frac{2m_{\Lambda}}{s_-}p_{\mu}-
\frac{2m_{\Lambda_b}}{s_-}k_{\mu}\big\}\Big]u_{\Lambda_b}(p, s_{\Lambda_b}).\label{HM2}
\end{align}
Additionally,
\begin{align}
\langle \Lambda(k, s_{\Lambda})| \bar{s}i\sigma_{\mu\nu}\gamma^{\nu}&b | \Lambda_b(p, s_{\Lambda_b})\rangle
=-\bar u_{\Lambda}(k, s_{\Lambda})\Big[f_0^T(q^2)\frac{q^2}{s_+}\big\{p_{\mu}+k_{\mu}-\frac{q_{\mu}}
{q^2}(m^2_{\Lambda_b}-m^2_{\Lambda})\big\}\notag
\\
&+f_{\perp}^T(q^2)(m_{\Lambda_b}+m_{\Lambda})\big\{\gamma_{\mu}-\frac{2m_{\Lambda}}{s_+}p_{\mu}-
\frac{2m_{\Lambda_b}}{s_+}k_{\mu}\big\}\Big]u_{\Lambda_b}(p, s_{\Lambda_b}),\label{HM3}
\\
\langle \Lambda(k, s_{\Lambda})| \bar{s}i\sigma_{\mu\nu}\gamma^{\nu}\gamma_5 &b| \Lambda_b(p, s_{\Lambda_b})\rangle
=-\bar u_{\Lambda}(k, s_{\Lambda})\gamma_5\Big[f_0^{T5}(q^2)\frac{q^2}{s_-}\big\{p_{\mu}+k_{\mu}-\frac{q_{\mu}}
{q^2}(m^2_{\Lambda_b}-m^2_{\Lambda})\big\}\notag
\\
&+f_{\perp}^{T5}(q^2)(m_{\Lambda_b}-m_{\Lambda})\big\{\gamma_{\mu}+\frac{2m_{\Lambda}}{s_-}p_{\mu}-
\frac{2m_{\Lambda_b}}{s_-}k_{\mu}\big\}\Big]u_{\Lambda_b}(p, s_{\Lambda_b}).\label{HM4}
\end{align}

\section{$\Lambda \rightarrow N \pi$ decay}
Considering Eq. (\ref{eq:cascade2}), and using Eqs. (\ref{eq:cascade4})--(\ref{eq:cascade7}), the corresponding amplitude squared for $\Lambda \rightarrow N \pi$ decay, with the unpolarized $\Lambda$ can be obtained as
\begin{eqnarray}
\Big|\mathcal{M}_2\Big|^2
=\frac{1}{2}\sum_{s_{\Lambda},\,s_{N}}|N_2|^2\Big|{H}_2\left(s_{\Lambda},s_N\right)\Big|^2 
=|N_2|^2 \Big(r_+|\omega|^2+r_-|\xi|^2\Big).\label{eq:cascadefullamp}
\end{eqnarray}
The decay rate is given by
\begin{equation}
	\Gamma\left(\Lambda\to N \pi\right) =  \frac{\sqrt{r_+r_-}}{16\pi m_\Lambda^3} |N_2|^2\bigg( r_+|\omega|^2+ r_-|\xi|^2   \bigg)\, .
\end{equation}
From Eq. (\ref{eq:FullFF2}), the computation of the four-fold angular distributions will require the interference terms between matrix elements with different $\Lambda$ spins, which can be defined as~\footnote{Note that the normalization factor for $\Gamma_2(s^{(a)}_{\Lambda},\,s^{(b)}_{\Lambda})$ is taken as the phase space part which is common in both $\Lambda\to N\pi$ and $\Lambda_b\to \Lambda\,(\to N\pi)\ell^+\ell^-$ decays. With this choice, our definition of the $\Gamma_{2}(s^{(a)}_{\Lambda},\,s^{(b)}_{\Lambda})$ is slightly different from the one given in \cite{Boer:2014kda}, and is such that the decay rate of $\Lambda\to N\pi$ reads as in Eq. (\ref{Gammma2combfull}).}
\begin{eqnarray}
\Gamma_{2}\Big(s^{(a)}_{\Lambda},\,s^{(b)}_{\Lambda}\Big)=\frac{\sqrt{r_+r_-}}{16\pi m_{\Lambda}^3}
\sum_{s_{N}}|N_2|^2{H}_2\left(s^{(a)}_{\Lambda},s_N\right) H_2^*\left(s^{(b)}_{\Lambda},s_N\right)
.\label{Amp1Gammma2}
\end{eqnarray}
The explicit expression of $\Gamma_{2}$ is written as
\begin{eqnarray}
\Gamma_{2}&=&\mathcal{B}(\Lambda\to N \pi)\Gamma_{\Lambda}\begin{pmatrix}
1+\alpha \cos{\theta_{\Lambda}} & -\alpha\sin\theta_{\Lambda} \\
-\alpha\sin\theta_{\Lambda} & 1-\alpha \cos{\theta_{\Lambda}} 
\end{pmatrix},\label{Gammma2comba}
\end{eqnarray}
where the rows and columns correspond to the values of $s^{(a)}_{\Lambda},s^{(b)}_{\Lambda}=1/2,-1/2$, respectively, and the parity violating parameter $\alpha$ is the same as given in \cite{Boer:2014kda,Das:2018iap}. The factor $\mathcal{B}(\Lambda\to N \pi)$ is the same as obtained in Eqs. (\ref{eq:finaltrig}), (\ref{eq:lon}), (\ref{eq:nor}), and (\ref{eq:trans}), with its form given as
\begin{equation}
	\mathcal{B}\left(\Lambda\to N \pi\right) =\frac{1}{\Gamma_\Lambda}  \frac{\sqrt{r_+r_-}}{16\pi m_\Lambda^3} |N_2|^2\bigg( r_+|\omega|^2+ r_-|\xi|^2   \bigg)\, .
\end{equation}
The decay rate of $\Lambda\to N\pi$ is also written as
\begin{eqnarray}
\Gamma(\Lambda\to N\pi)=\sum_{s_{\Lambda}}\frac{\Gamma_{2}\Big(s_{\Lambda},\,s_{\Lambda}\Big)}{2}=\frac{1}{2}\Big[\Gamma_{2}\left(+1/2,\,+1/2\right)+\Gamma_{2}\left(-1/2,\,-1/2\right)\Big].\label{Gammma2combfull}
\end{eqnarray}

\section{ Analytical expressions of the angular coefficients in terms of helicity amplitudes with unpolarized and polarized lepton}
\label{AppE}
The expressions of the unpolarized angular coefficients, in terms of the helicity amplitudes, read as
\begin{align}
        K_{1ss} &= |N_3|^2\Bigg[\frac{2m_{\ell}^2}{q^2} \bigg(\Big|H_{-}^{1}\left(-\tfrac{1}{2}, -\tfrac{1}{2}\right)\Big|^2+ \Big|  H_{+}^{1}\left(+\tfrac{1}{2}, +\tfrac{1}{2}\right)\Big|^2-2\Big|  H_{0}^{2}\left(-\tfrac{1}{2}, +\tfrac{1}{2}\right)\Big|^2 -2\Big|  H_{0}^{2}\left(+\tfrac{1}{2}, -\tfrac{1}{2}\right)\Big|^2- \Big|  H_{-}^{2}\left(-\tfrac{1}{2}, -\tfrac{1}{2}\right)\Big|^2\notag\\ &  \quad -\Big|  H_{+}^{2}\left(+\tfrac{1}{2}, +\tfrac{1}{2}\right)\Big|^2+2\Big|  H_{t}^{2}\left(-\tfrac{1}{2}, +\tfrac{1}{2}\right)\Big|^2 +2\Big|  H_{t}^{2}\left(+\tfrac{1}{2}, -\tfrac{1}{2}\right)\Big|^2\bigg)+\frac{1}{2}\bigg( 2  \Big|  H_{0}^{1}\left(-\tfrac{1}{2}, +\tfrac{1}{2}\right)\Big|^2+ 2\Big|  H_{0}^{1}\left(+\tfrac{1}{2}, -\tfrac{1}{2}\right)\Big|^2  \notag \\ &  \quad + \Big|  H_{-}^{1}\left(-\tfrac{1}{2}, -\tfrac{1}{2}\right)\Big|^2
        +\Big|  H_{+}^{1}\left(+\tfrac{1}{2}, +\tfrac{1}{2}\right)\Big|^2
        +2\Big|  H_{0}^{2}\left(-\tfrac{1}{2}, +\tfrac{1}{2}\right)\Big|^2+ 2 \Big|  H_{0}^{2}\left(+\tfrac{1}{2}, -\tfrac{1}{2}\right)\Big|^2+\Big|  H_{-}^{2}\left(-\tfrac{1}{2}, -\tfrac{1}{2}\right)\Big|^2\notag \\ & \quad+\Big|  H_{+}^{2}\left(+\tfrac{1}{2}, +\tfrac{1}{2}\right)\Big|^2\bigg)\Bigg],\label{U1}
\\
        K_{1cc}&=  |N_3|^2\Bigg[
        \frac{4 m_{\ell}^2}{q^2}\bigg( \Big|  H_{0}^{1}\left(-\tfrac{1}{2}, +\tfrac{1}{2}\right)\Big|^2 +\Big|  H_{0}^{1}\left(+\tfrac{1}{2}, -\tfrac{1}{2}\right)\Big|^2 -\Big|  H_{-}^{2}\left(-\tfrac{1}{2}, -\tfrac{1}{2}\right)\Big|^2-\Big|  H_{+}^{2}\left(+\tfrac{1}{2}, +\tfrac{1}{2}\right)\Big|^2 + \Big|  H_{t}^{2}\left(-\tfrac{1}{2}, +\tfrac{1}{2}\right)\Big|^2\notag \\ & \quad+\Big|  H_{t}^{2}\left(+\tfrac{1}{2}, -\tfrac{1}{2}\right)\Big|^2\bigg)+\bigg( \Big|  H_{-}^{1}\left(-\tfrac{1}{2}, -\tfrac{1}{2}\right)\Big|^2 +\Big|  H_{+}^{1}\left(+\tfrac{1}{2}, +\tfrac{1}{2}\right)\Big|^2  +\Big|  H_{-}^{2}\left(-\tfrac{1}{2}, -\tfrac{1}{2}\right)\Big|^2 + \Big|  H_{+}^{2}\left(+\tfrac{1}{2}, +\tfrac{1}{2}\right)\Big|^2\bigg)\Bigg],
\\ %
        K_{1c} &=
         -2\,|N_3|^2 \beta_{\ell}\,\mathcal{R}e\bigg(H_{-}^{1}\left(-\tfrac{1}{2}, -\tfrac{1}{2}\right)H_{-}^{2 \ast}\left(-\tfrac{1}{2}, -\tfrac{1}{2}\right)- H_{+}^{1}\left(+\tfrac{1}{2}, +\tfrac{1}{2}\right)H_{+}^{2 \ast}\left(+\tfrac{1}{2}, +\tfrac{1}{2}\right)\bigg),
        \label{Uk1c}
\\ %
        K_{2ss}&= |N_3|^2\alpha \Bigg[-\frac{2m_\ell^2}{q^2} \bigg(\Big|  H_{-}^{1}\left(-\tfrac{1}{2}, -\tfrac{1}{2}\right)\Big|^2 -\Big|  H_{+}^{1}\left(+\tfrac{1}{2}, +\tfrac{1}{2}\right)\Big|^2 + 2\Big|  H_{0}^{2}\left(-\tfrac{1}{2}, +\tfrac{1}{2}\right)\Big|^2 - 2\Big|  H_{0}^{2}\left(+\tfrac{1}{2}, -\tfrac{1}{2}\right)\Big|^2
        \notag\\ & \quad- \Big|H_{-}^{2}\left(-\tfrac{1}{2}, -\tfrac{1}{2}\right)\Big|^2  +\Big|H_{+}^{2}\left(+\tfrac{1}{2}, +\tfrac{1}{2}\right)\Big|^2  - 2\Big|  H_{t}^{2}\left(-\tfrac{1}{2}, +\tfrac{1}{2}\right)\Big|^2+2\Big|  H_{t}^{2}\left(+\tfrac{1}{2}, -\tfrac{1}{2}\right)\Big|^2\bigg)+\frac{1}{2}\bigg( 2\Big|  H_{0}^{1}\left(-\tfrac{1}{2}, +\tfrac{1}{2}\right)\Big|^2 \notag\\ & \quad-2\Big|  H_{0}^{1}\left(+\tfrac{1}{2}, -\tfrac{1}{2}\right)\Big|^2 -\Big|  H_{-}^{1}\left(-\tfrac{1}{2}, -\tfrac{1}{2}\right)\Big|^2+\Big|  H_{+}^{1}\left(+\tfrac{1}{2}, +\tfrac{1}{2}\right)\Big|^2+ 2\Big|  H_{0}^{2}\left(-\tfrac{1}{2}, +\tfrac{1}{2}\right)\Big|^2 -2\Big|  H_{0}^{2}\left(+\tfrac{1}{2}, -\tfrac{1}{2}\right)\Big|^2\notag \\ & \quad-\Big| H_{-}^{2}\left(-\tfrac{1}{2}, -\tfrac{1}{2}\right)\Big|^2+\Big|  H_{+}^{2}\left(+\tfrac{1}{2}, +\tfrac{1}{2}\right)\Big|^2\bigg)\Bigg],
        \\
        K_{2cc }&=|N_3|^2\alpha \Bigg[\frac{4m_\ell^2}{q^2} \bigg( \Big|  H_{0}^{1}\left(-\tfrac{1}{2}, +\tfrac{1}{2}\right)\Big|^2 -\Big|  H_{0}^{1}\left(+\tfrac{1}{2}, -\tfrac{1}{2}\right)\Big|^2 +\Big|  H_{-}^{2}\left(-\tfrac{1}{2}, -\tfrac{1}{2}\right)\Big|^2- \Big|  H_{+}^{2}\left(+\tfrac{1}{2}, +\tfrac{1}{2}\right)\Big|^2+\Big|  H_{t}^{2}\left(-\tfrac{1}{2}, +\tfrac{1}{2}\right)\Big|^2\notag \\ & \quad  -\Big|  H_{t}^{2}\left(+\tfrac{1}{2}, -\tfrac{1}{2}\right)\Big|^2\bigg) + \bigg(\Big|H_{+}^{1}\left(+\tfrac{1}{2}, +\tfrac{1}{2}\right)\Big|^2 -\Big|  H_{-}^{1}\left(-\tfrac{1}{2}, -\tfrac{1}{2}\right)\Big|^2+\Big|  H_{+}^{2}\left(+\tfrac{1}{2}, +\tfrac{1}{2}\right)\Big|^2 -\Big|  H_{-}^{2}\left(-\tfrac{1}{2}, -\tfrac{1}{2}\right)\Big|^2\bigg) \Bigg],
\\ %
        K_{2c} &=2\,|N_3|^2\alpha\, \beta_{\ell}\,\mathcal{R}e\bigg(H_{-}^{1}\left(-\tfrac{1}{2}, -\tfrac{1}{2}\right)H_{-}^{2 \ast}\left(-\tfrac{1}{2}, -\tfrac{1}{2}\right)+ H_{+}^{1}\left(+\tfrac{1}{2}, +\tfrac{1}{2}\right)H_{+}^{2 \ast}\left(+\tfrac{1}{2}, +\tfrac{1}{2}\right)\bigg),
\\
K_{3sc}&= \sqrt{2}\,|N_3|^2\alpha  \Bigg[-\frac{4m_\ell^2}{q^2}\,\mathcal{I}m\bigg(H_{-}^{1}\left(-\tfrac{1}{2}, -\tfrac{1}{2}\right)H_{0}^{1 \ast}\left(-\tfrac{1}{2}, +\tfrac{1}{2}\right)+ H_{+}^{1}\left(+\tfrac{1}{2}, +\tfrac{1}{2}\right)H_{0}^{1 \ast}\left(+\tfrac{1}{2}, -\tfrac{1}{2}\right)\notag\\&\quad+H_{-}^{2}\left(-\tfrac{1}{2}, -\tfrac{1}{2}\right)H_{0}^{2 \ast}\left(-\tfrac{1}{2}, +\tfrac{1}{2}\right) +H_{+}^{2}\left(+\tfrac{1}{2}, +\tfrac{1}{2}\right)H_{0}^{2\ast}\left(+\tfrac{1}{2}, -\tfrac{1}{2}\right)\bigg)+
        \mathcal{I}m\bigg(H_{-}^{1}\left(-\tfrac{1}{2}, -\tfrac{1}{2}\right)H_{0}^{1 \ast}\left(-\tfrac{1}{2}, +\tfrac{1}{2}\right)
        \notag\\&\quad+ H_{+}^{1}\left(+\tfrac{1}{2}, +\tfrac{1}{2}\right)H_{0}^{1 \ast}\left(+\tfrac{1}{2}, -\tfrac{1}{2}\right)+H_{-}^{2}\left(-\tfrac{1}{2}, -\tfrac{1}{2}\right)H_{0}^{2 \ast}\left(-\tfrac{1}{2}, +\tfrac{1}{2}\right) +H_{+}^{2}\left(+\tfrac{1}{2}, +\tfrac{1}{2}\right)H_{0}^{2\ast}\left(+\tfrac{1}{2}, -\tfrac{1}{2}\right)\bigg)\Bigg],
\\        
K_{3s}&= \sqrt{2}\,|N_3|^2\alpha\,\beta_{\ell}\, \mathcal{I}m\bigg(H_{+}^{2}\left(+\tfrac{1}{2}, +\tfrac{1}{2}\right)H_{0}^{1 \ast}\left(+\tfrac{1}{2}, -\tfrac{1}{2}\right)-H_{-}^{2}\left(-\tfrac{1}{2}, -\tfrac{1}{2}\right)H_{0}^{1\ast }\left(-\tfrac{1}{2}, +\tfrac{1}{2}\right)\notag \\&\quad +H_{+}^{1}\left(+\tfrac{1}{2}, +\tfrac{1}{2}\right)H_{0}^{2 \ast}\left(+\tfrac{1}{2}, -\tfrac{1}{2}\right)-H_{-}^{1}\left(-\tfrac{1}{2}, -\tfrac{1}{2}\right)H_{0}^{2\ast}\left(-\tfrac{1}{2}, +\tfrac{1}{2}\right)\bigg).\label{Uk2c}
        \end{align}

\begin{align}   
        K_{4sc}&= \sqrt{2}\,|N_3|^2\,\alpha \Bigg[-\frac{4m_\ell^2}{q^2}\,\mathcal{R}e\bigg(H_{-}^{1}\left(-\tfrac{1}{2}, -\tfrac{1}{2}\right)H_{0}^{1 \ast}\left(-\tfrac{1}{2}, +\tfrac{1}{2}\right)- H_{+}^{1}\left(+\tfrac{1}{2}, +\tfrac{1}{2}\right)H_{0}^{1 \ast}\left(+\tfrac{1}{2}, -\tfrac{1}{2}\right)\notag\\&\quad
        + H_{-}^{2}\left(-\tfrac{1}{2}, -\tfrac{1}{2}\right)H_{0}^{2 \ast}\left(-\tfrac{1}{2}, +\tfrac{1}{2}\right)-H_{+}^{2}\left(+\tfrac{1}{2}, +\tfrac{1}{2}\right)H_{0}^{2\ast}\left(+\tfrac{1}{2}, -\tfrac{1}{2}\right)\bigg)
        +\mathcal{R}e\bigg(H_{-}^{1}\left(-\tfrac{1}{2}, -\tfrac{1}{2}\right)H_{0}^{1 \ast}\left(-\tfrac{1}{2}, +\tfrac{1}{2}\right)\notag\\&\quad- H_{+}^{1}\left(+\tfrac{1}{2}, +\tfrac{1}{2}\right)H_{0}^{1 \ast}\left(+\tfrac{1}{2}, -\tfrac{1}{2}\right)+ H_{-}^{2}\left(-\tfrac{1}{2}, -\tfrac{1}{2}\right)H_{0}^{2 \ast}\left(-\tfrac{1}{2}, +\tfrac{1}{2}\right)-H_{+}^{2}\left(+\tfrac{1}{2}, +\tfrac{1}{2}\right)H_{0}^{2\ast}\left(+\tfrac{1}{2}, -\tfrac{1}{2}\right)\bigg)
        \Bigg],\label{Uk4s}
\\
        K_{4s}&=  -\sqrt{2}\,|N_3|^2\alpha\,\beta_{\ell}\,\mathcal{R}e\bigg(H_{+}^{2}\left(+\tfrac{1}{2}, +\tfrac{1}{2}\right)H_{0}^{1 \ast}\left(+\tfrac{1}{2}, -\tfrac{1}{2}\right)+H_{-}^{2}\left(-\tfrac{1}{2}, -\tfrac{1}{2}\right)H_{0}^{1 \ast}\left(-\tfrac{1}{2}, +\tfrac{1}{2}\right)
        \notag\\&\quad
        +H_{+}^{1}\left(+\tfrac{1}{2}, +\tfrac{1}{2}\right)H_{0}^{2 \ast}\left(+\tfrac{1}{2}, -\tfrac{1}{2}\right)+H_{-}^{1}\left(-\tfrac{1}{2}, -\tfrac{1}{2}\right)H_{0}^{2 \ast}\left(-\tfrac{1}{2}, +\tfrac{1}{2}\right)
        \bigg),\label{U2}
\end{align}
where $\beta_{\ell} = \sqrt{1-4m^2_\ell/q^2}$, and $N_3$ , is defined as,
\begin{eqnarray}\label{24abc}
N_3=V_{tb}V^{\ast}_{ts}\Bigg[\frac{G_{F}^2\alpha^2_{em}}{3\times 2^{11} \pi^5 m_{\Lambda_b}^{3}} q^2\sqrt{\lambda}\beta_\ell\Bigg]^{1/2}\;.
\end{eqnarray}
Here, we abbreviate $\lambda \equiv \lambda(m_{\Lambda_b}^2, m_{\Lambda}^2, q^2)$, with the Källén function $\lambda(a,b,c) = a^2 + b^2 + c^2 - 2(ab + ac + bc)$.

The expressions of $\xi_{L}$ dependent angular coefficients present in the L polarized four-fold angular decay distribution are given, in terms of the helicity amplitudes, as

\begin{align}
        \mathcal{K}_{1ss \text{L}}^\prime\ &=\frac{1}{2}|N_3|^2\,\beta_{\ell}\,\mathcal{R}e\bigg(2\, H_{0}^{1}(-\tfrac{1}{2}, +\tfrac{1}{2}) \,H_{0}^{2\ast}(-\tfrac{1}{2}, +\tfrac{1}{2})+ 2\,H_{0}^{1}(+\tfrac{1}{2}, -\tfrac{1}{2}) \,
         H_{0}^{2\ast}(+\tfrac{1}{2}, -\tfrac{1}{2})\notag\\&\quad
        + H^{1}_{-}\left(-\tfrac{1}{2}, -\tfrac{1}{2}\right)\,
        H^{2\ast}_{-}\left(-\tfrac{1}{2}, -\tfrac{1}{2}\right)
        +H^{1}_{+}\left(+\tfrac{1}{2}, +\tfrac{1}{2}\right)\, H^{2\ast}_{+}\left(+\tfrac{1}{2}, +\tfrac{1}{2}\right)\bigg),\label{L1}
\\     
        \mathcal{K}^{\prime}_{1cc \text{L}}&=|N_3|^2\beta_\ell \,\mathcal{R}e\bigg( H^{1}_{-}\left(-\tfrac{1}{2}, -\tfrac{1}{2}\right)\,
        H^{2\ast}_{-}\left(-\tfrac{1}{2}, -\tfrac{1}{2}\right)
        +H^{1}_{+}\left(+\tfrac{1}{2}, +\tfrac{1}{2}\right)\, H^{2\ast}_{+}\left(+\tfrac{1}{2}, +\tfrac{1}{2}\right)\bigg),    
\\ 
        \mathcal{\mathcal{K}}^{\prime}_{1c \text{L}} &=|N_3|^2\Bigg[
        \frac{2m^2_{\ell}}{q^2}\bigg(\left|H^{2}_{-}\left(-\tfrac{1}{2}, -\tfrac{1}{2}\right)\right|^2-\left|H^{2}_{+}\left(+\tfrac{1}{2}, +\tfrac{1}{2}\right)\right|^2+2\,\mathcal{R}e\Big(H^{1}_{0} \left(-\tfrac{1}{2}, +\tfrac{1}{2}\right) H^{2\ast}_{t} \left(-\tfrac{1}{2}, +\tfrac{1}{2}\right)\notag \\ & \quad
        + H^{1}_{0} \left(+\tfrac{1}{2}, -\tfrac{1}{2}\right) H^{2\ast}_{t} \left(+\tfrac{1}{2}, -\tfrac{1}{2}\right)\Big)\bigg)+\frac{1}{2}\bigg(\left|H^{1}_{+}\left(+\tfrac{1}{2}, +\tfrac{1}{2}\right)\right|^2-\left|H^{1}_{-}\left(-\tfrac{1}{2}, -\tfrac{1}{2}\right)\right|^2
        +\left|H^{2}_{+}\left(+\tfrac{1}{2}, +\tfrac{1}{2}\right)\right|^2\notag \\ & \quad-\left|H^{2}_{-}\left(-\tfrac{1}{2}, -\tfrac{1}{2}\right)\right|^2\bigg)\Bigg],
        \label{Lk1cp}
\\
        \mathcal{K}^{\prime}_{2ss \text{L}}&=\frac{1}{2}|N_3|^2\alpha\,\beta_\ell\,\mathcal{R}e\bigg( 2\,H_{0}^{1}(-\tfrac{1}{2}, +\tfrac{1}{2}) \,
        H_{0}^{2\ast}(-\tfrac{1}{2}, +\tfrac{1}{2})-2\,H_{0}^{1}(+\tfrac{1}{2}, -\tfrac{1}{2}) \,
        H_{0}^{2\ast}(+\tfrac{1}{2}, -\tfrac{1}{2})\notag\\&\quad
        -H_{-}^{1}(-\tfrac{1}{2}, -\tfrac{1}{2}) \,H_{-}^{2\ast}(-\tfrac{1}{2}, -\tfrac{1}{2})+H_{+}^{1}(+\tfrac{1}{2}, +\tfrac{1}{2}) \,H_{+}^{2\ast}(+\tfrac{1}{2}, +\tfrac{1}{2})\bigg),
\\
\mathcal{K}^{\prime}_{2cc \text{L} }&=|N_3|^2\alpha\,\beta_\ell\,\mathcal{R}e\bigg(H^{1}_{+}\left(+\tfrac{1}{2}, +\tfrac{1}{2}\right)\, H^{2\ast}_{+}\left(+\tfrac{1}{2}, +\tfrac{1}{2}\right)- H^{1}_{-}\left(-\tfrac{1}{2}, -\tfrac{1}{2}\right)\, H^{2\ast}_{-}\left(-\tfrac{1}{2}, -\tfrac{1}{2}\right)\bigg),
\\
        \mathcal{K}^{\prime}_{2c \text{L}} &=|N_3|^2\alpha\Bigg[
        -\frac{2m^2_{\ell}}{q^2}\bigg(\left|H^{2}_{-}\left(-\tfrac{1}{2}, -\tfrac{1}{2}\right)\right|^2+\left|H^{2}_{+}\left(+\tfrac{1}{2}, +\tfrac{1}{2}\right)\right|^2+2\,\mathcal{R}e\Big( H^{1}_{0} \left(+\tfrac{1}{2}, -\tfrac{1}{2}\right) H^{2\ast}_{t} \left(+\tfrac{1}{2}, -\tfrac{1}{2}\right)\notag\\&\quad-H^{1}_{0} \left(-\tfrac{1}{2}, +\tfrac{1}{2}\right) H^{2\ast}_{t} \left(-\tfrac{1}{2}, +\tfrac{1}{2}\right)\Big)\bigg)+\frac{1}{2}\bigg(\left|H^{1}_{+}\left(+\tfrac{1}{2}, +\tfrac{1}{2}\right)\right|^2+\left|H^{1}_{-}\left(-\tfrac{1}{2}, -\tfrac{1}{2}\right)\right|^2+\left|H^{2}_{+}\left(+\tfrac{1}{2}, +\tfrac{1}{2}\right)\right|^2\notag\\&\quad
        +\left|H^{2}_{-}\left(-\tfrac{1}{2}, -\tfrac{1}{2}\right)\right|^2\bigg)\Bigg],
        \label{Lk2cp}
\\
        \mathcal{K}^{\prime}_{3sc \text{L}}&= \frac{1}{\sqrt{2}}|N_3|^2\alpha\,\beta_\ell\,\mathcal{I}m\bigg(-H^{1}_{0}\left(-\tfrac{1}{2}, +\tfrac{1}{2}\right) \, H^{2\ast}_{-}\left(-\tfrac{1}{2}, -\tfrac{1}{2}\right)
        - H^{1}_{0}\left(+\tfrac{1}{2}, -\tfrac{1}{2}\right) \, H^{2\ast}_{+}\left(+\tfrac{1}{2}, +\tfrac{1}{2}\right) \notag\\&\quad+H^{1}_{-}\left(-\tfrac{1}{2}, -\tfrac{1}{2}\right) \, H^{2\ast}_{0}\left(-\tfrac{1}{2}, +\tfrac{1}{2}\right)+H^{1}_{+}\left(+\tfrac{1}{2}, +\tfrac{1}{2}\right) \, H^{2\ast}_{0}\left(+\tfrac{1}{2}, -\tfrac{1}{2}\right)\bigg),
\end{align}
\begin{align}
        \mathcal{K}^{\prime}_{3s \text{L}}&=\sqrt{2}\,|N_3|^2\, \alpha\Bigg[ -\frac{2m^2_\ell}{q^2}\,\mathcal{I}m\bigg( H^{1}_{-}\left(-\tfrac{1}{2}, -\tfrac{1}{2}\right)\, H^{2\ast}_{t}\left(-\tfrac{1}{2}, +\tfrac{1}{2}\right)
        + \, H^{1}_{+}\left(+\tfrac{1}{2}, +\tfrac{1}{2}\right)\,H^{2\ast}_{t}\left(+\tfrac{1}{2}, -\tfrac{1}{2}\right) \notag\\&\quad -H^{2}_{-}\left(-\tfrac{1}{2},-\tfrac{1}{2}\right) \, H^{2\ast}_{0}\left(-\tfrac{1}{2}, +\tfrac{1}{2}\right)+ H^{2}_{+}\left(+\tfrac{1}{2}, +\tfrac{1}{2}\right) \, H^{2\ast}_{0}\left(+\tfrac{1}{2}, -\tfrac{1}{2}\right)\bigg)-\frac{1}{2}\,\mathcal{I}m\bigg(H^{1}_{-}\left(-\tfrac{1}{2}, -\tfrac{1}{2}\right)\,H^{1\ast}_{0}\left(-\tfrac{1}{2}, +\tfrac{1}{2}\right)\notag\\&\quad - H^{1}_{+}\left(+\tfrac{1}{2}, +\tfrac{1}{2}\right)\,H^{1\ast}_{0}\left(+\tfrac{1}{2}, -\tfrac{1}{2}\right)+H^{2}_{-}\left(-\tfrac{1}{2}, -\tfrac{1}{2}\right)\,H^{2\ast}_{0}\left(-\tfrac{1}{2}, +\tfrac{1}{2}\right)-H^{2}_{+}\left(+\tfrac{1}{2}, +\tfrac{1}{2}\right)\,H^{2\ast}_{0}\left(+\tfrac{1}{2}, -\tfrac{1}{2}\right)\bigg)\Bigg],
\\
        \mathcal{K}^{\prime}_{4sc \text{L}}&= \frac{1}{\sqrt{2}}|N_3|^2\alpha\,\beta_\ell\,\mathcal{R}e\bigg(H^{1}_{0}\left(-\tfrac{1}{2}, +\tfrac{1}{2}\right) \, H^{2\ast}_{-}\left(-\tfrac{1}{2}, -\tfrac{1}{2}\right)
        - H^{1}_{0}\left(+\tfrac{1}{2}, -\tfrac{1}{2}\right) \, H^{2\ast}_{+}\left(+\tfrac{1}{2}, +\tfrac{1}{2}\right)\notag\\&\quad +H^{1}_{-}\left(-\tfrac{1}{2}, -\tfrac{1}{2}\right) \, H^{2\ast}_{0}\left(-\tfrac{1}{2}, +\tfrac{1}{2}\right)-H^{1}_{+}\left(+\tfrac{1}{2}, +\tfrac{1}{2}\right) \, H^{2\ast}_{0}\left(+\tfrac{1}{2}, -\tfrac{1}{2}\right) \bigg),
        \label{L2}
\\
        \mathcal{K}^{\prime}_{4s \text{L}}&=\sqrt{2}\,|N_3|^2\, \alpha\Bigg[ \frac{2m^2_\ell}{q^2}\,\mathcal{R}e\bigg(-H^{1}_{-}\left(-\tfrac{1}{2}, -\tfrac{1}{2}\right) \, H^{2\ast}_{t}\left(-\tfrac{1}{2}, +\tfrac{1}{2}\right)
        + H^{1}_{+}\left(+\tfrac{1}{2}, +\tfrac{1}{2}\right) \, H^{2\ast}_{t}\left(+\tfrac{1}{2}, -\tfrac{1}{2}\right) \notag\\&\quad +H^{2}_{-}\left(-\tfrac{1}{2}, -\tfrac{1}{2}\right)\,H^{2\ast}_{0}\left(-\tfrac{1}{2}, +\tfrac{1}{2}\right) + H^{2}_{+}\left(+\tfrac{1}{2}, +\tfrac{1}{2}\right)\,H^{2\ast}_{0}\left(+\tfrac{1}{2}, -\tfrac{1}{2}\right)\bigg)-\frac{1}{2}\,\mathcal{R}e\bigg( H^{1}_{-}\left(-\tfrac{1}{2}, -\tfrac{1}{2}\right)\,H^{1\ast}_{0}\left(-\tfrac{1}{2}, +\tfrac{1}{2}\right)\notag\\&\quad +H^{1}_{+}\left(+\tfrac{1}{2}, +\tfrac{1}{2}\right)\,H^{1\ast}_{0}\left(+\tfrac{1}{2}, -\tfrac{1}{2}\right)+H^{2}_{-}\left(-\tfrac{1}{2}, -\tfrac{1}{2}\right)\,H^{2\ast}_{0}\left(-\tfrac{1}{2}, +\tfrac{1}{2}\right) +H^{2}_{+}\left(+\tfrac{1}{2}, +\tfrac{1}{2}\right)\,H^{2\ast}_{0}\left(+\tfrac{1}{2}, -\tfrac{1}{2}\right)\bigg)\Bigg].
        \label{Lk4sp} 
\end{align}

The expressions of $\xi_{N}$ dependent angular coefficients present in the N polarized four-fold angular decay distribution are given, in terms of the helicity amplitudes, as    
\begin{align}
        \mathcal{K}_{1s \text{N}}&= |N_3|^2 \,\beta_{\ell} \frac{m_{\ell}}{\sqrt{q^2}}\, \mathcal{I}m \bigg( H_{+}^{1}\left(+\tfrac{1}{2}, +\tfrac{1}{2}\right)\,H_{+}^{2\ast}\left(+\tfrac{1}{2}, +\tfrac{1}{2}\right) -H_{-}^{1}\left(-\tfrac{1}{2}, -\tfrac{1}{2}\right)\,
        H_{-}^{2\ast}\left(-\tfrac{1}{2}, -\tfrac{1}{2}\right)
        \notag\\&\quad -2\,
        H_{0}^{2}\left(-\tfrac{1}{2}, +\tfrac{1}{2}\right)\,H_{t}^{2\ast}\left(-\tfrac{1}{2}, +\tfrac{1}{2}\right)-2\,
        H_{0}^{2}\left(+\tfrac{1}{2}, -\tfrac{1}{2}\right)\,H_{t}^{2\ast}\left(+\tfrac{1}{2}, -\tfrac{1}{2}\right)\bigg),
        \label{NewN1}
\\
        \mathcal{K}_{2s \text{N}}&= |N_3|^2\,\alpha\, \beta_{\ell}\frac{ m_{\ell}}{\sqrt{q^2}}\, \mathcal{I}m \bigg( H_{+}^{1}\left(+\tfrac{1}{2}, +\tfrac{1}{2}\right)\,H_{+}^{2\ast}\left(+\tfrac{1}{2}, +\tfrac{1}{2}\right) +H_{-}^{1}\left(-\tfrac{1}{2}, -\tfrac{1}{2}\right)\,
        H_{-}^{2\ast}\left(-\tfrac{1}{2}, -\tfrac{1}{2}\right)
        \notag\\&\quad -2\,
        H_{0}^{2}\left(-\tfrac{1}{2}, +\tfrac{1}{2}\right)\,H_{t}^{2\ast}\left(-\tfrac{1}{2}, +\tfrac{1}{2}\right)+2\,
        H_{0}^{2}\left(+\tfrac{1}{2}, -\tfrac{1}{2}\right)\,H_{t}^{2\ast}\left(+\tfrac{1}{2}, -\tfrac{1}{2}\right)\bigg),
\\
        \mathcal{K}_{3c \text{N}}&= \sqrt{2}\,|N_3|^2\,\alpha\,\beta_{\ell}\,\frac{m_{\ell}}{\sqrt{q^2}}\,\mathcal{R}e\bigg( H^{1}_{0}\left(-\tfrac{1}{2}, +\tfrac{1}{2}\right)\, H^{2\ast}_{-}\left(-\tfrac{1}{2}, -\tfrac{1}{2}\right)- H^{1}_{0}\left(+\tfrac{1}{2}, -\tfrac{1}{2}\right)\, H^{2\ast}_{+}\left(+\tfrac{1}{2}, +\tfrac{1}{2}\right) \notag\\
        &\quad +H^{2}_{-}\left(-\tfrac{1}{2}, -\tfrac{1}{2}\right)\, H^{2\ast}_{t}\left(-\tfrac{1}{2}, +\tfrac{1}{2}\right)+H^{2}_{+}\left(+\tfrac{1}{2}, +\tfrac{1}{2}\right)\, H^{2\ast}_{t}\left(+\tfrac{1}{2}, -\tfrac{1}{2}\right)\bigg),
\\
        \mathcal{K}_{3\text{N}}&= -\sqrt{2}\,|N_3|^2\,\alpha\, \frac{ m_{\ell}}{\sqrt{q^2}}\,\mathcal{R}e\bigg(H^{1}_{-}\left(-\tfrac{1}{2}, -\tfrac{1}{2}\right)\,H^{1\ast}_{0}\left(-\tfrac{1}{2}, +\tfrac{1}{2}\right)
        +H^{1}_{+}\left(+\tfrac{1}{2}, +\tfrac{1}{2}\right)H^{1 \ast}_{0}\left(+\tfrac{1}{2}, -\tfrac{1}{2}\right)\notag\\&\quad+H^{1}_{-}\left(-\tfrac{1}{2}, -\tfrac{1}{2}\right)\, H^{2\ast}_{t}\left(-\tfrac{1}{2}, +\tfrac{1}{2}\right)-H^{1}_{+}\left(+\tfrac{1}{2}, +\tfrac{1}{2}\right)\, H^{2\ast}_{t}\left(+\tfrac{1}{2}, -\tfrac{1}{2}\right) \bigg),
\\        
        \mathcal{K}_{4c \text{N}}&=  \sqrt{2}\,|N_3|^2\,\alpha\, \beta_{\ell}\frac{ m_{\ell}}{\sqrt{q^2}}\,\mathcal{I}m\bigg(H^{1}_{0}\left(-\tfrac{1}{2}, +\tfrac{1}{2}\right)\,H^{2 \ast}_{-}\left(-\tfrac{1}{2}, -\tfrac{1}{2}\right)
        +H^{1}_{0}\left(+\tfrac{1}{2}, -\tfrac{1}{2}\right) \, H^{2\ast}_{+}\left(+\tfrac{1}{2}, +\tfrac{1}{2}\right) \notag\\&\quad -H^{2}_{-}\left(-\tfrac{1}{2}, -\tfrac{1}{2}\right)\, H^{2\ast}_{t}\left(-\tfrac{1}{2}, +\tfrac{1}{2}\right)+H^{2}_{+}\left(+\tfrac{1}{2}, +\tfrac{1}{2}\right)H^{2\ast}_{t}\left(+\tfrac{1}{2}, -\tfrac{1}{2}\right)\bigg),
\\
\mathcal{K}_{4\text{N}}&= \sqrt{2}\,|N_3|^2\,\alpha\,\frac{ m_{\ell}}{\sqrt{q^2}}\,\mathcal{I}m\bigg( H^{1}_{-}\left(-\tfrac{1}{2}, -\tfrac{1}{2}\right)\,H^{1\ast}_{0}\left(-\tfrac{1}{2}, +\tfrac{1}{2}\right)
        - H^{1}_{+}\left(+\tfrac{1}{2}, +\tfrac{1}{2}\right)H^{1 \ast}_{0}\left(+\tfrac{1}{2}, -\tfrac{1}{2}\right)\notag\\&\quad+H^{1}_{-}\left(-\tfrac{1}{2}, -\tfrac{1}{2}\right)\, H^{2\ast}_{t}\left(-\tfrac{1}{2}, +\tfrac{1}{2}\right)+H^{1}_{+}\left(+\tfrac{1}{2}, +\tfrac{1}{2}\right)\, H^{2\ast}_{t}\left(+\tfrac{1}{2}, -\tfrac{1}{2}\right)\bigg). 
        \label{NewN2}
\end{align}
The expressions of $\xi_{T}$ dependent angular coefficients present in the T polarized four-fold angular decay distribution are given, in terms of the helicity amplitudes, as     
\begin{align}
        \mathcal{K}_{1sc \text{T}}&=-|N_3|^2\,\beta_{\ell}\,\frac{m_{\ell}}{\sqrt{q^2}}\,\mathcal{R}e\bigg( 2H_{0}^{1 }\left(-\tfrac{1}{2}, +\tfrac{1}{2}\right)\,
        H_{0}^{2\ast}\left(-\tfrac{1}{2}, +\tfrac{1}{2}\right)
        +2H_{0}^{1}\left(+\tfrac{1}{2}, -\tfrac{1}{2}\right)\,
        H_{0}^{2 \ast}\left(+\tfrac{1}{2}, -\tfrac{1}{2}\right)\notag \\ & \quad
        -H_{-}^{1 }\left(-\tfrac{1}{2}, -\tfrac{1}{2}\right)\,
        H_{-}^{2\ast}\left(-\tfrac{1}{2}, -\tfrac{1}{2}\right) -H_{+}^{1}\left(+\tfrac{1}{2}, +\tfrac{1}{2}\right)\,
        H_{+}^{2 \ast}\left(+\tfrac{1}{2}, +\tfrac{1}{2}\right) \bigg),\label{ZiTexpF}
\end{align}        
\begin{align}
        \mathcal{K}_{1s \text{T}}&=|N_3|^2\,\frac{m_{\ell}}{\sqrt{q^2}}\,\bigg(\big| H^{1}_{+}\left(+\tfrac{1}{2}, +\tfrac{1}{2}\right)\big|^2 -\big| H^{1}_{-}\left(-\tfrac{1}{2}, -\tfrac{1}{2}\right)\big|^2
        +2\,\mathcal{R}e\Big( H^{1}_{0}\left(-\tfrac{1}{2}, +\tfrac{1}{2}\right)\, H^{2\ast}_{t}\left(-\tfrac{1}{2}, +\tfrac{1}{2}\right)\notag\\&
        + H^{1}_{0}\left(+\tfrac{1}{2}, -\tfrac{1}{2}\right)\, H^{2 \ast}_{t}\left(+\tfrac{1}{2}, -\tfrac{1}{2}\right)\Big) \bigg),
        \label{NewT1}
\\
        \mathcal{K}_{2sc \text{T}}&=-|N_3|^2\,\alpha\, \beta_{\ell}\frac{ m_{\ell}}{\sqrt{q^2}}\,\mathcal{R}e\bigg( 2H_{0}^{1 }\left(-\tfrac{1}{2}, +\tfrac{1}{2}\right)\,
        H_{0}^{2\ast}\left(-\tfrac{1}{2}, +\tfrac{1}{2}\right)
        -2H_{0}^{1}\left(+\tfrac{1}{2}, -\tfrac{1}{2}\right)\,
        H_{0}^{2 \ast}\left(+\tfrac{1}{2}, -\tfrac{1}{2}\right)\notag \\ & \quad
        +H_{-}^{1 }\left(-\tfrac{1}{2}, -\tfrac{1}{2}\right)\,
        H_{-}^{2\ast}\left(-\tfrac{1}{2}, -\tfrac{1}{2}\right) -H_{+}^{1}\left(+\tfrac{1}{2}, +\tfrac{1}{2}\right)\,
        H_{+}^{2 \ast}\left(+\tfrac{1}{2}, +\tfrac{1}{2}\right) \bigg),
\\
        \mathcal{K}_{2s \text{T}}&=|N_3|^2\,\alpha\, \frac{ m_{\ell}}{\sqrt{q^2}}\bigg(\big| H^{1}_{+}\left(+\tfrac{1}{2}, +\tfrac{1}{2}\right)\big|^2+\big| H^{1}_{-}\left(-\tfrac{1}{2}, -\tfrac{1}{2}\right)\big|^2
        +2\mathcal{R}e\Big( H^{1}_{0}\left(-\tfrac{1}{2}, +\tfrac{1}{2}\right)\, H^{2\ast}_{t}\left(-\tfrac{1}{2}, +\tfrac{1}{2}\right)\notag\\&
        - H^{1}_{0}\left(+\tfrac{1}{2}, -\tfrac{1}{2}\right)\, H^{2 \ast}_{t}\left(+\tfrac{1}{2}, -\tfrac{1}{2}\right)\Big)\bigg),
\\
        \mathcal{K}_{3ss\text{T}}&=\sqrt{2}\,|N_3|^2\,\alpha\, \beta_{\ell}\frac{ m_{\ell}}{\sqrt{q^2}}\,\mathcal{I}m\bigg(H_{-}^{1}\left(-\tfrac{1}{2}, -\tfrac{1}{2}\right)\,H_{0}^{2\ast}\left(-\tfrac{1}{2}, +\tfrac{1}{2}\right)+H_{+}^{1}\left(+\tfrac{1}{2}, +\tfrac{1}{2}\right)\,H_{0}^{2\ast}\left(+\tfrac{1}{2}, -\tfrac{1}{2}\right)\bigg),
\\
        \mathcal{K}_{3cc\text{T}}&=\sqrt{2}\,|N_3|^2\,\alpha\, \beta_{\ell}\frac{ m_{\ell}}{\sqrt{q^2}}\,\mathcal{I}m\bigg(
        H_{0}^{1}\left(-\tfrac{1}{2}, +\tfrac{1}{2}\right)H_{-}^{2\ast}\left(-\tfrac{1}{2}, -\tfrac{1}{2}\right)+H_{0}^{1}\left(+\tfrac{1}{2}, -\tfrac{1}{2}\right)H_{+}^{2 \ast}\left(+\tfrac{1}{2}, +\tfrac{1}{2}\right)\bigg),
\\
        \mathcal{K}_{3c \text{T}}&=\sqrt{2}\,|N_3|^2\,\alpha\,\frac{ m_{\ell}}{\sqrt{q^2}}\,\mathcal{I}m\bigg(H^{1}_{-}\left(-\tfrac{1}{2}, -\tfrac{1}{2}\right)\,H^{1\ast}_{0}\left(-\tfrac{1}{2}, +\tfrac{1}{2}\right)
        -H^{1}_{+}\left(+\tfrac{1}{2}, +\tfrac{1}{2}\right)H^{1\ast}_{0}\left(+\tfrac{1}{2}, -\tfrac{1}{2}\right)
        \notag\\&\quad+H^{1}_{-}\left(-\tfrac{1}{2}, -\tfrac{1}{2}\right)\, H^{2\ast}_{t}\left(-\tfrac{1}{2}, +\tfrac{1}{2}\right)+ H^{1}_{+}\left(+\tfrac{1}{2}, +\tfrac{1}{2}\right)\, H^{2\ast}_{t}\left(+\tfrac{1}{2}, -\tfrac{1}{2}\right)\bigg),
\\
        \mathcal{K}_{3 \text{T}}&= - \sqrt{2}\,|N_3|^2\,\alpha\, \beta_{\ell}\frac{ m_{\ell}}{\sqrt{q^2}}\,\mathcal{I}m\bigg( H_{-}^{2}\left(-\tfrac{1}{2}, -\tfrac{1}{2}\right)\, H_{t}^{2\ast}\left(-\tfrac{1}{2}, +\tfrac{1}{2}\right)
        -H_{+}^{2}\left(+\tfrac{1}{2}, +\tfrac{1}{2}\right)\,
        H_{t}^{2 \ast}\left(+\tfrac{1}{2}, -\tfrac{1}{2}\right)
        \bigg),
\\
        \mathcal{K}_{4ss\text{T}}&=\sqrt{2}\,|N_3|^2\,\alpha\, \beta_{\ell}\frac{ m_{\ell}}{\sqrt{q^2}}\,\mathcal{R}e\bigg(H_{-}^{1}\left(-\tfrac{1}{2}, -\tfrac{1}{2}\right)\,H_{0}^{2\ast}\left(-\tfrac{1}{2}, +\tfrac{1}{2}\right)-H_{+}^{1}\left(+\tfrac{1}{2}, +\tfrac{1}{2}\right)\,H_{0}^{2\ast}\left(+\tfrac{1}{2}, -\tfrac{1}{2}\right)\bigg),
\\
        \mathcal{K}_{4cc\text{T}}&=-\sqrt{2}\,|N_3|^2\,\alpha\, \beta_{\ell}\frac{ m_{\ell}}{\sqrt{q^2}}\,\mathcal{R}e\bigg(H_{0}^{1}\left(-\tfrac{1}{2}, +\tfrac{1}{2}\right)\,H_{-}^{2\ast}\left(-\tfrac{1}{2}, -\tfrac{1}{2}\right)-H_{0}^{1}\left(+\tfrac{1}{2}, -\tfrac{1}{2}\right)H_{+}^{2\ast}\left(+\tfrac{1}{2}, +\tfrac{1}{2}\right)\bigg),
        \label{NewT2}
\\
        \mathcal{K}_{4c \text{T}}&=\sqrt{2}\,|N_3|^2\,\alpha\,\frac{ m_{\ell}}{\sqrt{q^2}}\,\mathcal{R}e\bigg(H^{1}_{-}\left(-\tfrac{1}{2}, -\tfrac{1}{2}\right)\,H^{1\ast}_{0}\left(-\tfrac{1}{2}, +\tfrac{1}{2}\right)
        +H^{1}_{+}\left(+\tfrac{1}{2}, +\tfrac{1}{2}\right)\,H^{1\ast}_{0}\left(+\tfrac{1}{2}, -\tfrac{1}{2}\right)
        \notag\\&\quad +H^{1}_{-}\left(-\tfrac{1}{2}, -\tfrac{1}{2}\right)\, H^{2\ast}_{t}\left(-\tfrac{1}{2}, +\tfrac{1}{2}\right)-H^{1}_{+}\left(+\tfrac{1}{2}, +\tfrac{1}{2}\right)\, H^{2\ast}_{t}\left(+\tfrac{1}{2}, -\tfrac{1}{2}\right)\bigg),
\\
        \mathcal{K}_{4 \text{T}}&=  - \sqrt{2}\,|N_3|^2\,\alpha\, \beta_{\ell}\frac{ m_{\ell}}{\sqrt{q^2}}\,\mathcal{R}e\bigg( H_{-}^{2}\left(-\tfrac{1}{2}, -\tfrac{1}{2}\right)\,H_{t}^{2\ast}\left(-\tfrac{1}{2}, +\tfrac{1}{2}\right)
        +H_{+}^{2}\left(+\tfrac{1}{2}, +\tfrac{1}{2}\right)\,
        H_{t}^{2 \ast}\left(+\tfrac{1}{2}, -\tfrac{1}{2}\right)
        \bigg).\label{ziTexpL}
\end{align}

\section{ Analytical expressions of the angular coefficients in terms of transversity amplitudes with unpolarized and polarized lepton}
\label{AppF}
The transversity amplitudes can be obtained from the helicity amplitudes given in Eqs. (\ref{Hamp002})--(\ref{Hamp006}). we first define the helicity amplitudes
\begin{eqnarray}
H_{\pm}^{L,\,R}\left(\pm \frac{1}{2}, \pm\frac{1}{2}\right) &=&\left[ H_{\pm}^{1}\left(\pm\frac{1}{2}, \pm\frac{1}{2}\right)\mp H_{\pm}^{2}\left(\pm\frac{1}{2}, \pm\frac{1}{2}\right)\right],\\
H_{0}^{L,\,R}\left(\pm \frac{1}{2}, \mp\frac{1}{2}\right) &=& \left[H_{0}^{1}\left(\pm\frac{1}{2}, \mp\frac{1}{2}\right)\mp H_{0}^{2}\left(\pm\frac{1}{2}, \mp\frac{1}{2}\right)\right],
\label{LRtransamp}    
\end{eqnarray}
where $L$ and $R$ correspond to left and right-handed chiralities of the dilepton system, respectively. As there are no $H_{t}^{1}\left(\pm \frac{1}{2}, \mp \frac{1}{2}\right)$ contributions, therefore we do not have separate left and right-handed parts for $H_{t}$. Hence, the two time-like transversity amplitudes are obtained as
\begin{eqnarray}
A_{\perp_t} = \sqrt{2} \bigg[ H^{2}_t\left(- \frac{1}{2}, +\frac{1}{2}\right) - H^{2}_t\left(+ \frac{1}{2}, -\frac{1}{2}\right) \bigg],\\
A_{\|_t} = \sqrt{2} \bigg[ H^{2}_t\left(- \frac{1}{2}, +\frac{1}{2}\right) + H^{2}_t\left(+ \frac{1}{2}, -\frac{1}{2}\right) \bigg].
\end{eqnarray}
The other transversity amplitudes are obtained as
\begin{align}\label{eq:transampfll}
& A_{\perp_1}^{L(R)} = \frac{1}{\sqrt{2}} \bigg[H^{L(R)}_+\left(+ \frac{1}{2}, +\frac{1}{2}\right) - H^{L(R)}_-\left(- \frac{1}{2}, -\frac{1}{2}\right) \bigg],\\
& A_{\|_1}^{L(R)} = \frac{1}{\sqrt{2}} \bigg[H^{L(R)}_+\left(+ \frac{1}{2}, +\frac{1}{2}\right) + H^{L(R)}_-\left(- \frac{1}{2}, -\frac{1}{2}\right) \bigg],\\
& A_{\perp_0}^{L(R)} = \frac{1}{\sqrt{2}}\bigg[H^{L(R)}_0\left(- \frac{1}{2}, +\frac{1}{2}\right) - H^{L(R)}_0\left(+ \frac{1}{2}, -\frac{1}{2}\right)  \bigg],\\
& A_{\|_0}^{L(R)} = \frac{1}{\sqrt{2}}\bigg[H^{L(R)}_0\left(- \frac{1}{2}, +\frac{1}{2}\right) + H^{L(R)}_0\left(+ \frac{1}{2}, -\frac{1}{2}\right)  \bigg].
\end{align} 

The expressions of the unpolarized angular coefficients, in terms of the transversity amplitudes are obtained as
\begin{align}
        K_{1ss}\ &=|N_3|^2\Bigg[\frac{m_{\ell}^2}{q^2}\Bigg\{ -\bigg(\big|A_{ \parallel 0}^{L}\big|^2+\big|A_{ \perp 0}^{L}\big|^2+\{L\leftrightarrow R\}\bigg)+2\mathcal{R}e\bigg(A_{ \parallel 0}^{L}A_{ \parallel 0}^{\ast R}+A_{\parallel 1}^{L}A_{\parallel 1}^{\ast R}+\{\parallel \leftrightarrow \perp \}\bigg)\notag \\ & +\bigg(\big|A_{\parallel t}\big|^2+\{\parallel \leftrightarrow \perp \}\bigg)\Bigg\} +\frac{1}{4}\bigg(2\big|A_{ \parallel 0}^{L}\big|^2+2\big|A_{ \perp 0}^{L}\big|^2+\big|A_{\parallel 1}^{L}\big|^2+\big|A_{ \perp 1}^{L}\big|^2+\{L\leftrightarrow R\}\bigg)\Bigg],
        \label{TU1}
\\ %
        K_{1cc}&=|N_3|^2\Bigg[\frac{m_{\ell}^2}{q^2}\Bigg\{\bigg(\big|A_{ \parallel 0}^{L}\big|^2+\big|A_{ \perp 0}^{L}\big|^2-\big|A_{\parallel 1}^{L}\big|^2-\big|A_{\perp 1}^{L}\big|^2+\{L\leftrightarrow R\}\bigg)+\bigg(\big|A_{\parallel t}\big|^2+\{\parallel \leftrightarrow \perp \}\bigg)\notag \\ & 
        +2\mathcal{R}e\bigg(A_{ \parallel 0}^{L}A_{ \parallel 0}^{\ast R}+A_{\parallel 1}^{L}A_{\parallel 1}^{\ast R}+\{\parallel \leftrightarrow \perp \}\bigg)\Bigg\}+\frac{1}{2}\bigg(\big|A_{\parallel 1}^{L}\big|^2+\big|A_{ \perp 1}^{L}\big|^2+\{L\leftrightarrow R\}\bigg)\Bigg],
\\ %
        K_{1c } &=-|N_3|^2\beta_\ell\mathcal{R}e\bigg(A_{\parallel 1}^{L}A_{ \perp 1}^{\ast L}-\{L\leftrightarrow R\}\bigg),
        \label{TUk1c}
\\ %
        K_{2ss } &=|N_3|^2\,\alpha\Bigg[\frac{2m_{\ell}^2}{q^2}\mathcal{R}e\bigg(-A_{ \parallel 0}^{L}A_{ \perp 0}^{\ast L}+A_{ \parallel 0}^{L}A_{ \perp 0}^{\ast R}+A_{\parallel 1}^{L}A_{ \perp 1}^{\ast R}+ \{L \leftrightarrow R \}+A_{\parallel t}A^\ast_{\perp t}\bigg)\notag \\ & +\frac{1}{2}\,\mathcal{R}e\bigg(2A_{ \parallel 0}^{L}A_{ \perp 0}^{\ast L}+A_{\parallel 1}^{L}A_{ \perp 1}^{\ast L}+\{L\leftrightarrow R\}\bigg)\Bigg],
\\ %
        K_{2cc}&=|N_3|^2\,\alpha\Bigg[\frac{2m_{\ell}^2}{q^2}\mathcal{R}e\bigg(A_{ \parallel 0}^{L}A_{ \perp 0}^{\ast L}+A_{ \parallel 0}^{L}A_{ \perp 0}^{\ast R}+A_{\parallel 1}^{L}A_{ \perp 1}^{\ast R}-A_{\parallel 1}^{L}A_{ \perp 1}^{\ast L}+ \{L \leftrightarrow R \}+A_{\parallel t}A^\ast_{\perp t}\bigg)\notag \\ & +\mathcal{R}e\bigg(A_{\parallel 1}^{L}A_{ \perp 1}^{\ast L}+\{L\leftrightarrow R\}\bigg)\Bigg],
\\ %
        K_{2c } &=-\frac{1}{2}|N_3|^2\alpha\beta_\ell\bigg(\big|A_{\parallel 1}^{L}\big|^2+\big|A_{ \perp 1}^{L}\big|^2-\{L\leftrightarrow R\}\bigg),
        \label{TUk2c}
\\
        K_{3sc } &=\frac{1}{\sqrt{2}}|N_3|^2\,\alpha\Bigg[\frac{-4m_{\ell}^2}{q^2}\mathcal{I}m\bigg(A_{\parallel 1}^{L}A_{ \parallel 0}^{\ast L}-A_{\perp 1}^{L}A_{\perp 0}^{\ast L}+\{L\leftrightarrow R\}\bigg)\notag \\ & + \mathcal{I}m\Big(A_{\parallel 1}^{L}A_{ \parallel 0}^{\ast L}-A_{\perp 1}^{L}A_{\perp 0}^{\ast L}+\{L\leftrightarrow R\}\Big)\Bigg],
\\ %
        K_{3s}&=\frac{1}{\sqrt{2}} |N_3|^2\alpha\,\beta_\ell\,\mathcal{I}m\bigg(-A_{\perp 1}^{L}A_{ \parallel 0}^{\ast L}+A_{ \parallel 1}^{L}A_{\perp 0}^{\ast L}-\{L\leftrightarrow R\}\bigg),    
\\ %
        K_{4sc} &=\frac{1}{\sqrt{2}}|N_3|^2\,\alpha\Bigg[\frac{4m_{\ell}^2}{q^2}\mathcal{R}e\bigg(A_{ \perp 1}^{L}A_{\parallel 0}^{\ast L}-A_{ \parallel 1}^{L}A_{\perp 0}^{\ast L}+\{L\leftrightarrow R\}\bigg)\notag \\ & +\mathcal{R}e\bigg(-A_{\perp 1}^{L}A_{\parallel 0}^{\ast L}+A_{\parallel 1}^{L}A_{\perp 0}^{\ast L}+\{L\leftrightarrow R\}\bigg)\Bigg],
        \label{TU2}
\\ %
        K_{4s }&=\frac{1}{\sqrt{2}}|N_3|^2\alpha\,\beta_\ell\, \mathcal{R}e\bigg(A_{ \parallel 1}^{L}A_{\parallel 0}^{\ast L}-A_{ \perp 1}^{L}A_{\perp 0}^{\ast L}
        -\{L\leftrightarrow R\}\bigg).
        \label{TUk4s}    
\end{align}

The expressions of $\xi_{L}$ dependent angular coefficients present in the L polarized four-fold angular decay distribution are given, in terms of the transversity amplitudes, as
\begin{align}
        \mathcal{K}_{1ss \text{L}}^\prime\ &=-\frac{1}{8}|N_3|^2\beta_\ell\bigg(2\big|A_{ \parallel 0}^{L}\big|^2+2\big|A_{ \perp 0}^{L}\big|^2+\big|A_{\parallel 1}^{L}\big|^2+\big|A_{ \perp 1}^{L}\big|^2-\{L\leftrightarrow R\}\bigg),
        \label{TL1}
\\
        \mathcal{K}_{1cc \text{L}}^\prime\ &=-\frac{1}{4}|N_3|^2\beta_\ell\bigg(\big|A_{\parallel 1}^{L}\big|^2+\big|A_{ \perp 1}^{L}\big|^2-\{L\leftrightarrow R\}\bigg),
\\
        \mathcal{K}_{1c \text{L}}^\prime\ &=|N_3|^2\Bigg[\frac{m^2_\ell}{q^2}\mathcal{R}e\bigg(A_{\parallel t}A_{ \parallel 0}^{\ast L}+A_{\perp t}A_{ \perp 0}^{\ast L}-A_{\parallel 1}^{L}A_{ \perp 1}^{\ast L}+A_{\parallel 1}^{L}A_{ \perp 1}^{\ast R}+\{L\leftrightarrow R\}\bigg)+\frac{1}{2}\mathcal{R}e\bigg(A_{\parallel 1}^{L}A_{ \perp 1}^{\ast L}+\{L\leftrightarrow R\}\bigg)\Bigg],
        \label{TLk1cp}
\\
        \mathcal{K}_{2ss \text{L}}^\prime\ &=-\frac{1}{4}|N_3|^2\alpha\, \beta_\ell\,\mathcal{R}e\bigg(2A_{ \parallel 0}^{L}A_{ \perp 0}^{\ast L}+A_{\parallel 1}^{L}A_{ \perp 1}^{\ast L}-\{L\leftrightarrow R\}\bigg),
\\
        \mathcal{K}_{2cc \text{L}}^\prime\ &=-\frac{1}{2}|N_3|^2\alpha\, \beta_\ell\,\mathcal{R}e\bigg(A_{\parallel 1}^{L} A_{ \perp 1}^{\ast L}-\{L\leftrightarrow R\}\bigg),
\\
        \mathcal{K}_{2c \text{L}}^\prime\ &=|N_3|^2\alpha\Bigg[\frac{m^2_\ell}{q^2}\Bigg\{\mathcal{R}e\bigg(A_{\parallel 1}^{L}A_{\parallel 1}^{\ast R}+A_{ \perp 1}^{L}A_{ \perp 1}^{\ast R}+\Big(
        A_{\perp t}A_{ \parallel 0}^{\ast L}+A_{\parallel t}A_{ \perp 0}^{\ast L}+\{L\leftrightarrow R\}\Big)\bigg)\notag \\ &-\frac{1}{2}\bigg(\big|A_{\parallel 1}^{L}\big|^2+\big|A_{ \perp 1}^{L}\big|^2+\{L\leftrightarrow R\}\bigg)\Bigg\}+\frac{1}{4}\bigg(\big|A_{\parallel 1}^{L}\big|^2+\big|A_{ \perp 1}^{L}\big|^2+\{L\leftrightarrow R\}\bigg)\Bigg],
        \label{TLk2cp}     
\\
        \mathcal{K}_{3sc \text{L}}^\prime\ &=\frac{1}{2\sqrt{2}}|N_3|^2\alpha\,\beta_\ell\,\mathcal{I}m\bigg(-A_{\parallel 1}^{L}A_{ \parallel 0}^{\ast L}+A_{ \perp 1}^{L}A_{\perp 0}^{\ast L}-\{L\leftrightarrow R\}\bigg),
\\
        \mathcal{K}_{3s \text{L}}^\prime\ &=\frac{1}{\sqrt{2}}|N_3|^2\alpha\Bigg[\frac{m^2_\ell}{q^2}\,\mathcal{I}m\bigg(-A_{ \perp 1}^{L}A_{ \parallel 0}^{\ast L}+A_{\parallel 1}^{L}A_{ \perp 0}^{\ast L}+A_{\parallel t}A_{\parallel 1}^{\ast L}-A_{\perp t}A_{\perp 1}^{\ast L}+A_{ \perp 1}^{R}A_{ \parallel 0}^{\ast L}-A_{\parallel 1}^{R}A_{ \perp 0}^{\ast L}+\{L\leftrightarrow R\}\bigg)\notag \\ &+\frac{1}{2}\mathcal{I}m\bigg(A_{\perp 1}^{L}A_{ \parallel 0}^{\ast L}-A_{ \parallel 1}^{L}A_{\perp 0}^{\ast L}+\{L\leftrightarrow R\}\bigg)\Bigg],    
\\
        \mathcal{K}_{4sc \text{L}}^\prime\ &=\frac{1}{2\sqrt{2}}|N_3|^2\alpha\,\beta_\ell\,\mathcal{R}e\bigg(A_{ \perp 1}^{L}A_{ \parallel 0}^{\ast L}-A_{\parallel 1}^{L}A_{\perp 0}^{\ast L}-\{L\leftrightarrow R\}\bigg),
        \label{TL2}
\\
        \mathcal{K}_{4s \text{L}}^\prime\ &=\frac{1}{\sqrt{2}}|N_3|^2\alpha\Bigg[\frac{m^2_\ell}{q^2}\,\mathcal{R}e\bigg(A_{ \parallel 0}^{L}A_{\parallel 1}^{\ast L}-A_{\perp 0}^{L}A_{ \perp 1}^{\ast L}
        -A_{\perp t}A_{\parallel 1}^{\ast L}+A_{\parallel t}A_{\perp 1}^{\ast L}-A_{\parallel 0}^{L}A_{\parallel 1}^{\ast R}+A_{\perp 0}^{L}A_{ \perp 1}^{\ast R}+\{L\leftrightarrow R\}\bigg)\notag \\ &+\frac{1}{2}\mathcal{R}e\bigg(-A_{ \parallel 1}^{L}A_{\parallel 0}^{\ast L}+A_{ \perp 1}^{L}A_{\perp 0}^{\ast L}
        +\{L\leftrightarrow R\}\bigg)\Bigg].
        \label{TLk4sp}    
\end{align}

Similarly, $\xi_{N}$ dependent angular coefficients, in terms of the transversity amplitudes, take the form
\begin{align}
        \mathcal{K}_{1s \text{N}}&=-\frac{1}{2}|N_3|^2 \,\beta_{\ell} \frac{m_{\ell}}{\sqrt{q^2}}\,\mathcal{I}m\bigg(A_{\parallel t}A_{\parallel0}^{\ast L}+A_{\perp t}A_{\perp 0}^{\ast L}-A_{\parallel1}^{L}A_{\perp 1}^{\ast R}-\{L\leftrightarrow R\}\bigg),
\\
        \mathcal{K}_{2s \text{N}}&=\frac{1}{2}|N_3|^2\,\alpha\, \beta_{\ell}\frac{ m_{\ell}}{\sqrt{q^2}}\,\mathcal{I}m\bigg(A_{\parallel 1}^{L}A_{\parallel 1}^{\ast R}+A_{\perp 1}^{L}A_{\perp 1}^{\ast R}-\Big(A_{\parallel t}A_{\perp 0}^{\ast L}+A_{\perp t}A_{\parallel 0}^{\ast L}-\{L\leftrightarrow R\}\Big)\bigg),
\\  
        \mathcal{K}_{3c \text{N}}&=\frac{1}{2\sqrt{2}}|N_3|^2\,\alpha\,\beta_{\ell}\,\frac{m_{\ell}}{\sqrt{q^2}}\,\mathcal{R}e\bigg(A_{\perp 1}^{ L}A_{\parallel0}^{\ast L}-A_{\parallel 1}^{L}A_{\perp 0}^{\ast L}-A_{\parallel t}A_{\parallel1}^{\ast L}+A_{\perp t}A_{\perp 1}^{\ast L}-A_{\parallel0}^{L}A_{\perp 1}^{\ast R}+A_{\perp0}^{L}A_{\parallel 1}^{\ast R}-\{L\leftrightarrow R\}\bigg),
\\
        \mathcal{K}_{3 \text{N}}&=-\frac{1}{2\sqrt{2}}|N_3|^2\,\alpha\, \frac{ m_{\ell}}{\sqrt{q^2}}\,\mathcal{R}e\bigg(A_{\parallel 1}^{L}A_{\parallel 0}^{\ast L}-A_{\perp 1}^{ L}A_{\perp 0}^{\ast L}-A_{\parallel t}A_{\perp 1}^{\ast L}+A_{\perp t}A_{\parallel 1}^{\ast L}+A_{\parallel 0}^{L}A_{\parallel 1}^{\ast R}-A_{\perp 0}^{L}A_{\perp 1}^{\ast R}+\{L\leftrightarrow R\}\bigg),
\\
        \mathcal{K}_{4c \text{N}}&=\frac{1}{2\sqrt{2}}|N_3|^2\,\alpha\, \beta_{\ell}\frac{ m_{\ell}}{\sqrt{q^2}}\,\mathcal{I}m\bigg(A_{\parallel 1}^{L}A_{\parallel 0}^{\ast L}-A_{\perp 1}^{L}A_{\perp 0}^{\ast L}-A_{\perp t}A_{\parallel 1}^{\ast L}+A_{\parallel t}A_{\perp 1}^{\ast L}+A_{\parallel 0}^{ L}A_{\parallel 1}^{\ast R}-A_{\perp 0}^{L}A_{\perp 1}^{\ast R}-\{L\leftrightarrow R\}\bigg),
\\
        \mathcal{K}_{4 \text{N}}&= -\frac{1}{2\sqrt{2}}|N_3|^2\,\alpha\,\frac{ m_{\ell}}{\sqrt{q^2}}\mathcal{I}m\bigg(A_{\perp 1}^{L}A_{\parallel 0}^{\ast L}-A_{\parallel 1}^{L}A_{\perp 0}^{\ast L}+A_{\parallel t}A_{\parallel 1}^{\ast L}-A_{\perp t}A_{\perp 1}^{\ast L}-A_{\parallel 0}^{ L}A_{\perp 1}^{\ast R}+A_{\perp 0}^{L}A_{\parallel 1}^{\ast R}+\{L\leftrightarrow R\}\bigg).
\end{align}

Also, $\xi_{T}$ dependent angular coefficients, in terms of the transversity amplitudes, are obtained as
\begin{align}
        \mathcal{K}_{1sc \text{T}}&=\frac{1}{4}|N_3|^2\,\beta_{\ell}\,\frac{m_{\ell}}{\sqrt{q^2}}\,
        \Bigg(2\big|A_{\parallel 0}^{L}\big|^2+2\big|A_{\perp 0}^{L}\big|^2-\big|A_{\parallel 1}^{L}\big|^2-\big|A_{\perp 1}^{L}\big|^2-\{L\leftrightarrow R\}\bigg),
\\
        \mathcal{K}_{1s \text{T}}&=\frac{1}{2}|N_3|^2\,\frac{m_{\ell}}{\sqrt{q^2}}\,\mathcal{R}e\bigg(A_{\parallel t}A_{\parallel 0}^{\ast L}+A_{\perp t}A_{\perp 0}^{\ast L}+A_{\perp 1}^{ L}A_{\parallel 1}^{\ast L}+A_{\parallel 1}^{L}A_{\perp 1}^{\ast R}+\{L\leftrightarrow R\}\bigg),
\\  
        \mathcal{K}_{2sc \text{T}}&=\frac{1}{2}|N_3|^2\,\alpha\, \beta_{\ell}\frac{ m_{\ell}}{\sqrt{q^2}}\,\mathcal{R}e\bigg(2A_{\parallel 0}^{L}A_{\perp 0}^{\ast L}-A_{\parallel 1}^{L}A_{\perp 1}^{\ast L}-\{L\leftrightarrow R\}\bigg),
\\
        \mathcal{K}_{2s \text{T}}&=\frac{1}{2}|N_3|^2\,\alpha\, \frac{ m_{\ell}}{\sqrt{q^2}}\,\Bigg[\mathcal{R}e\bigg(A_{\parallel 1}^{L}A_{\parallel 1}^{\ast R}+A_{\perp 1}^{L}A_{\perp 1}^{\ast R}+\Big(A_{\parallel t}A_{\perp 0}^{\ast L}+A_{\perp t}A_{\parallel 0}^{\ast L}+\{L\leftrightarrow R\}\Big)\bigg)\notag \\ &+\frac{1}{2}\bigg(\big|A_{\parallel 1}^{L}\big|^2+\big|A_{\perp 1}^{L}\big|^2+\{L\leftrightarrow R\}\bigg)\Bigg],
\\
        \mathcal{K}_{3ss \text{T}}&= \frac{1}{2\sqrt{2}}|N_3|^2\,\alpha\, \beta_{\ell}\frac{ m_{\ell}}{\sqrt{q^2}}\,\mathcal{I}m\bigg(-A_{\parallel1}^{L}A_{\parallel 0}^{\ast L}+A_{\perp 1}^{L}A_{\perp 0}^{\ast L}+A_{\parallel 0}^{L}A_{\parallel1}^{\ast R}-A_{\perp 0}^{L}A_{\perp 1}^{\ast R}-\{L\leftrightarrow R\}\bigg),
\\
        \mathcal{K}_{3cc \text{T}}&= \frac{1}{2\sqrt{2}}|N_3|^2\,\alpha\, \beta_{\ell}\frac{ m_{\ell}}{\sqrt{q^2}}\,
        \mathcal{I}m\bigg(A_{\parallel 1}^{L}A_{\parallel 0}^{\ast L}-A_{\perp 1}^{L}A_{\perp 0}^{\ast L}+A_{\parallel 0}^{ L}A_{\parallel 1}^{\ast R}-A_{\perp 0}^{L}A_{\perp 1}^{\ast R}-\{L\leftrightarrow R\}\bigg),
\\  
        \mathcal{K}_{3c \text{T}}&=-\frac{1}{2\sqrt{2}}|N_3|^2\,\alpha\,\frac{ m_{\ell}}{\sqrt{q^2}}\,
        \mathcal{I}m\bigg(A_{\perp 1}^{L}A_{\parallel 0}^{\ast L}-A_{\parallel 1}^{L}A_{\perp 0}^{\ast L}+A_{\parallel t}A_{\parallel 1}^{\ast L}-A_{\perp t}A_{\perp 1}^{\ast L}-A_{\parallel 0}^{L}A_{\perp 1}^{\ast R}+A_{\perp 0}^{L}A_{\parallel 1}^{\ast R}+\{L\leftrightarrow R\}\bigg),
 \\
        \mathcal{K}_{3 \text{T}}&=\frac{1}{2\sqrt{2}}|N_3|^2\,\alpha\, \beta_{\ell}\frac{ m_{\ell}}{\sqrt{q^2}}\,
        \mathcal{I}m\bigg(A_{\parallel t}A_{\perp 1}^{\ast L}-A_{\perp t}A_{\parallel 1}^{\ast L}-\{L\leftrightarrow R\}\bigg),
\\
        \mathcal{K}_{4ss \text{T}}&= -\frac{1}{2\sqrt{2}}|N_3|^2\,\alpha\, \beta_{\ell}\frac{ m_{\ell}}{\sqrt{q^2}}\,
        \mathcal{R}e\bigg(A_{\parallel 1}^{L}A_{\perp 0}^{\ast L}-A_{\perp 1}^{L}A_{\parallel0}^{\ast L}-A_{\parallel0}^{L}A_{\perp 1}^{\ast R}+A_{\perp 0}^{L}A_{\parallel 1}^{\ast R}-\{L\leftrightarrow R\}\bigg),
\\
        \mathcal{K}_{4cc \text{T}}&= \frac{1}{2\sqrt{2}}|N_3|^2\,\alpha\, \beta_{\ell}\frac{ m_{\ell}}{\sqrt{q^2}}\,
        \mathcal{R}e\bigg(-A_{\perp 1}^{L}A_{\parallel 0}^{\ast L}+
        A_{\parallel 1}^{L}A_{\perp 0}^{\ast L}+A_{\parallel 0}^{L}A_{\perp 1}^{\ast R}-A_{\perp 0}^{L}A_{\parallel 1}^{\ast R}-\{L\leftrightarrow R\}\bigg),      
\\
        \mathcal{K}_{4c \text{T}}&=\frac{1}{2\sqrt{2}}|N_3|^2\,\alpha\,\frac{ m_{\ell}}{\sqrt{q^2}}\,
        \mathcal{R}e\bigg(A_{\parallel 1}^{L}A_{\parallel0}^{\ast L}-A_{\perp 1}^{L}A_{\perp 0}^{\ast L}-A_{\parallel t}A_{\perp 1}^{\ast L}+A_{\perp t}A_{\parallel 1}^{\ast L}
        +A_{\parallel 0}^{L}A_{\parallel 1}^{\ast R}-A_{\perp 0}^{L}A_{\perp 1}^{\ast R}+\{L\leftrightarrow R\}\bigg),
\\
        \mathcal{K}_{4 \text{T}}&= \frac{1}{2\sqrt{2}}|N_3|^2\,\alpha\, \beta_{\ell}\frac{ m_{\ell}}{\sqrt{q^2}}\,
        \mathcal{R}e\bigg(A_{\parallel t}A_{\parallel1}^{\ast L}-A_{\perp t}A_{\perp 1}^{\ast L}-\{L\leftrightarrow R\}\bigg).
        \label{T-expressions}
\end{align}

\section{Binned predictions of physical observables}
\label{AppG}
In this appendix, we give the SM and NP binned predictions for the different physical observables in $q^2=[0.045-8.0]$ GeV$^2$, and $q^2=[15.0-20.27]$ GeV$^2$ bins.
\begin{table}[ht!]
\centering
\caption{The SM binned predictions in $q^2=[0.045-8.0]$ GeV$^2$, and $q^2=[15.0-20.27]$ GeV$^2$ bins, for the observables shown in Figs.~\ref{Longitudinal} and \ref{Trans}. $\left\langle d\mathcal{B}/dq^2\right\rangle_{p\pi^-}=\left\langle d\mathcal{B}\left(\Lambda_b \rightarrow \Lambda(\rightarrow  p \pi^-)\mu^{+}\mu^{-}\right)/dq^2\right\rangle$, and $\left\langle d\mathcal{B}/dq^2\right\rangle_{n\pi^0}=\left\langle d\mathcal{B}\left(\Lambda_b \rightarrow \Lambda(\rightarrow  n \pi^0)\mu^{+}\mu^{-}\right)/dq^2\right\rangle$. The differential branching ratios are given in units of $10^{-7}$ GeV$^{-2}$. The listed errors arise mainly due to the uncertainties in the form factors.}\label{Bin1-analysis001}
\renewcommand{\arraystretch}{1.3}
\resizebox{\textwidth}{!}{%
\begin{tabular}{|c| c c c c c|}
\hline\hline  
&$\left\langle{d\mathcal{B}}/{dq^2}\right\rangle_{p\pi^-}$ & 
$\left\langle\frac{d\mathcal{B}_{\,\text{L}}\left(\vec{s}_{\ell^-}=+\mathbf{\hat{e}}_L\right)}{dq^{2}}\right\rangle_{p\pi^-}$ & $\left\langle\frac{d\mathcal{B}_{\,\text{L}}\left(\vec{s}_{\ell^-}=-\mathbf{\hat{e}}_L\right)}{dq^{2}}\right\rangle_{p\pi^-}$ & $\left\langle\frac{d\mathcal{B}_{\,\text{T}}\left(\vec{s}_{\ell^-}=+\mathbf{\hat{e}}_T\right)}{dq^{2}}\right\rangle_{p\pi^-}$  & $\left\langle\frac{d\mathcal{B}_{\,\text{T}}\left(\vec{s}_{\ell^-}=-\mathbf{\hat{e}}_T\right)}{dq^{2}}\right\rangle_{p\pi^-}$
\\[1.5ex] 
\hline \hline 
$[0.045-8.0]$  &$0.1350\pm0.0812$&$0.0052\pm0.0006$&$0.1298\pm0.0806$&$0.0605\pm0.0406$&$0.0745\pm0.0406$  \\
$[15.0-20.27]$   &$0.5475\pm0.0425$&$0.0014\pm0.0004$&$0.5461\pm0.0425$&$0.2617\pm0.0317$&$0.2858\pm0.0317$ \\ \hline

&$\left\langle{d\mathcal{B}}/{dq^2}\right\rangle_{n \pi^0}$ & 
$\left\langle\frac{d\mathcal{B}_{\,\text{L}}\left(\vec{s}_{\ell^-}=+\mathbf{\hat{e}}_L\right)}{dq^{2}}\right\rangle_{n \pi^0}$ & $\left\langle\frac{d\mathcal{B}_{\,\text{L}}\left(\vec{s}_{\ell^-}=-\mathbf{\hat{e}}_L\right)}{dq^{2}}\right\rangle_{n \pi^0}$ & $\left\langle\frac{d\mathcal{B}_{\,\text{T}}\left(\vec{s}_{\ell^-}=+\mathbf{\hat{e}}_T\right)}{dq^{2}}\right\rangle_{n \pi^0}$  & $\left\langle\frac{d\mathcal{B}_{\,\text{T}}\left(\vec{s}_{\ell^-}=-\mathbf{\hat{e}}_T\right)}{dq^{2}}\right\rangle_{n \pi^0}$ \\[1.5ex] 
\hline \hline
$[0.045-8.0]$  &$0.0693\pm0.0341$&$0.0003\pm0.0001$&$0.0690\pm0.0340$&$0.0344\pm0.0170$&$0.0349\pm0.0170$  \\
$[15.0-20.27]$   &$0.2973\pm0.0277$&$0.0002\pm0.0001$&$0.2971\pm0.0277$&$0.1481\pm0.0105$&$0.1491\pm0.0105$ \\ \hline

&$\mathcal{A}^{\ell}_{\text{FB}}$ & $\mathcal{A}_{\text{FBL}}^{\ell}\left(\vec{s}_{\ell^-}=+\mathbf{\hat{e}}_L\right)$ & $\mathcal{A}_{\text{FBL}}^{\ell}\left(\vec{s}_{\ell^-}=-\mathbf{\hat{e}}_L\right)$ & $\mathcal{A}_{\text{FBT}}^{\ell}\left(\vec{s}_{\ell^-}=+\mathbf{\hat{e}}_T\right)$ & $\mathcal{A}_{\text{FBT}}^{\ell}\left(\vec{s}_{\ell^-}=-\mathbf{\hat{e}}_T\right)$ \\
\hline \hline
$[0.045-8.0]$&$0.0283\pm0.0411$&$-0.0951\pm0.0271$&$0.1234\pm0.0239$&$0.0351\pm0.0202$&$-0.0068\pm0.0215$ \\
$[15.0-20.27]$   &$0.3443\pm0.0351$&$-0.0036\pm0.0004$&$0.3480\pm0.0351$&$0.1736\pm0.0143$ &$0.1707\pm0.0143$  \\ \hline 
&$\mathcal{A}^{\ell \Lambda}_{\text{FB}}$ & $\mathcal{A}_{\text{FBL}}^{\ell\Lambda}\left(\vec{s}_{\ell^-}=+\mathbf{\hat{e}}_L\right)$ & $\mathcal{A}_{\text{FBL}}^{\ell\Lambda}\left(\vec{s}_{\ell^-}=-\mathbf{\hat{e}}_L\right)$ & $\mathcal{A}_{\text{FBT}}^{\ell\Lambda}\left(\vec{s}_{\ell^-}=+\mathbf{\hat{e}}_T\right)$ & $\mathcal{A}_{\text{FBT}}^{\ell\Lambda}\left(\vec{s}_{\ell^-}=-\mathbf{\hat{e}}_T\right)$ \\
\hline \hline
$[0.045-8.0]$ &$-0.0097\pm0.0054$&$0.0306\pm0.0049$&$-0.0402\pm0.0051$&$-0.0113\pm0.0027$& $0.0016\pm0.0026$\\
$[15.0-20.27]$   &$-0.1396\pm0.0010$&$0.0013\pm0.0001$&$-0.1410\pm0.0008$&$-0.0704\pm$0.0008& $-0.0693\pm0.0114$\\ \hline 
&$\mathcal{A}^{\Lambda}_{\text{FB}}$ & $\mathcal{A}_{\text{FBL}}^{\Lambda}\left(\vec{s}_{\ell^-}=+\mathbf{\hat{e}}_L\right)$ & $\mathcal{A}_{\text{FBL}}^{\Lambda}\left(\vec{s}_{\ell^-}=-\mathbf{\hat{e}}_L\right)$ & $\mathcal{A}_{\text{FBT}}^{\Lambda}\left(\vec{s}_{\ell^-}=+\mathbf{\hat{e}}_T\right)$ & $\mathcal{A}_{\text{FBT}}^{\Lambda}\left(\vec{s}_{\ell^-}=-\mathbf{\hat{e}}_T\right)$ \\
\hline \hline
$[0.045-8.0]$&$-0.3119\pm0.0124$&$-0.0443\pm0.0021$& $-0.2676\pm0.0215$&$-0.1276\pm0.0025$&$-0.1843\pm0.0026$\\
$[15.0-20.27]$  & $-0.2675\pm0.0267$&$-0.0022\pm0.0002$&$-0.2654\pm0.0267$&$-0.1272\pm0.0187$&$-0.1404\pm0.0191$  \\ \hline \hline
\end{tabular}}
\end{table}

\begin{table}[ht!]
\centering
\caption{The SM binned predictions in $q^2=[0.045-8.0]$ GeV$^2$, and $q^2=[15.0-20.27]$ GeV$^2$ bins, for the normalized angular coefficients shown in FIG.~\ref{Longitudinal}. The listed errors arise mainly due to the uncertainties in the form factors.}\label{Bin1-analysis002}
\renewcommand{\arraystretch}{1.3}
\resizebox{\textwidth}{!}{%
\begin{tabular}{|c| c c c c c c|}
\hline\hline  
&$\hat{K}_{\text{2cc}}$ & 
$\hat{\mathcal{K}}_{2cc\text{L}}\left(\vec{s}_{\ell^-}=+\mathbf{\hat{e}}_L\right)$ & $\hat{\mathcal{K}}_{2cc\text{L}}\left(\vec{s}_{\ell^-}=-\mathbf{\hat{e}}_L\right)$ & 
$\hat{K}_{\text{2ss}}$ & 
$\hat{\mathcal{K}}_{2ss\text{L}}\left(\vec{s}_{\ell^-}=+\mathbf{\hat{e}}_L\right)$ & $\hat{\mathcal{K}}_{2ss\text{L}}\left(\vec{s}_{\ell^-}=-\mathbf{\hat{e}}_L\right)$\\
\hline \hline
$[0.045-8.0]$ &$-0.0953\pm0.0215$&$-0.0397\pm0.0211$&$-0.0556\pm0.0230$&$-0.2646\pm0.0159$&$-0.0239\pm0.0019$ &$-0.2407\pm0.0421$ \\
$[15.0-20.27]$ &$-0.1378\pm0.0012$&$-0.0013\pm0.0001$&$-0.1364\pm0.0010$&$-0.1726\pm0.0013$&$-0.0012\pm0.0001$&$-0.1714\pm0.0013$ \\ \hline 

&$\hat{K}_{\text{4s}}$ & $\hat{\mathcal{K}}_{4s\text{L}}\left(\vec{s}_{\ell^-}=+\mathbf{\hat{e}}_L\right)$ & $\hat{\mathcal{K}}_{4s\text{L}}\left(\vec{s}_{\ell^-}=-\mathbf{\hat{e}}_L\right)$ & 
$\hat{K}_{\text{4sc}}$ & $\hat{\mathcal{K}}_{4sc\text{L}}\left(\vec{s}_{\ell^-}=+\mathbf{\hat{e}}_L\right)$ & $\hat{\mathcal{K}}_{4sc\text{L}}\left(\vec{s}_{\ell^-}=-\mathbf{\hat{e}}_L\right)$ \\
\hline \hline
$[0.045-8.0]$& $-0.0209\pm0.0321$&$-0.0010\pm0.0001$&$-0.0199\pm0.0321$&$0.0056\pm0.0344$&$0.0006\pm0.0001$& $0.0050\pm0.0345$\\
$[15.0-20.27]$   &$0.1049\pm0.0231$&$-0.0008\pm0.0001$&$0.1058\pm0.0231$&$-0.0197\pm0.0121$&$0.0001\pm0.0001$&$-0.0198\pm0.0121$  \\ \hline \hline
\end{tabular}}
\end{table}

\begin{table}[ht!]
\centering
\caption{SM and NP binned predictions in $q^2=[0.045-8.0]$ GeV$^2$, and $q^2=[15.0-20.27]$ GeV$^2$ bins, for the asymmetry observables in the case of longitudinally polarized lepton, as shown in FIG.~\ref{PrimeK}. For each observable, the first and the second row predictions correspond to $q^2=[0.045-8.0]$ GeV$^2$, and $q^2=[15.0-20.27]$ GeV$^2$ bin, respectively. The listed errors arise mainly due to the uncertainties in the form factors.}\label{Bin1-analysis003}
\renewcommand{\arraystretch}{1.5}
\resizebox{\textwidth}{!}{%
\begin{tabular}{|M{1.8cm}|M{2.9cm}M{2.9cm}M{2.9cm}M{2.9cm}M{2.9cm}|}
 \hline\hline
Observables & SM & S1 & S2 & S3 & S4  \\
 \hline\hline

 \multirow{2}{2.5em}{$\left\langle P_L\right\rangle$} & $-0.7262\pm 0.1220$& $-0.3833\pm0.1692$ & $-0.4135\pm0.1846$ & $-0.4673\pm0.1806 $&$-0.5933\pm0.1572$ \\
 
 &$-0.9845\pm0.0061$ & $-0.7115\pm0.0145 $&$-0.8027\pm0.0135$  &$-0.8557\pm0.0112$  &$-0.9323\pm 0.0065$ \\
 \hline

 \multirow{2}{2em}{$\mathcal{\overline{A}}^{\,\ell}_{\text{FBL}}$}
 &$ -0.2226 \pm0.0942$&$ -0.2556\pm 0.1019$&$ -0.2748\pm 0.1024$&$ -0.2703\pm0.1017 $&$-0.2453\pm0.0987$  \\
 
 & $-0.3223\pm0.0339$ & $-0.3256\pm0.0325 $& $-0.3237\pm0.0333$ & $-0.3228\pm0.0336$ &$-0.3316\pm0.0340 $ \\
 \hline

 \multirow{2}{2em}{$\mathcal{\overline{A}}^{\,\ell \Lambda }_{\text{FBL}}$}
 & $0.0722\pm0.0302$ &$ 0.0825\pm0.0302$ & $0.0889\pm0.0328$ & $0.0872\pm0.0325 $&$0.0792\pm0.0315 $ \\
 
 & $0.1462\pm 0.0085$&$ 0.1466\pm0.0082$ &$0.1468\pm0.0084 $ & $0.1463\pm0.0085$ &$0.1452\pm 0.0084$ \\
 \hline

 \multirow{2}{2em}{$\mathcal{\overline{A}}^{\Lambda }_{\text{FBL}}$}
 &$0.2245\pm0.0411 $ & $0.1165\pm0.0542$ & $0.1256\pm0.0592 $& $0.1438\pm 0.0580$&$0.1803\pm0.0517$  \\
 
 & $ 0.2373\pm0.0123$&$0.1716\pm 0.0096$ &$ 0.1983\pm 0.0107$ & $0.2052\pm0.0112$&$0.2291\pm0.0111$  \\
 \hline

 \multirow{2}{2em}{$\mathcal{\overline{K}}_{\text{2ccL}}$}
 & $ 0.0136\pm0.0362$&$ -0.0228\pm0.0326$ & $-0.0249\pm0.0357 $&$-0.0194\pm0.0362$  &$-0.0046\pm0.0361$ \\
 
 & $ 0.1309\pm0.0093$& $ 0.0953\pm0.0105$&$0.1063\pm0.0108$  &$0.1136\pm0.0107$  & $0.1267\pm0.0102$ \\
 \hline

 \multirow{2}{2em}{$\mathcal{\overline{K}}_{\text{2ssL}}$}
 &$ 0.2166\pm0.0398$ & $0.1264\pm0.0467$ &$0.1369\pm0.0508 $ &$0.1511\pm0.0504 $ &$0.1813\pm0.0506$  \\
 
 &$0.1703\pm0.0123$  & $0.1233\pm0.0102$ & $0.1396\pm0.0113$ &  $0.1486\pm0.0117$& $0.1647\pm0.105$ \\
 \hline

 \multirow{2}{2em}{$\mathcal{\overline{K}}_{\text{4sL}}$}
 &$0.0189\pm0.0738 $ &$0.0205\pm 0.0624$ &$0.0163\pm0.0639$&$0.0156\pm0.0658$  &$0.0255\pm 0.0690$ \\
 
 &  $-0.1275\pm0.0178$& $-0.1261\pm0.0167$ &$-0.1272\pm0.0173 $&$-0.1276\pm0.0176$  &$-0.1207\pm0.0182 $ \\
 \hline

 \multirow{2}{2em}{$\mathcal{\overline{K}}_{\text{4scL}}$}
 & $-0.0031\pm0.0741 $& $-0.0073\pm0.0608$ &$-0.0071\pm0.0667$  & $-0.0075\pm0.0692 $&$-0.0096\pm 0.0718$ \\
 
 &$0.0203\pm0.0057$  & $0.0182\pm0.0072$ & $0.0211\pm0.0061$ &  $0.0211\pm0.0059$&$0.0216\pm0.0059$  \\
 \hline\hline
\end{tabular}}
\end{table}

\begin{table}[ht!]
\centering
\caption{SM and NP binned predictions in $q^2=[0.045-8.0]$ GeV$^2$, and $q^2=[15.0-20.27]$ GeV$^2$ bins, for the asymmetry observables in the case of transversely polarized lepton, as shown in FIG.~\ref{TNPplot}. For each observable, the first and the second row predictions correspond to $q^2=[0.045-8.0]$ GeV$^2$, and $q^2=[15.0-20.27]$ GeV$^2$ bin, respectively. The listed errors arise mainly due to the uncertainties in the form factors.}\label{Bin1-analysis004}
\renewcommand{\arraystretch}{1.5}
\resizebox{\textwidth}{!}{%
\begin{tabular}
{|M{1.8cm}|M{2.9cm}M{2.9cm}M{2.9cm}M{2.9cm}M{2.9cm}|}
 \hline\hline
Observables & SM & S1 & S2 & S3 & S4  \\
 \hline\hline
\multirow{2}{2em}{$\left\langle P_T\right\rangle$} 
 & $-0.1794\pm0.0036$ &$ -0.1416\pm 0.0033$&  $-0.1705\pm0.0030$& $-0.1782\pm 0.0030$&$-0.1774\pm0.0038$\\
 
 &  $-0.0309\pm0.0004$&  $-0.0185\pm0.0003$& $-0.0217\pm 0.0004$&$-0.0237\pm0.0004$  &$-0.0277\pm0.0004$  \\
 \hline

 \multirow{2}{2em}{$\mathcal{\overline{A}}^{\,\ell}_{\text{FBT}}$}
 & $0.0392\pm0.0010$ & $0.0294\pm0.0010$ & $0.0316\pm 0.0010$& $0.0338\pm0.0010$ &$0.0365\pm0.0010$ \\
 
 & $0.0024\pm0.0002$ & $0.0016\pm0.0002 $& $0.0029\pm0.0002 $&
 $0.0027\pm0.0002$ &
 $ 0.0024\pm0.0002 $\\
 \hline

 \multirow{2}{2em}{$\mathcal{\overline{A}}^{\,\ell \Lambda }_{\text{FBT}}$}
 &$-0.0121\pm0.0035$  & $ -0.0084\pm0.0026$& $-0.0092\pm0.0029$ &$-0.0109\pm0.0030 $ &  $-0.0115\pm0.0033$\\
 
 &$-0.0008\pm0.0004$  &$-0.0007\pm0.0003$ &
 $ -0.0008\pm0.0004$ &   $-0.0008\pm0.0004$&
 $ -0.0009\pm0.0004 $\\
 \hline

 \multirow{2}{2em}{$\mathcal{\overline{A}}^{\Lambda }_{\text{FBT}}$}
 &$0.0573\pm0.0203$  &$ 0.0452\pm0.0181$& $0.0544\pm0.0198 $&$ 0.0576\pm0.0201$ &  $0.0567\pm0.0200$\\
 
 & $0.0145\pm0.0008$ & $0.0153\pm0.0006$ &$0.0156\pm 0.0007$ &$0.0157\pm 0.0007$ & $0.0157\pm0.0008$ \\
 \hline\hline
\end{tabular}}
\end{table}

\begin{table}[htbp!]
\centering
\caption{The SM binned predictions in $q^2=[0.045-8.0]$ GeV$^2$, and $q^2=[15.0-20.27]$ GeV$^2$ bins, for the polarization asymmetries of the new real angular coefficients for the normally polarized muon ($\overline{\mathcal{K}}_{\text{3cN}}$, $\overline{\mathcal{K}}_{\text{3N}}$), and the transversely polarized muon ($\overline{\mathcal{K}}_{\text{4ssT}}$, $\overline{\mathcal{K}}_{\text{4ccT}}$, $\overline{\mathcal{K}}_{\text{4cT}}$, $\overline{\mathcal{K}}_{\text{4T}}$), as shown in Figs.~\ref{Nspin} and \ref{Tspin}, respectively. The listed errors arise mainly due to the uncertainties in the form factors.}\label{Bin1-analysis005}
\renewcommand{\arraystretch}{1.3}
\resizebox{\textwidth}{!}{%
\begin{tabular}{|c| c c c c c c|}
\hline \hline
&$\mathcal{\overline{K}}_{3\text{cN}}$ & $\mathcal{\overline{K}}_{3\text{N}}$ & $\mathcal{\overline{K}}_{4\text{ssT}}$ & $\mathcal{\overline{K}}_{4\text{ccT}}$  & $\mathcal{\overline{K}}_{4\text{cT}}$ & $\mathcal{\overline{K}}_{4\text{T}}$    \\
\hline\hline 
$[0.045-8.0]$   & $-0.0004\pm0.0075$   &$ -0.0019\pm0.0083  $ &  $-0.0019\pm0.0003$   & $-0.0005\pm0.0041$  &$0.0014\pm0.0132$   &$0.0009\pm0.0045$   \\
$[15.0-20.27]$   & $0.0006\pm0.0008$  &
$-0.0067\pm0.0002 $    & $0.0004\pm0.0001$   & $-0.0004\pm0.0006$ & $0.0064\pm0.0008$& $-0.0002\pm0.0004$  \\
\hline \hline
\end{tabular}}
\end{table}

\clearpage

\bibliographystyle{refstyle}
\bibliography{references}
\end{document}